\documentclass{article}

\usepackage{url}
\usepackage{algorithm}
\usepackage[noend]{algpseudocode}
\usepackage{amsmath}
\usepackage{amsfonts}
\usepackage{amssymb}
\usepackage{latexsym}
\usepackage{graphicx,geometry}
\usepackage{subfigure}

\usepackage{color}

\def\VIZ#1{(\ref{#1})}      % use for references to formulae

\newcommand{\rightbararrow}{\mbox{\setlength{\unitlength}{0.1em}%
		\begin{picture}(16,10)%
		\put(14,0){\line(0,1){6}}%
		\put(14,3){\line(-1,0){10}}%
		\end{picture}
	}
}

\title{A Unified Approach for Sparse Dynamical System Inference from Temporal Measurements}

\author{Yannis Pantazis\thanks{Institute of Applied and Computational Mathematics, FORTH, Heraklion, 70013, Greece} \and Ioannis Tsamardinos\footnotemark[1] \thanks{Computer Science Department, University of Crete, Heraklion, 70013, Greece and Gnosis Data Analysis PC, Palaiokapa 64, 71305, Heraklion, Greece}}

\begin{document}

\maketitle

\abstract{\textbf{Motivation:} Temporal variations in biological systems and more generally in natural sciences are typically modelled as a set of Ordinary, Partial, or Stochastic Differential or Difference Equations. Algorithms for learning the structure and the parameters of a dynamical system are distinguished based on whether time is discrete or continuous, observations are time-series or time-course, and whether the system is deterministic or stochastic, however, there is no approach able to handle the various types of dynamical systems simultaneously. \\
\textbf{Results:} In this paper, we present a unified approach to infer both the structure and the parameters of nonlinear dynamical systems of any type under the restriction of being linear with respect to the unknown parameters. Our approach, which is named Unified Sparse Dynamics Learning (USDL), constitutes of two steps. First, an atemporal system of equations is derived through the application of the weak formulation. Then, assuming a sparse representation for the dynamical system, we show that the inference problem can be expressed as a sparse signal recovery problem, allowing the application of an extensive body of algorithms and theoretical results.  Results on simulated data demonstrate the efficacy and superiority of the USDL algorithm under multiple interventions and/or stochasticity. Additionally, USDL's accuracy significantly correlates with theoretical metrics such as the exact recovery coefficient. On real single-cell data, the proposed approach is able to induce high-confidence subgraphs of the signaling pathway.\\
%\textbf{Availability:} Source code is available at  \textit{Bioinformatics} online. USDL algorithm has been also integrated in SCENERY (\url{http://scenery.csd.uoc.gr/}); an online tool for single-cell mass cytometry analytics. \\
%\textbf{Contact:} \href{pantazis@iacm.forth.gr}{pantazis@iacm.forth.gr} and \href{tsamard@csd.uoc.gr}{tsamard@csd.uoc.gr} \\
%\textbf{Supplementary information:} Supplementary materials are available at \textit{Bioinformatics} online.
 }

\maketitle

\section{Introduction}

The great majority of biological processes are time-varying requiring the use of dynamical models for their quantitative description. Examples range from macroscopic processes studied for instance in epidemiology and population dynamics to microscopic processes such as biochemical reactions and gene regulation in a living cell, all of which are modeled as time-varying dynamical systems of complex interactions \cite{Newman2014}. Learning the structural form and the parameters of a dynamical system allows one to predict not only the evolution of the system but also the effects of manipulation and perturbation.
Depending on the characteristics of the biological system under study as well as on the available measurements, a palette of dynamical model formalisms have been successfully applied. For deterministic processes typical types of models include difference equations (when time is modeled as discrete), ordinary differential equations (ODEs) for continuous time, and their space-extended counterpart, partial differential equations. For stochastic processes, typical types of models include Markov Chains such as auto-regressive and moving averages (for discrete time), and, stochastic differential equations for continuous time. Due to the absence of a unifying mechanism, inferring the structure and the parameters of a dynamical system depends on the underlying formalism requiring specialized techniques and algorithms.

Several approaches that learn the structural form of an ODE system have been presented in the literature. One of the first attempts was the Inferelator \cite{Bonneau2006} which is an algorithm for 
de novo learning of parsimonious regulatory networks from systems-biology data sets using a shrinkage operator that induces sparsity. ODEion \cite{Gennemark2014} searches over the model space and performs an optimization via simulation approach while SINDy is another sparsity-induced algorithm \cite{Brunton2016,Mangan2016}. Sparse regression dedicated on single-cell data have been also proposed \cite{Klimovskaia2016}. Generally, sparse solutions are obtained through convex relaxation approaches \cite{Tropp2006} such as linear programming or Lasso \cite{Tibshirani1994} and, not surprisingly, such optimization programs have been studied extensively in structure inference and network reconstruction \cite{Gustafsson2009,August2009,Friedman2008,Charbonnier2010,Bolstad2011}.
Bayesian inference approaches \cite{Friston2003,Daniels2015} have been also utilized for structure learning and phenomenological modeling of deterministic dynamical systems. Bayesian methods can tackle unmeasured (latent) variables on the cost of increased computational demands due to the substantial sampling of the model space. Moreover,  theoretical guarantees on the performance is hard to obtain for Bayesian methods. Studies on stochastic dynamical systems prove under appropriate assumptions that the true sparse stochastic dynamics is guaranteed to be inferred \cite{Bolstad2011,Bento2010}. The prominent feature of these studies is that the dynamical system is linear with respect to both the unknown parameters and the state variables. Given the structure of the dynamical system, parameter estimation algorithms are also partitioned according to the type of the dynamical system \cite{DiStefano2015}.

In this paper, we present USDL (Unified Sparse Dynamics Learning) algorithm which is a novel approach to infer both the structure and the parameters of any dynamical system from temporal measurements.
First, by employing the {\it weak formulation} \cite{Strang2008,Davis1984},  the problem of inducing the structure of a dynamical model is transformed into an equivalent yet atemporal learning problem.  The weak formulation can be intuitively understood as a projection operator that multiplies the dynamical system's equation by an arbitrary function, called a test function, and, then integrates over time and/or space. The weak formulation can be also thought as a type of feature extraction. %This is similar to other methods that fit trajectories on the temporal data.
The weak formulation has several advantages: (a) by using integration by parts, the weak formulation does not require the computation of the derivatives of the trajectories; in contrast, other methods compute numerically the derivatives thus amplifying the noise and deteriorating the reconstruction accuracy, (b) by suitable definition of the test functions, the same algorithm can be applied to almost any type of dynamical system, thus, {\it unification} across different families of dynamical models is achieved, and, (c) the weak formulation transforms the dynamical system into a {\it linear system of equations} where the time and/or space dimensions have been completely eliminated.

Second, we assume sparsity of the solution which results --in combination with the weak formulation-- into a well-posed, well-studied problem in computer science, namely sparse signal recovery \cite{Candes2006,Bruckstein2009} also known as compressed sensing \cite{Donoho2006,Foucart2013}. Sparsity in our context means that the dynamics of each state variable are typically driven by a relatively small number of variables. Sparsity is critical for learning large systems from finite data and constitutes a form of complexity penalization and regularization thus favoring simpler solutions. 
The reformulation of the problem as a sparse signal recovery problem allows us to straightforwardly apply a large vividly-evolving body of theoretical and algorithmic results. Specifically, we choose the Orthogonal Matching Pursuit algorithm \cite{Pati1993,Davis1997,Tropp2007} which is a greedy and fast algorithm for recovering the sparse solution while  theoretical guarantees on the correctness of the learned solution are provided based on the mutual incoherence parameter \cite{Donoho2001} and the exact recovery coefficient \cite{Tropp2006}. The presented examples reveal that the above (in)coherency metrics and especially the latter are informative indicators of the reconstruction accuracy as measured by the precision and recall curves.
Furthermore, multiple interventions typically convey crucial information about the true structure of a biological dynamical system \cite{Sachs2005}. The proposed USDL algorithm is capable of handling not only observational but also interventional data, a property that distinguishes it from the existing sparsity-induced approaches.

Finally, we compare USDL algorithm with SINDy \cite{Brunton2016} and despite the fact that SINDy occasionally converges faster than USDL with respect to the amount of data as in Lorenz96 model \cite{Lorenz1996}, we demonstrate that our approach performs better in terms of accuracy in both interventional time-course data and stationary stochastic time-series as shown in the Results section by the respective protein networks and multidimensional Ornstein-Uhlenbeck process given sufficient amount of temporal data. The merit of the weak formulation is notably highlighted at the stationary regime of the Ornstein-Uhlenbeck process where an educated guess of test functions resulted in perfect reconstruction of the stochastic dynamical system showing the plasticity and generality of USDL algorithm which stems from the plethora of choices for the test functions.

%\vspace*{-12pt}

%\begin{methods}
\section{Methods}

\subsection{Weak Formulation}
%\label{weak:form:sec}

The weak formulation has been employed primarily in the field of applied mathematics where it provides a rigorous theoretical framework to define solutions that are not necessarily differentiable \cite{Evans1998}. Another important application of the weak formulation is found in numerical analysis. Particularly, the finite elements method which is a numerical technique for estimating approximate solutions of dynamical systems is based on it \cite{Strang2008}. In our setting, the solutions are given as measurements while the system of differential equations is unknown and has to be inferred. Thus, instead of applying the weak formulation to approximate solutions, we use it to transform the problem of learning the structure of a dynamical system into an equivalent yet atemporal learning problem.

For clarity purposes, the weak formulation is presented for ODEs which is a particular example of dynamical systems. Let $x=x(t)\in\mathbb R^N$ be an $N$-dimensional vector function of time which represents the state variables while $N$ is the number of state variables. Following physics notation (i.e., $\dot{x}:=\frac{dx}{dt}$), a system of ODEs with linear parameters is defined as $\dot{x} = A \psi(x)$, $x(t_0)=x_0$ where $t_0$ is the starting time instant with initial value $x_0\in\mathbb R^N$. $A\in\mathbb R^{N\times Q}$ is the unknown and usually sparse connectivity (or coefficient or parameter) matrix to be estimated. Dictionary, $\psi(\cdot)$, is a (given) $Q$-dimensional vector-valued vector function, $\psi:\mathbb R^N\to\mathbb R^Q$ which contains all the predetermined candidate functions that might drive the dynamics. Candidate functions are usually powers, cross-products, fractions, trigonometric, exponential or logarithmic functions of the state variables leading to nonlinear dynamical systems.  The completion of the dictionary, which could be guided by the presented incoherence metrics, is typically an application- and/or user-specific problem. %The highlighted incoherence metrics offer an evaluation of the dictionary in terms of their reconstruction accuracy.
Element-wise, the ODE system can be equivalently rewritten in a non-matrix form as
\begin{equation}
	\dot{x}_n = \sum_{q=1}^Q a_{nq} \psi_q(x)\ , \ \ \ x_n(t_0)=x_{0n}\ , \ \ \ n=1,...,N .
	\label{ODE:eq2}
\end{equation}

An example from the theory of biochemical reaction networks \cite{Lente2015} is presented in Table~\ref{rxn:net:psi:table}. Under mass action kinetics law and depending on the number of reactants in a chemical reaction network, two candidate dictionaries are shown. If the user assumes that only single-reactant reactions occur then the dictionary is simply the identity function (i.e., $\psi(x)=x$) since the reaction rates are linear with respect to the state variables for this case. However, if the user assumes that two-reactant reactions occur then the dictionary is augmented with all quadratic terms as shown in the last row of Table~\ref{rxn:net:psi:table}.
%Notice that explicit solution of \VIZ{ODE:eq2} does exist for the case of single-reactant reactions since the ODE system is linear and the solution can be obtained using Laplace transform and eigenvalue analysis. However, when there are more than one reactants, the resulted ODE system is nonlinear and no explicit solution is available.

\begin{table}
	\begin{center}
		\caption{Different choices for the dictionary, $\psi(x)$, according to the allowed chemical reaction types under mass action kinetics law. Symbols $X_1$--$X_N$ correspond to the state variables (a.k.a. (reaction) species in chemistry and systems biology). The row vector $x_{i:j}$ is defined as $x_{i:j} := [x_i, x_{i+1}, ..., x_j]$. For the two-reactant case, the dynamical system is nonlinear with respect to the state variables.
			%		In SI, particular ODE systems are provided for each case.
		} {\small
		\begin{tabular}{|c|c|c|} \hline
			Unknown reactions  & Dictionary, $\psi(x)$ & Size, $Q$ \\ \hline \hline
			$X_i \to X_{i'}$ & $x$ & $N$ \\ \hline
			{ $ X_i+X_j\to X_{i'}+X_{j'}$} & { $ [x, x_1x_{1:N}, x_2x_{2:N},...,x_N^2]^T$} & {$ (N+3)N/2$} \\ \hline
		\end{tabular}}
		\label{rxn:net:psi:table}
	\end{center}
\end{table}

In order to derive the (finite) weak formulation, a set of $M$ test functions denoted by $\{{\phi}_m(t)\}_{m=1}^{M}$ has to be specified. The test functions are smooth, not necessary orthogonal and can be chosen from a large repository of functions. Typical examples are polynomials, splines, Fourier modes (i.e., sines and cosines with varying frequency) or kernel functions from other integral transforms.  Fourier modes are preferred when trajectories exhibit periodicities while splines are more appropriate when localized phenomena have to be highlighted. As we will show later, customized families of test functions may be required in order to reveal imperceptible variable interactions. Proceeding, let $T$ be the final time and without loss of generality assume that $t_0=0$. Denoting by $\langle f,g \rangle := \int_{0}^{T} f(t)g(t) dt$ the inner product between two functions $f$ and $g$ belonging to the $L^2$ function space, define the $M$-dimensional vector $z_n$ whose $m$-th element is given by $z_{nm}:=\langle\dot{x}_n, \phi_m\rangle$, the $M\times Q$ dictionary matrix $\Psi$ whose $(m,q)$-th element is given by $\Psi_{mq}:=\langle\psi_q(x),{\phi}_m\rangle$ and let $a_n$ be a $Q$-dimensional vector which corresponds to the $n$-th row of matrix $A$. Then, the weak form of the ODE system in eq. \VIZ{ODE:eq2} is
% (for a detailed derivation we refer to the Supplementary Materials)
\begin{equation}
	z_n=\Psi a_n\ , \ \ \ n=1,...,N .
	\label{weak:ODE:eq2}
\end{equation}
In Appendix A, we provide the detailed derivation of the weak formulation for ODEs as well as for other types of dynamical systems.
%Explicit families of test functions are also discussed there.

The intuition behind weak formulation is that it projects the solution of a dynamical system to a finite-dimensional vector (i.e., to a set of linear functionals in mathematical language) whose elements are defined from the inner product between the solution and the test functions.
A key advantage of the weak formulation is that appropriate test functions for various dynamical systems such as partial and/or stochastic differential equations exist and can be utilized  transforming again the dynamical inference problem to an atemporal/aspatial problem similar to eq. \VIZ{weak:ODE:eq2}. Thus, {\em unification} of the structure inference problem for various dynamical systems is achieved. Moreover, the original problem where time is continuous is transformed from an infinite dimensional (meaning that eq. \VIZ{ODE:eq2} should be satisfied for all $t\in[0,T]$) to a finite dimensional one. 
% This is achieved by projecting the dynamical system's solution to the set of test functions.  
Additionally, and more importantly from a practical viewpoint, {\em there is no need to numerically estimate the time derivatives of the state variables}. Indeed, exploiting the integration by parts formula, it is straightforward to obtain that $$z_{nm}=x_n(t){\phi}_m(t)\Big|_{0}^{T} - \langle x_n, \dot{{\phi}}_m\rangle \ ,$$
and since test functions, ${\phi}_m$, are explicitly know, their differentiation is exact.
In general, differentiation amplifies the noise of a signal resulting in high variance estimates of the derivatives deteriorating the performance of any inference approach which necessitates the use of derivative approximation and regularization. Finally, when new trajectories (or time-series) from different initial conditions are obtained, they can be easily incorporated into the formulation by straightforward concatenation. Indeed, if $P$ trajectories, $\big\{x^{(p)}(t)\big\}_{p=1}^P$, are provided and expanding both 
$$z_n = \begin{bmatrix}
z_n^{(1)} \\ \vdots \\ z_n^{(P)}
\end{bmatrix}
\in\mathbb R^{MP} \ \ \text{and} \ \
\Psi = \begin{bmatrix}
\Psi^{(1)} \\ \vdots \\ \Psi^{(P)}
\end{bmatrix}
\in\mathbb R^{MP\times Q}$$
%with typically $MP\gg Q$
with $z_{nm}^{(p)}=\langle\dot{x}_n^{(p)}, \phi_m\rangle$ and $\Psi_{mq}^{(p)}=\langle\psi_q(x^{(p)}),{\phi}_m\rangle$, then, eq. \VIZ{weak:ODE:eq2} is still valid.

\subsection{Type of Measurements}

In the weak formulation, the estimation of the integrals (i.e., the inner products between functions) from the temporal data is required. There are two major categories of temporal data that we consider depending whether the same object is repeatedly measured or not. For the case of repeated measurements, the same object is measured sequentially over a time interval hence a time-series is constructed at the sampling points. When the sampling frequency is high enough, the collected time-series can be considered as continuous over time. For repeated measurements and deterministic systems, standard techniques such as trapezoidal rule, Simpson's rule are utilized for the numerical integration. When measurements are far from each other, interpolation between the time points can be applied. Additionally, there is no need for equispaced sampling since these methods can handle uneven sampling. For stochastic integrals, numerical integration requires different treatment since the definition of the integrals is different (e.g., Ito integral), nevertheless, numerical methods do also exist for this case \cite{Oksendal1985}.

Non-repeated measurements --we also refer to them as time-course data-- measure at each time instant a different object. This may happen because the object is destroyed during the measurement process as, for instance, in mass cytometry (see the demonstration examples below) and the same object cannot be measured more than once. However, it is assumed that all the measured objects are drawn from the same distribution. For time-course data, time-series cannot be directly constructed from the data. In order to create time-series from the time-course data, the collocation method \cite{Ramsay2007} in conjunction with a trajectory smoothing penalty \cite{Zhan2011,Craven1978} are utilized. In the collocation method, a time-series is approximated by a weighted sum of basis functions.
%The underlying assumption is that the overall noise is Gaussian thus a least-squares problem is defined which results in a linear optimization problem for the weight vector.

Moreover, time-course data from different experiments might be available. Each experiment might perform interventions to some or all variables hence each variable has its own time-series which resembles the so called multiple shooting method \cite{Peifer2007} and therefore each variable has its own trajectory per experiment. Interventional data are important in structure inference because they often reveal critical information about the biological system under study. Details on the collocation method and  how to deal with interventions (e.g., inhibition) using the multiple shooting method can be found in Appendix B and C, respectively.

\subsection{Sparse Signal Recovery}
%\label{ssr:sec}

The weak formulation enables us to transform a dynamical system inference problem to the identification of the sparse solution of eq. \VIZ{weak:ODE:eq2} which belongs to the well-studied field of sparse signal recovery (SSR). In particular, the structure inference is transformed to the problem of finding the support of the sparse solution. In SSR, the goal is to minimize the $l_0$ quasi-norm of the coefficient vector, $a$, given that $||z-\Psi a||_2 < \epsilon$ where $\epsilon$ is a predefined tolerance\footnote{For the shake of simplicity, we drop the dependence on $n$ for the rest of the section.}. Performing the minimization is computationally feasible in small scale systems but, in general, it is an intractable problem since it grows exponentially, hence, alternative approximation methods have been developed. One type of approximation is based on the so-called convex relaxation approach where the above optimization is replaced by a convex program \cite{Tropp2006}. Convex relaxation is appealing because the optimization can be completed in polynomial time using standard software. General conditions under which the convex relaxation program returns the right answer has been presented in the literature \cite{Tropp2006}. In the presence of noise, additional assumptions are imposed on the strength of the components' coefficient in order to achieve perfect reconstruction with high probability. In Appendix D, we present two commonly used convex relaxation formulations namely $l_1$ error where instead of minimizing the $l_0$ norm, the $l_1$ norm is minimized and Lasso where an $l_1$ norm regularization term is added to the quadratic cost functional. Another family of techniques that solves the SSR problem is greedy algorithms such as matching pursuit \cite{Mallat1993} and orthogonal matching pursuit (OMP) \cite{Pati1993,Davis1997,Tropp2007}. For these greedy algorithms, the correct support of the signal is recovered under suitable assumptions which are similar to the assumptions of convex relaxation. Details on OMP as well on known theoretical results that assert under which conditions these algorithms infer the true sparse representation are provided in Appendix D.

In order for any sparse signal identification algorithm to perform perfect reconstruction both the degree of collinearity among the columns of $\Psi$ and the signal-to-noise ratio have to be properly controlled. Mutual incoherence parameter (MIP) first introduced in \cite{Donoho2001} which is defined by
\begin{equation}
	\mu(\Psi) := \max_{1\le q,q' \le Q, q\ne q'} \frac{|\psi_{q}^T\psi_{q'}|}{||\psi_{q}||_2||\psi_{q'}||_2}
\end{equation}
is a measure of similarity between the columns of $\Psi$. In the extreme cases, $\mu(\Psi)=0$ when the matrix is orthogonal while $\mu(\Psi)=1$ when for instance there are two columns of the matrix which are collinear. In Appendix D, theorems that determine under which conditions on MIP the SSR algorithms are guaranteed to return the correct solution are presented. However, MIP can be unimportantly conservative because it penalizes for the correlation of features that may not participate in the solution. Indeed, if two columns of $\Psi$ are highly similar (i.e., collinear or strongly dependent) but not part of the solution then MIP is close to one but the SSR algorithms are still expected to estimate the right solution. This can be circumvented with exact recovery coefficient (ERC) which does not measure the collinearity between two columns in general but measures only the collinearity of the subspace defined by the solution with respect to each column not in the solution. The definition of ERC given a set of indices, $\mathcal S \subset \{1,...,Q\}$, is \cite{Tropp2004}
\begin{equation}
	ERC(\mathcal S) := 1 - \max_{q'\notin \mathcal S} || (\bar{\Psi}_{\mathcal S}^T \bar{\Psi}_{\mathcal S})^{-1} \bar{\Psi}_{\mathcal S}^T \bar{\psi}_{q'}||_1
\end{equation}
where $\bar{\psi}_q := \frac{\psi_q}{||\psi_q||_2}$ are the normalized dictionary atoms (i.e., normalized columns of $\Psi$) while $\bar{\Psi}_{\mathcal S}$ is the matrix that contains only the columns of $\bar{\Psi}:=[\bar{\psi}_1|...|\bar{\psi}_Q]$ that are indexed by the set $\mathcal S$.
Letting $\mathcal T$ be the true index set (i.e., the support of the true signal, or, in our case, the true structure of the dynamical system), a necessary condition for the convex relaxation algorithms or for the OMP to correctly solve SSR is $ERC(\mathcal T)>0$. The difficulty for estimating ERC arises from the fact that the true index set, $\mathcal T$, is not known a priori. Nevertheless, it can be used as an a posteriori indicator of accuracy. Under noise, OMP algorithm with the standard stopping criterion returns the true solution, if it additionally holds that \cite{Cai2011}
\begin{equation}
	|a_q|\ge \frac{2}{SNR(q) ERC(\mathcal T) \lambda_{\min}(\mathcal T)} \ ,\ \ \text{for all} \ q=1,...,Q
\end{equation}
where $SNR(q) = \frac{||\psi_q||_2}{||e||_2}$ is the signal-to-noise ratio between the $q$-th dictionary atom and the error/noise vector $e:=z-\Psi a$ while $\lambda_{\min}(\mathcal T)$ is the smallest eigenvalue of the matrix $\bar{\Psi}_{\mathcal T}^T \bar{\Psi}_{\mathcal T}$. Overall, {\it monitoring these quantities are extremely informative on determining the quality and the confidence of the learned structure.} %  SM provides several known theoretical results on the conditions of these algorithms.

{\bf Remark 1:} There is a difference between our sparse identification problem and typical SSR. In standard SSR, the number of dictionary atoms (columns of $\Psi$)  is usually larger than the number of measurements (rows of $\Psi$) while the opposite is true here since multiple experiments and multiple interventions are typically performed and measured.

{\bf Remark 2:} SSR algorithms have one hyperparameter that needs to be fine-tuned. We incorporate a selection process using the F1 score, which is the harmonic mean of precision and recall, as a performance metric. We always select the value that maximizes the F1 score.

{\bf Remark 3:} Restricted isometry property \cite{Candes2005} like MIP is another a priori metric that ensures perfect reconstruction, however, like MIP, it suffers from the same problem of being too conservative. Additionally, it is computationally expensive thus we choose not to present it in detail.

%\end{methods}

\subsection{Algorithmic Summary}
Before proceeding with the demonstration examples, a coarse summary of the proposed dynamical system inference algorithm is presented below as pseudo-code.

\begin{algorithm}
	\caption{USDL - Unified Sparse Dynamics Learning}
	\begin{algorithmic}[1]
		\State {\bf Input:} Time-series or time-course measurements, dictionary, $\psi(x)$, and set of test functions, $\{\phi_m\}$.
		%		\State Choose a dictionary, $\psi(x)$.
		%		\State Choose a set of test functions, $\{\phi_m(t)\}$.
		\If {time-course measurements}
		\State Apply the Collocation method. \Comment{Time-series interpolation}
		\EndIf
		\State Compute $\Psi$ and $z_n,\ n=1,...,N$. \Comment{Weak formulation}
		\State Estimate MIP from $\Psi$.
		\For {$n=1,...,N$} \Comment{For each row of $A$}
		\State $\hat{a}_n = \text{SSR}(z_n, \Psi)$ \Comment{Solve SSR problem}
		\State Estimate $ERC(\hat{\mathcal T}_n)$.
		\EndFor
		\State {\bf Output:} $\hat{A}$, MIP and ERC.
	\end{algorithmic}
	\label{alg:summary}
\end{algorithm}
\vspace*{-12pt}

\begin{figure*}[ht]
	\centering
	\includegraphics[width=0.9\textwidth]{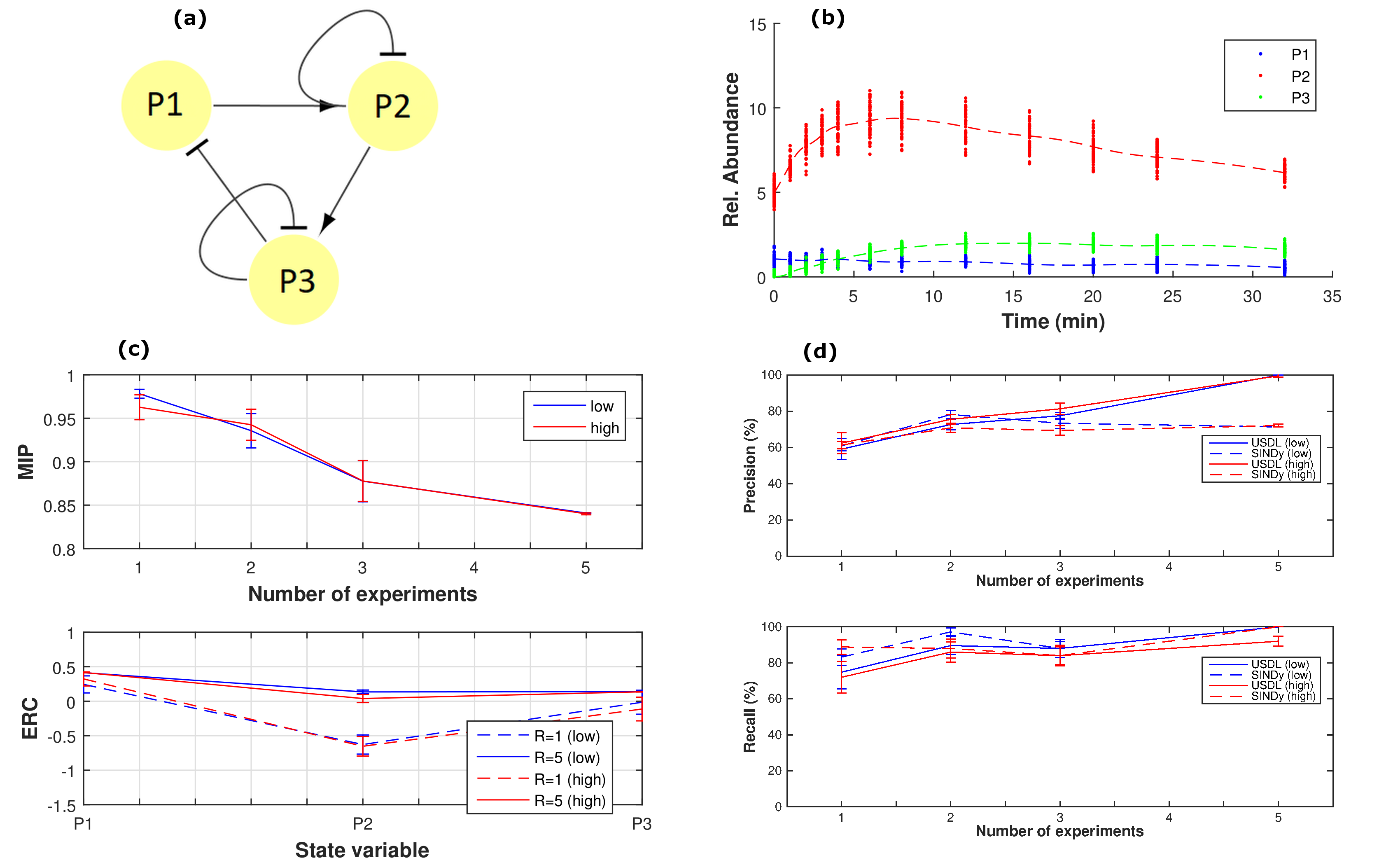}
	\caption{(a) The network of interactions between the three species (P1, P2 and P3). This graph is a coarse high-level representation and it should not be confused with the detailed biochemical reaction network which is given in Appendix E. (b) Time-course measurements (dots) and the estimated smoothed trajectories (dashed curves). The collocation method in conjunction with smoothing penalty is used for the estimation of the time-series. Notice that time-course data from the low noise case are shown. (c) MIP and ERC for P1--P3 as a function of the number of experiments under low (blue) and high (red) measurement uncertainty. Standard deviation of the various stochastic terms in high uncertainty regime is twice as much compared to the low uncertainty regime. From the lower plot, it is evident from the negativity of ERC that the problematic variable is P2 when only one experiment is used. (d) Precision and recall curves under low (blue) and high (red) measurement noise as a function of the number of experiments. Results from both USDL (solid lines) and SINDy (dashed lines) algorithms are presented. Perfect inference is achieved only with USDL under five experimental interventions and the low noise case. Precision (solid lines) seems to be insensitive to higher noise levels, however, recall slightly degrades.}
	\label{gene:reg:fig}
\end{figure*}

\section{Results}
The inference capabilities of the proposed approach for several classes of dynamical systems is presented. Source code that produces the figures is available (S1 Code).
Among the algorithms that are capable of solving SSR, we adopt OMP due to the following reasons. First, it is computationally more efficient since a forward selection of the components is performed. Second, the theoretical justifications between the various methods are very similar making the algorithms almost equivalent. Third, the hyperparameter of OMP is more intuitive compared for instance with Lasso since it is the energy of the noise term (i.e., $||e||_2$). Indeed, we approximate the noise energy as $(1+\alpha)$ times the $l_2$-norm of the residual between the signal and the complete Least Squares solution with $\alpha$ being usually a small positive number. OMP stops when relative residual energy becomes smaller than $\alpha$, therefore, OMP returns sparser solutions as we increase the value of $\alpha$ (see Appendix D for more details). Lastly, it is straightforward to incorporate prior knowledge by adding any known contribution to the dynamics, hence, it is straightforward to include data with one or more interventions with known effects.

\subsection{Protein interaction network}
%\label{protein:reg:sec}

The first demonstration is a simulation of mass cytometry measurements with a three-species prototypical protein interaction network with cycles. The complete biochemical reaction network is given in Appendix E and it has been simulated with standard ODE solver. The ground truth of interactions are shown in Fig.~\ref{gene:reg:fig}(a) where the arrow means that the source variable up-regulates the target variable while the vertical bar means that the source variable down-regulates the target variable. The experimental setup assumes that P1, P2 and P3 are measured at specific time points and every sample is destroyed during the measurement (see dots in Fig.~\ref{gene:reg:fig}(b)). In order to apply the weak formulation, time-series have to be constructed, thus, we first apply the collocation method with smoothing penalty. Dashed curves in Fig.~\ref{gene:reg:fig}(b) correspond to the estimated time-series. If, additionally, multiple experimental interventions are performed, the multiple shooting method is applied and one time-series per experiment is obtained.

Proceeding, despite the fact that the complete biochemical reaction network is nonlinear\footnote{As explained in Appendix E, the complete reaction network is unfortunately unidentifiable making it an improper example for testing the learning algorithms.} with respect to the state variables, the assumed model for inference is linear, 
$$ \dot{x} = A x \ ,$$
where vector $x(t)\in\mathbb R^3$ contains the abundance of P1--P3 at time instant $t$. Connectivity matrix, $A$, encodes the direct causal interactions within the set of state variables. Indeed, if element $a_{nq}$ is zero then no direct causal interaction exists from species $x_q$ to $x_n$. If $a_{nq}$ is positive then an increase of variable $x_q$ implies an increase of the rate of $x_n$ which results in increasing the concentration of $x_n$ thus $x_q$ activates or up-regulates $x_n$ and it is denoted with an arrow from $x_q$ to $x_n$. On the contrary, if $a_{nq}$ is negative then $x_q$ inhibits or down-regulates $x_n$ and it is denoted with a vertical bar. Thus, the structure of the network can be induced from the matrix $A$. Indeed, both the strength and the type of an interaction is inferred from the absolute value and the sign of the corresponding element of $A$, respectively.
Furthermore, it is noteworthy that several sources of error exist in this benchmark example. First, there is measurement error related to the machine limitations and it is assumed to be additive. Second, there is uncertainty error due to the fact that each measurement comes from a different cell and each cell has different concentrations of the measured quantities as Fig.~\ref{gene:reg:fig}(b) demonstrates. % Details on the amount of noise level can be found in SM.
Additional sources of error stems from the facts that (i) the complete reaction network is nonlinear for the state variables while the assumed model is not and, (ii) not all species are measured since the compound P1P3 is not quantified resulting in the existence of latent confounding variables.

\begin{figure*}[ht]
	\centering
	\includegraphics[width=0.98\textwidth]{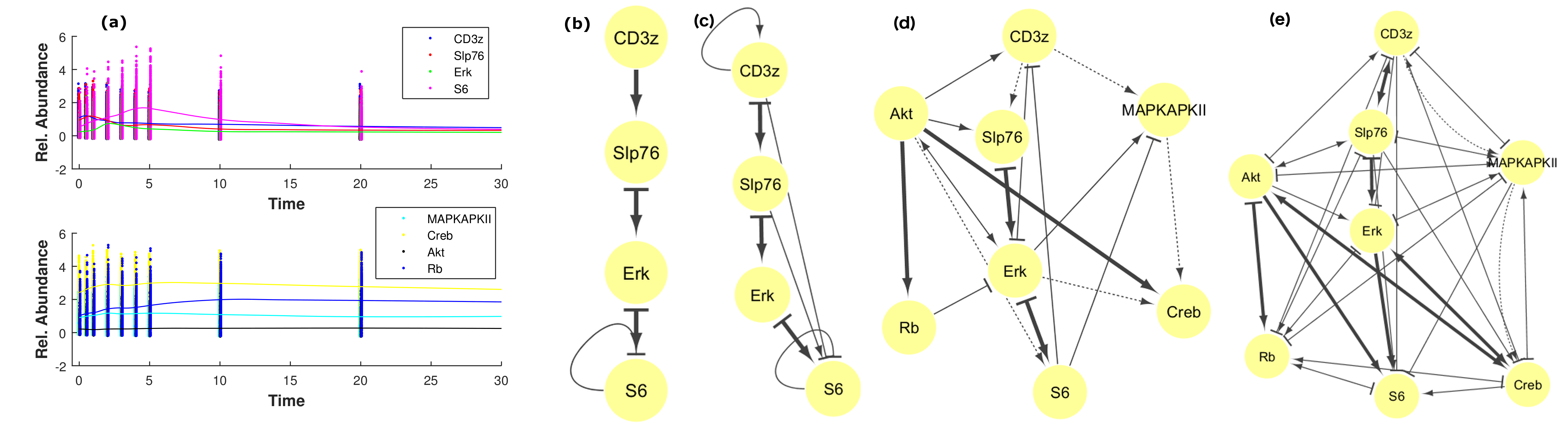}
	\caption{Network reconstruction of protein interactions from temporal mass cytometry data. (a) Time-course measurements (dots) and the estimated smoothed trajectories (solid lines). The collocation method in conjunction with smoothing penalty was used for the estimation of the time-series. Observe the high level of stochasticity of the time-course mass cytometry data. (b) The reconstructed subnetwork with four proteins using the USDL algorithm. Bold arrows (true positives) indicate that the true network of interactions is inferred. (c) Similar to (b) for SINDy. (d) The reconstructed network with eight proteins with non-bold arrows correspond to false positives while dotted arrows correspond to false negatives. Dynamics for the additional proteins vary less over time as it is evident from the lower panel of (a). Nevertheless, most of the interactions are directly (such as Akt $\to$ Rb or Erk $\to$ S6) or indirectly (like Akt $\to$ Erk $\to$ S6 instead of Akt $\to$ S6) inferred. (e) Similar to (d) for SINDy.}
	\label{dremi:fig}
\end{figure*}

Fig.~\ref{gene:reg:fig}(c) presents the MIP as a function of the number of experiments (upper plot) as well as the ERC for each state variable (lower plot). Even though MIP drops as the number of experiments is increased, the decrease is not significant making the correct inference of the interaction network a priori less certain. ERC which is an a posteriori metric, asserts that the problematic variable  is P2 when only one experiment is fed to the inference algorithm (dashed lines in lower plot of Fig.~\ref{gene:reg:fig}(c)) while ERC is positive when all five interventions are used for the inference making one crucial assumption for perfect reconstruction true. However, this is not always enough as shown be the performance of the algorithms when the measurement noise is doubled (red solid lines in Fig.~\ref{gene:reg:fig}(c)\&(d)).

Indeed, the precision-recall curves of Fig.~\ref{gene:reg:fig}(d) assert that the network is partially reconstructed when USDL (solid lines) fed with data from only one intervention is applied. In contrast, the true network is reconstructed when all five interventions are taken into account. Note that $41$ Fourier modes which correspond to a constant function, 20 sines and 20 cosines were used as test functions in USDL algorithm. Additionally, Fig.~\ref{gene:reg:fig}(d) presents the reconstruction accuracy for the SINDy algorithm (dashed lines). The hyperparameter value for both approaches is optimally selected by maximizing the F1 score which is shown in Figure 1(a) of Appendix E. When one intervention is used, the performance of SINDy is slightly better to USDL, however, SINDy does not improve its accuracy as the number of experiments increases. Evidently, SINDy is not capable of handling datasets that have multiple complex interventions. Moreover, we evaluate the forecast capabilities of the inferred model on new experiments. Prediction accuracy (Figure 1(b) in Appendix E) is in accordance with the precision-recall curves (i.e., Fig.~\ref{gene:reg:fig}(d)) in all cases revealing once again that the problematic variable is P2. Overall, this demonstration example shows the necessity of designing and executing several experiments for guaranteed perfect reconstruction of a network from non-repeated time-course measurements. In Appendix E, we present additional case studies on the performance of the proposed approach when the number of sampling points is reduced as well as when different weights for the smoothing penalty in the collocation method are applied.

{\bf Remark 4:} For small systems, a brute force alternative is tractable. A complete search of all possible solutions when the non-zero components are less than ten is computationally feasible for dictionary size up to twenty atoms. However, such an approach will provide little or no information on how to design a new experiment or a new data acquisition policy compared with greedy algorithms or convex relaxation methods where metrics such as MIP and ERC can guide the experimental designer.

\subsection{Protein network inference from mass cytometry data}

The second demonstration is the inference of protein interactions from publicly available mass cytometry data \cite{Krishnaswamy2014}. Single-cell analysis and particularly mass cytometry widely opens new directions for understanding cellular responses to perturbations and cellular functionalities due to the capability of measuring tens of proteins in each cell. Moreover, it can be multiplexed resulting in studying the cells under different conditions and time points in a relatively cheap and fast way \cite{Bodenmiller2012,Krishnaswamy2014}. Given the high resolution of single cell analysis it is expected to become a standard technique in medical sciences in the near future. In \cite{Krishnaswamy2014}, thirteen time points are sampled and measurements are separated into three subpopulations, namely, CD4+, CD8+ and Effector/Memory. Two activation cocktails which stimulate the receptors CD3/CD28 and CD3/CD28/CD4 were applied, respectively. Each experiment was repeated twice with different activation levels. Reconstructing the signaling pathway upon activation is a non-trivial task because few proteins inside the cells are measured and on top of that many interfering mechanisms with different rate are also occurring. Both result in a large number of latent confounding factors. 
Thus, it is very hard to reconstruct directly the complete system of interactions. However, network reconstruction would be more successful if restricted to subnetworks.

We perform network inference for two subnetworks with the first being  a cascade of CD3z, SLP76, Erk and S6 proteins while the second is enriched with MAPKAPKII, Creb, Akt and Rb. Fig.~\ref{dremi:fig}(a) presents the trajectories of the proteins estimated from the mass cytometry data using the collocation method with smoothing penalty. The multiple shooting method is utilized for each subpopulation and each experiment. 
We note that the collocation method assumes that the overall measurement noise is Gaussian, however, we observed that the noise in the mass cytometry data is sometimes skewed and/or multimodal potentially deteriorating the quality of the estimated trajectories. Furthermore, time-series adjustment is performed by subtracting the minimum value of each trajectory which merely corresponds to the state of no activity. As it is evident from the figure, the level of stochasticity is high making the dense sampling of the signaling phenomenon necessary.

Figs.~\ref{dremi:fig}(b)\&(c) present the reconstruction of the smaller subnetwork for USDL and SINDy algorithm, respectively, while Figs.~\ref{dremi:fig}(d)\&(e) present the reconstruction results for the larger subnetwork. We consider a linear ODE model ($\dot{x}=Ax$) excluding more complicated protein interactions. USDL algorithm utilizes 21 Fourier modes while the hyperparameter for each approach is fine-tuned based on the F1 score estimated on an independent subset of the data (see Figure 10 in Appendix E) using the KEGG database \cite{Kanehisa2000} as ground truth. The semantics of arrow and bar edges are the same as in the previous example. An edge is bold when it is also found in the KEGG database, regular if found with USDL (or SINDy) but it is not found in KEGG database while it is dotted if it is in the KEGG database but not found. Inference is repeated 100 times using a portion of the available data in each iteration and the reported edges are the ones that are found at least in 80\% of the times. Concentrated in the case with four proteins, the subnetwork is correctly reconstructed with USDL algorithm while SINDy infers two additional edges. The bar edges in Fig~\ref{dremi:fig}(b), which imply down-regulation, are explained as a mechanism to model the degradation of each variable over time. 
When four more proteins are added, the network reconstruction becomes harder. Nevertheless, USDL using CD4+ subpopulation and CD3/CD28 activator (see Fig.~\ref{dremi:fig}(d)) was able to recover half of the known edges. The cascade Slp76 $\to$ Erk $\to$ S6 is still correctly inferred. However, CS3z was replaced by Akt in the phosphorylation of Slp76. Additionally, the proposed algorithm correctly assesses the influence of Akt to the phosphorylation of both Rb and Creb showing that Akt plays a central role in pathway signaling of T cells. In contrast, SINDy with the optimal hyper-parameter returns an almost fully-connected graph resulting in a non-sparse solution.
% Are the missing edges primarily a result of having a strict threshold on the sparsity of OMP?

The difficulty in inferring the KEGG-based network of protein interactions is reflected on both MIP and ERC (reported in Appendix E). Both incoherency metrics deteriorate when more proteins are added to the analysis making the assumptions for perfect reconstruction less valid. Thus, more experiments are required in order to improve the accuracy of the network inference algorithm. These experiments can be guided by metrics such as MIP or ERC. Additional demonstrations can be found in Appendix E. Particularly, the reconstructed networks when additional activators and/or subpopulations are considered are presented as well as how important is the high sampling rate for protein interactions inference through mass cytometry data. For instance, the reconstruction accuracy is severely reduced when removing half of the time-points as it is shown in Appendix E.
Finally, we have integrated the USDL algorithm in SCENERY (\url{http://scenery.csd.uoc.gr/}) \cite{Papoutsoglou2017} which is an web tool for single-cell cytometry analysis. SCENERY provides a comprehensive and easy-to-use graphical user interface where users may upload their data and perform various types of protein network reconstruction. The incorporation of USDL algorithm into SCENERY aims to increase its reusability and, hopefully, its popularity.

\subsection{Multidimensional Ornstein-Uhlenbeck process}

The last example is a multidimensional Ornstein-Uhlenbeck process which is a system of linear stochastic differential equations with additive noise. It has applications in evolutionary biology where multiple traits are modelled over time with multidimensional Ornstein-Uhlenbeck process \cite{Bartoszek2012} as well as in particle physics \cite{Gardiner2004} and finance \cite{Gardiner2009}. Mathematically, the driving system of equations is given by 
\begin{equation}
	\dot{x} = -A x + \sigma \dot{B} \ ,
	\label{OU:proc:eq}
\end{equation}
where $x(t)\in\mathbb R^N$ is the stochastic process, connectivity matrix $A\in\mathbb R^{N\times N}$ determines the interactions between the variables, $\sigma\in\mathbb R$ corresponds to the noise level while $B(t) \in\mathbb R^N$ is an $N$-dimensional standard Brownian motion. Intuitively, the derivative of a Brownian motion can be understood as a continuous-time zero-mean white noise with variance one. We set $N=20$ while matrix $A$ is defined as the graph Laplacian with random edges and maximum outgoing degree equal to 3.

\begin{figure*}[ht]
	\centering
	\includegraphics[width=0.9\textwidth]{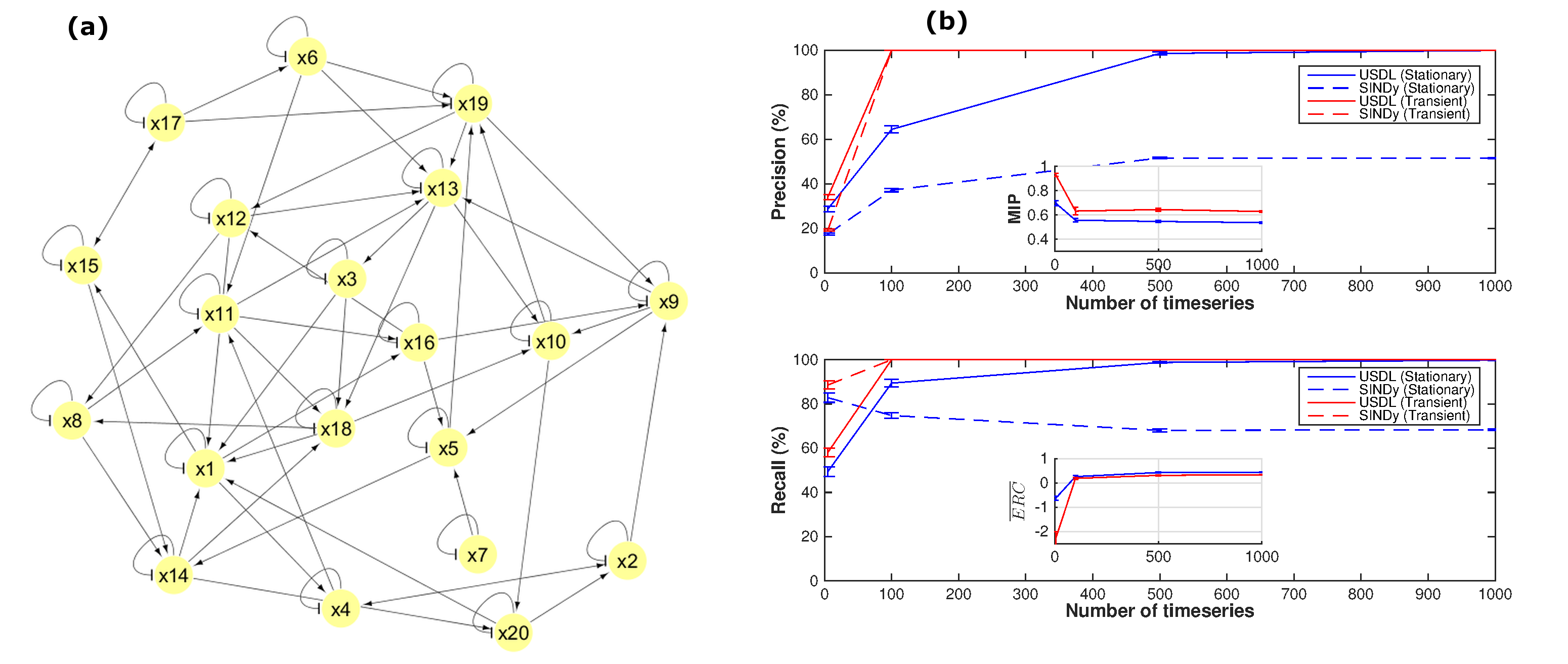}
	\caption{Performance analysis and comparison between USDL and SINDy algorithms for Ornstein-Uhlenbeck stochastic process. (a) The connectivity graph for each variable of the Ornstein-Uhlenbeck process. The edges, their direction as well as the type of interaction are determined by the non-zero elements of connectivity matrix $A$. (b) Precision and recall are shown as functions of the number of measured time-series in two different regimes; stationary (blue) and transient (red). Both USDL (solid) and SINDy (dashed) algorithms achieve perfect reconstruction of the dynamical system for the transient regime and when enough time-series are measured. For the stationary regime, perfect reconstruction is possible only for USDL and a special type of test functions (peaky Fourier modes) while SINDy (blue dashed) fails to recover completely the dynamical system in this regime due to the high stochasticity.}
	\label{OUprocess:fig}
\end{figure*}

Fig.~\ref{OUprocess:fig}(a) presents the graph of interactions for each variable of a randomly-drawn instance of $A$. We distinguish between two regimes, namely, the stationary (or equilibrium) regime and the transient regime. At stationarity, the driving force is primarily the stochastic or diffusion term (second summand in the r.h.s. of eq. \VIZ{OU:proc:eq}) with the deterministic or drift term (first summand in the r.h.s. of eq. \VIZ{OU:proc:eq}) acting as a stabilizer. In the transient regime, the dynamics are primarily driven by the drift term. This separation is of great importance because the signal in the former case is buried under the noise (i.e., both have approximately the same energy or, in other words, the signal-to-noise ratio is approximately one) while the signal is stronger compared to the stochastic term in the transient regime (see the simulated trajectories for both regimes in Appendix E, Figure 16(a)). As performance measures show in Fig.~\ref{OUprocess:fig}(b), both USDL and SINDy are able to infer the correct structure of $A$ (i.e., the correct graph of interactions) at the transient regime (red curves) when enough --approximately $P=100$-- time-series are provided. We remark here that, as in the previous examples, the hyperparameter values of both algorithms are selected based on the maximum F1 score. Averaged ERC is positive for this setup as inset plot reveals and signal-to-noise ratio is high enough to theoretically guarantee the perfect reconstruction of the dynamical system.

In the stationary regime, the driving force is the noise and positive (averaged) ERC is not enough to guarantee perfect reconstruction and control of the signal-to-noise ratio is required for true structure learning. We tested a series of typical test functions such as Fourier modes and splines as well as we varied the number of test functions, however, we were not able to perfectly reconstruct the dynamical system with the accuracy hitting a plateau (figures in Appendix E) despite the fact that theoretical results \cite{Bento2010} suggests that true recovery is possible. The problematic cases arise from variables whose time cross-correlations have similar shape and they are close to each other thus we need to define another type of test functions with the property of having sharp changes which assist the separation between small time-differences.  A well-educated choice of test functions, which we named peaky Fourier modes (details in Appendix E) result in perfect reconstruction of the connectivity matrix for the stationary regime when USDL is applied (blue solid lines in Fig.~\ref{OUprocess:fig}(b)). In contrast, SINDy algorithm (dashed blue lines) is incapable of inferring the structure of the dynamical system because it is not designed for stochastic systems and thus it is unable to handle the high intensity of the noise which is observed in the stationary regime.

Moreover, we measured the root mean squared error (RMSE) on the elements of the connectivity matrix $A$ and assess the parameter estimation behavior of the inference approaches. RMSE results (Figure 17(b) in Appendix E) reveal that two orders of magnitude lower RMSE is reported in the transient regime relative to the stationary regime. In the same figure, it is shown that the RMSE for USDL is lower than the RMSE for SINDy which is in accordance with the structure inference performance (i.e., the precision-recall curves).

%\vspace*{-12pt}

\section{Discussion}
The weak formulation enables the transformation of any spatio-temporal dynamical system that is linear with respect to its parameters into a linear system of equations. The unification through the weak formulation creates the foundations for general dynamical system inference in biological applications. For instance, in mass cytometry and more generally in single-cell analysis, not trajectories but the distribution of the species populations is measured over time. Thus, the structure learning from partial differential equations such as the Fokker-Planck or the master equation \cite{Gardiner2004} which both describe the evolution of the probability distribution of the measured quantities can be transformed into a linear set of equations. Moreover, the avoidance of differentiation through the integration-by-parts trick could benefit the already existing dynamical inference algorithms, especially, in adverse, noisy conditions. Additionally, the transformed structure learning problem can be considered not only as an SSR problem but also as a feature selection problem \cite{Guyon2003}, a subfield of machine learning and statistics. The extensive body of work on feature selection could be also employed and, therefore, boost the accuracy of the overall inference.

SSR literature offers an arsenal of theoretical indicators and metrics that we showed correlate well with the performance as quantified by the precision and recall curves. Even though there is a growing research area for dynamical system inference algorithms, limited number of attempts to compute and exploit such metrics can be found in biological studies. The presented examples revealed that the values of these metrics could be easily lay far from the theoretically desirable. For instance, MIP took values closer to one rather than to zero in the mass cytometry example necessitating the design of additional experiments or the elimination of some problematic proteins from the dictionary. Actually, the determination of the dictionary is crucial in biological inverse problem inference. Quantities that are constant over time can imperil the accuracy of a structure learning algorithm because of the addition of collinearities especially when quadratic terms are considered in the dictionary. Thus, it is preferable to remove some or all of the constant-over-time variables from the dictionary and attempt to infer the structure of the quantities that are time-varying. Both incoherence metrics can serve as a guideline for the construction of a dictionary with high potential for true recovery.

Dynamics in biological processes contain critical information about the underlying reaction mechanisms between molecules. Current technologies are able to measure several time-points increasing the possibility of inducing the interactions between the measured quantities. However, the shape characteristics of the biological dynamics is usually simple. Prominent examples are impulsive patterns, which are either up-regulating or down-regulating excitations followed by a return to their basis values, and sustained patterns where the measured quantity remains over-expressed or under-expressed after the excitation \cite{Bar-Joseph2012}. A cascade of four impulsive responses for protein signaling whose interactions were correctly inferred is shown in the upper plot of Fig.~\ref{dremi:fig}(a). Thus, the dynamical system that can be potentially identified correctly from relatively simple trajectories should not have complex driving forces. This is the primary reason why in our experiments we chose a linear dictionary. Adding more complex interactions to the dictionary will only result in less identifiable inference problems.

Finally, the presented examples assumed a linear with respect to the state variables dynamical model and, thus, a linear dictionary is constructed. However, the proposed inference algorithm is not restricted to linear differential equations. In Appendix F, we apply the proposed dynamical inference algorithm to a nonlinear and chaotic system from climate science, namely, Lorenz96 \cite{Lorenz1996} where comparisons with SINDy are also performed. The precision-recall results indicate that both methods are able to achieve perfect reconstruction. However, SINDy requires 3-4 times less data in order to succeed it for the case of moderate chaotic behaviour. Generally, chaotic systems enjoy richer and more complex dynamics which actually assist the structural learning of the differential equations as both the incoherence metrics and the obtained results on the accuracy revealed.

\section{Conclusions}
In this paper, we present the USDL algorithm, a generic and unified approach to solve the sparse dynamical structure inference problem from temporal data. It is based on the weak formulation of differential equations where the dimension of time is eliminated. Several properties of weak formulation such as being derivative free are useful and have high practical value. The transformed system is a set of linear equations whose sparse solution can be found using SSR algorithms. Convex relaxation methods as well as greedy algorithms such as OMP can be used and theoretical guarantees can be computed. To this end, a priori metrics such as MIP and a posteriori metrics such as ERC are computed and the satisfiability of the assumptions are checked. These metrics might also serve as candidates for optimization in experimental design. Moreover, various dynamical models can be transformed with the weak formulation to an SSR problem making our approach model independent. We test and compare USDL against SINDy on a wide range of dynamical models and show that, under high stochasticity, USDL achieves perfect reconstruction given enough data while SINDy fails for the same amount of data. This is notably evident for the stationary OU process where noise is prevalent revealing the generality and robustness of the proposed approach. 

%\vspace*{-12pt}

\section*{Acknowledgements}
We sincerely thank George Papoutsoglou for collecting and preprocessing the mass cytometry data and reading early drafts of the manuscript.
%\vspace*{-12pt}

\section*{Funding}
The research leading to these results has received funding from the European Research Council under the European Union's Seventh Framework Programme (FP/2007-2013) / ERC Grant Agreement n. 617393.
%\vspace*{-12pt}

\bibliographystyle{unsrt}
%\bibliographystyle{natbib}
%\bibliographystyle{achemnat}
%\bibliographystyle{plainnat}
%\bibliographystyle{abbrv}
%\bibliographystyle{bioinformatics}

%\bibliography{biblio}

\appendix

\section{Weak formulation for various differential equations}

The theory of the weak formulation has been developed in applied mathematics and particularly in the theory as well as in the numerical analysis of partial differential equations (PDEs). However, any dynamical system can be written in a weak form. This section presents the derivation of the weak formulation for ODEs, PDEs, stochastic differential equations (SDEs) and multivariate autoregressive models (MARs). Thus, a wide range of dynamical systems is covered. We also discuss various choices for the test functions. As we see in Appendix E, the choice of test functions can play a critical role in the perfect reconstruction of a dynamical system.

\subsection{Ordinary differential equations}
As presented in the main text, an ODE system with linear parameters is given in a non-matrix form by
\begin{equation}
\dot{x}_n = \sum_{q=1}^Q a_{nq} \psi_q(x)\ , \ \ \ x_n(0)=x_{0n}\ , \ \ \ n=1,...,N \ .
\label{ODE:eq}
\end{equation}
where $x_n(t)$ is the $n$-th state variable of the system while $\psi_q(x)$ is the $q$-th candidate function (or atom or element) of the dictionary.

In order to define the weak formulation, a set of test functions indexed by $m=1,...,M$ and denoted by $\phi_m(t)$ has to be specified. The number of test functions, $M$ can be infinite but for practical purposes it is always finite. For the rest of the paper, we will assume that it is finite. Then, by multiplying \VIZ{ODE:eq} with $\phi_m(t)$ and integrating from $0$ to $T$, we get for each $n$ that
\begin{equation}
\int_{0}^{T} \dot{x}_n(t)\phi_m(t) dt = \int_{0}^{T} \sum_{q=1}^Q a_{nq} \psi_q(x(t))\phi_m(t) dt\ .%, \ \ \ n=1,...,N \ .
\label{weak:ODE:eq}
\end{equation}
It is rewritten as
\begin{equation}
\langle \dot{x}_n, \phi_m\rangle =  \sum_{q=1}^Q a_{nq} \langle \psi_q(x),\phi_m\rangle \ ,%, \ \ \ n=1,...,N \ ,
%\label{weak:ODE:eq2}
\end{equation}
where $\langle f,g \rangle = \int_{0}^{T} f(t)g(t) dt$ is the inner product between two functions $f$ and $g$ in the $L_2$ function space. Notice that the infinite number of equations (one for each time instant $t\in[0,T]$) in \VIZ{ODE:eq} is transformed into $M$ equations (one for each test function). The above set of equations is an atemporal system of equations whose matrix form is given in eq. (2) of the main text.

Moreover, we assume that the test functions are smooth, having at least first derivative, hence, the differentiation of the state variables can be avoided using integration by parts. Indeed, it holds for each $n$ that
\begin{equation}
x_n(t)\phi_m(t)\Big|_{0}^{T} - \langle x_n, \dot{\phi}_m\rangle = \sum_{q=1}^Q a_{nq} \langle \psi_q(x),\phi_m\rangle \ .%, \ \ \ n=1,...,N \ .
\label{weak:ODE:eq3}
\end{equation}
This is a key feature of weak formulation both theoretically and practically. From a practical point of view, there is no need to numerically estimate the time derivatives of the state variables. As discussed, differentiation in general amplifies the noise of a signal resulting in high variance estimates of the derivatives deteriorating the performance of any inference approach which necessitates derivative approximation. The weak formulation avoids differentiating the noisy state variables by exploiting the integration by parts theorem as shown above.

\subsubsection{Test functions}
\begin{figure}%[tbhp]
	\centering
	\includegraphics[width=.8\linewidth]{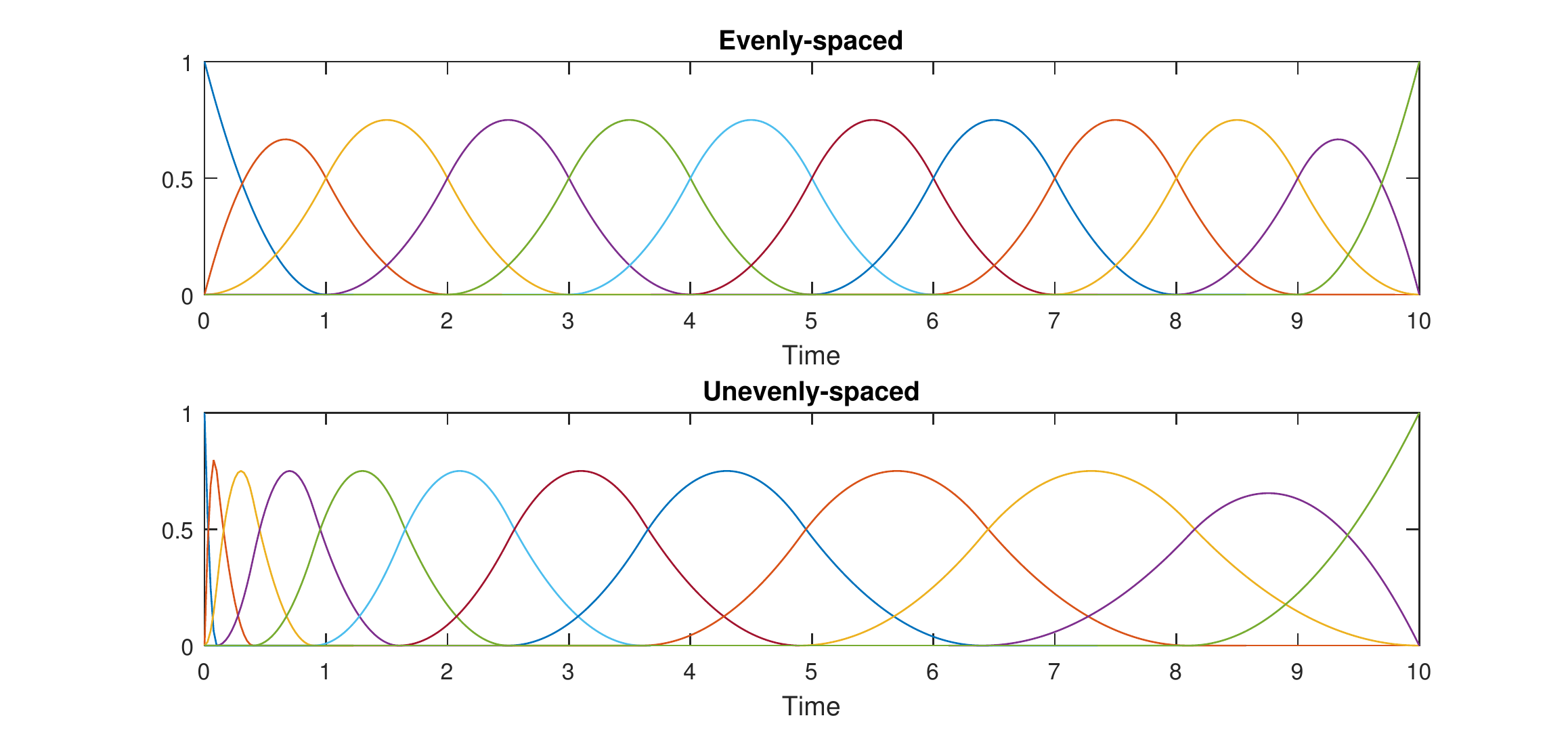}
	\caption{Upper plot: A set of 12 evenly-spaced B-splines in the interval $[0,10]$. Lower plot: A set of 12 unevenly-spaced B-splines in the interval $[0,10]$ where more resolution is put on the early times.}
	\label{b:splines:fig}
\end{figure}

The test functions utilized in weak formulation are smooth, not necessary orthogonal and can be chosen from a huge repository of functions. Examples are polynomials, Fourier/Laplace basis functions, B-splines, etc. Furthermore, the number of test functions should be minimum so as to reduce the computational complexity of the overall optimization problem. The choice of test functions could affect the performance of the inference procedure, hence, careful and educated selection is needed for optimal performance. Of course, the optimal selection of test functions depends on the specific problem at hand and there is no family of functions that will perform optimally for all cases. The most suitable test functions for phenomena that has periodicities are Fourier modes. A set of $M$ Fourier modes is defined by
\begin{equation}
\phi_1(t) = 1\ \ ,\ \ \ 
\phi_{2m-1}(t) = \cos\left(\frac{2\pi m t}{T}\right)\ \ \ \text{and} \ \ \ 
\phi_{2m}(t) = \sin\left(\frac{2\pi m t}{T}\right)\ , \ \ \ m=1,...,(M-1)/2
\end{equation}
with $T$ being the final time. In this paper, Fourier modes are used as the default choice of test functions since they form a complete basis of functions in the $L_2$ function space. Another important class of test functions are the B-spline functions. B-splines with or without equally-spaced knots constitute another set of test functions that we explore in our demonstration examples. The advantage of using unequally-spaced knots is that specific areas of interest can be indicated and exploited. For instance in cellular protein signaling, if the dynamical phenomenon is stronger at the beginning then more knots will be assigned during the early times. Figure~\ref{b:splines:fig} demonstrates a set of 12 B-spline functions both with evenly-spaced knots (upper panel) and with unevenly-spaced knots (lower panel).

A third class of test functions are data-dependent test functions which can capture optimally the variations of the measured time-series. Similar to Karhunen-Lo\`{e}ve expansion, spectral analysis of the measurement matrix reveals the principal components that explains the time-series. The principal components with the highest energy (i.e., eigenvalues) are used as test functions. The principal components are numerically computed through the use of singular value decomposition. We remark that this approach is known as principal component analysis or proper orthogonal decomposition.

\subsection{Partial Differential Equations}
PDEs constitute another class of dynamical systems where the variables depend not only on time but also on space. One of the simplest and well studied PDEs is heat equation which reads in 1D $x_t - x_{uu} = 0$ with some initial and boundary conditions. We denote with $x_t = \frac{\partial x}{\partial t}$ and $x_{uu} = \frac{\partial^2 x}{\partial u^2}$ the first order time-derivative and second-order space-derivative, respectively. For our purposes, we consider systems which are linear with respect to the unknown parameters. Restricting to one space dimension and one variable, the equation can be of the general form
\begin{equation}
\sum_{q=1}^{Q} a_{q} \psi_q(x,x_t,x_u,x_{tt},x_{tu},x_{uu},...) = 0
\label{pde:eq}
\end{equation}
where the driving functions, $\{\psi_q\}$, depend in the physical or engineering application. Several equations fall in the above general form. To name a few, transport equation (first order), heat equation, Laplace's equation (both parabolic PDEs) and wave equations with or without shocks, with or without interactions (hyperbolic PDEs). Additionally, special but general cases from reaction-diffusion equations (Fisher-Kolmogorov equation), chemotaxis, master equation as well as Fokker-Planck equation (with the last two belonging to the class of Kolmogorov equations) can be written in the general form given by \VIZ{pde:eq}.

Typically, the inference (or inverse) problems in PDEs try to estimate the boundaries or some other parameters of the PDE and not the driving forces that governs the system since the kinetic laws are assumed known. Nevertheless, in this paper, we want to show the generality of the weak formulation approach to transform a PDE into an atemporal system of equations. Interestingly, the weak formulation is best applied to finding the solution of PDEs. Proceeding, the test functions are now both time and space dependent and the integration is performed in both dimensions. So, let $\{\phi_m(t,x)\}_{m=1}^M$ be a set of $M$ test functions, then, the weak form of \VIZ{pde:eq} is
\begin{equation}
\sum_{q=1}^{Q} a_{nq} \langle\psi_q(x,x_t,x_u,x_{tt},x_{tu},x_{uu},...), \phi_m \rangle = 0
\end{equation}
where $\langle f, g \rangle = \int_{\mathcal D}\int_0^T f(t,u) g(t,u) dt du$ is the inner product adapted for functions with both time and space arguments. $\mathcal D$ is the space domain of the free variable. Notice that the integration by parts trick can be applied for both time-derivatives and space-derivatives. Thus, the need for numerical estimation of the derivatives is minimized. Moreover, a standard choice for time-space test functions is to assume that time and space are separated. Indeed, the test functions are defined as $\phi_m(t,u) = \phi_{m,1}(t)\phi_{m,2}(u)$ where $\phi_{m,1}$ and $\phi_{m,2}$ are test functions as in the previous subsection.

Finally, we would like to notice that our primary intention is to show that weak formulation is also applicable to PDEs without any difficulty in the derivation of the atemporal system. However, the study of particular PDEs may necessitate special treatment especially on the sampling scheme of the solution as well as on the suitable selection of test functions. An extensive discussion on PDE dynamical system inference is beyond the scope of this publication.

\subsection{Stochastic Differential Equations}
A stochastic process $x(t,\omega)$ is a function of two variables, time $t$ and random outcome $\omega$. For a fixed random variable $\omega$, it is a function of time, $x(t):=x(t,\omega)$ which is called a realization, time-series or trajectory of the process. In the following, we suppress the dependence on the random element and keep only the dependence on time. Proceeding, consider an $N$-dimensional stochastic process, $x(t)$ which is driven by an SDE with additive noise
\begin{equation}
\dot{x}_n = \sum_{q=1}^Q a_{nq} \psi_q(x) + \sigma \dot{B}_n\ , \ \ \ n=1,...,N ,
\label{SDE:eq}
\end{equation}
where $B_n(t)$ are independent Brownian motions for each $n=1,...,N$. Brownian motion as a function of time is nowhere differentiable thus its time derivative can be rigorously defined only through the weak formulation. However, and, in an intuitive level, the time derivative of a Brownian motion is known as the white noise, and, it is also part of the driving forces of this stochastic dynamical system. The more standard notation for SDEs is based on differentials and it is given by the following formula
\begin{equation}
dx_n(t) = \sum_{q=1}^Q a_{nq} \psi_q(x(t)) dt + \sigma d B_n(t)\ , \ \ \ n=1,...,N.
\label{SDE:eq2}
\end{equation}
As before the weak formulation is obtained by multiplication of the test function and integration. Thus, for $m = 1,...,M$, we get
\begin{equation}
\int_0^T \phi_m(t) dx_n(t) = \sum_{q=1}^Q a_{nq} \int_0^T \phi_m(t) \psi_q(x(t)) dt + \sigma \int_0^T \phi_m(t) d B_n(t)\ , \ \ \ n=1,...,N \ .
\label{SDE:weak}
\end{equation}
However there is an important difference between the standard Riemann integration used in ODEs and PDEs and stochastic integrals. The interpretation of the integrals is different which results in different approaches when numerical estimation is performed. Here, the stochastic integrals are Ito integrals \cite{Mikosch2003} which are defined in a similar manner to the Riemann-Stieltjes integrals. Numerically, Ito integral is approximated by the Riemann sum
\begin{equation}
\int_0^T f(t) dx_t \approx \sum_{i=0}^{K-1} f(t_{i}) \big( x(t_{i+1})-x(t_{i}) \big)
\end{equation}
where $\{t_i\}_{i=0}^K$ denotes a partition of the interval $[0,T]$. Notice that there exist a variant of the integration by parts theorem which holds for Ito integrals. However, we did not utilize it in the demonstrated SDE example.

\subsection{Multivariate Autoregressive Model}
The weak formulation can be also applied on discrete-time dynamical systems. The approach is similar to the continuous time except the definition of inner products where the integration is replaced by summation. We demonstrate the weak form of discrete-time dynamical systems with multivariate autoregressive (MAR) model which additionally is stochastic. In MAR models, the current measurement is a linear combination of the previous measurements plus a stochastic white noise (i.e., Gaussian and independent). The mathematical formula for MAR model is
\begin{equation}
x(t) = \sum_{\tau=1}^{\mathcal T} A(\tau) x(t-\tau) + e(t)\ ,\ \ t=0,1,...,T
\end{equation}
where $x(t)\in\mathbb R^N$ is the discrete-time $N$-dimensional MAR process, $A(\tau)$ is the connectivity matrix for the $\tau$-th previous measurement while $e(t)$ is the driving noise term. $\mathcal T$ determines the order of MAR.

The weak formulation for a discrete-time system is defined for a set of test functions $\{\phi_m\}_{m=1}^M$ whose domain is the set of natural numbers. The new atemporal system is given by
\begin{equation}
\langle x_n,\phi_m\rangle = \sum_{\tau=1}^{\mathcal T} \sum_{q=1}^Q a_{nq}(\tau) \langle x_q(t-\tau), \phi_m\rangle + \langle e_n,\phi_m\rangle
\end{equation}
where the inner product is now defined as $\langle f,g\rangle = \sum_{t=0}^T f(t)g(t)$. Defining the cross-correlation function between two functions as $C_{fg}(\tau)=\sum_{t=0}^T f(t) g(t+\tau)$, the system of weak-form equations becomes
\begin{equation}
C_{x_n\phi_m}(0) = \sum_{\tau=1}^{\mathcal T} \sum_{q=1}^Q a_{nq}(\tau) C_{x_q\phi_m}(-\tau) + C_{e_n\phi_m}(0) \ ,
\end{equation}
which is a linear system of equations. Test functions can be discrete versions of Fourier modes or B-spline functions. Using as test functions, the Dirac delta function (i.e., $\phi_m(t) = \delta(t-m)$) then the weak formulation falls back into the ``temporal'' formulation studied in \cite{Bolstad2011}.

\section{Time-series Data Fitting a.k.a. Collocation Method}

A sample of a time-course dataset (i.e., a non-repeated measurement) is denoted as
\begin{equation}
y_{nkp} = y_n^{(p)}(t_k)
\label{measure:def}
\end{equation}
where $n=1,...,N$ with $N$ being the number of species, $k=1,...,K$ with $K$ being the total number of sampling points while $p=1,...,P$ with $P$ being the number of measurements. The interpretation is that $y_{nkp}$ is the $p$-th measurement of the $n$-th species at time instant, $t_k$. We set $t_0=0$ and $t_K=T$ be the initial and final time points, respectively. Notice that the number of measurements may depend on the time index $k$ (i.e., $P=P(k)$) but for simplicity reasons we assume that it is constant. It is also useful to define the $P$-dimensional measurement vector, $y_n$, as
\begin{equation}
y_n = [y_n^{(1)}(t_1), ..., y_n^{(1)}(t_{K}), ..., y_n^{(P)}(t_1), ..., y_n^{(P)}(t_{K})]^T
\label{measure:time:course:def}
\end{equation}

The weak formulation does not apply directly when time-course data are provided. Thus, time-series interpolation is utilized. This is achieved by assuming that the state variables over time can be approximated by a linear combination of $L$ basis functions of time, denoted by $\{\bar{\phi}_l(t)\}_{l=1}^{L}$. Thus, the $n$-th state variable is written as
\begin{equation}
x_n(t) = \sum_{l=1}^L c_{nl} \bar{\phi}_l(t) = c_{n}^T \bar{\phi}(t) \ , \ \ \ n=1,...,N \ ,
\label{basis:func:expansion}
\end{equation}
where $\bar{\phi}(t) = [\bar{\phi}_1(t),...,\bar{\phi}_L(t)]^T\in\mathbb R^L$ is an $L$-dimensional vector whose elements are the basis functions. Typical choice for the basis functions are $B$-splines which are local in time and are defined by two parameters; the degree of the interpolating polynomials and the set of knots. Knots in $B$-spline construction are usually dictated by the sampling time points, $t_0,...,t_K$, but they can be enriched or reduced in a user-specific basis. If necessary, they can be also optimized through a cross-validation procedure.

In order to formulate the data fitting optimization problem, define the $L\times (K+1)P$ matrix, $\tilde{\phi}$, as
\begin{equation}
\tilde{\phi} = \begin{bmatrix}
\bar{\phi}_1(t_0) & ... & \bar{\phi}_1(t_{0}) & ... & \bar{\phi}_1(t_K) & ... & \bar{\phi}_1(t_{K}) \\
\vdots & & \vdots &  & \vdots & & \vdots \\
\bar{\phi}_L(t_0) & ... & \bar{\phi}_L(t_{0}) & ... & \bar{\phi}_L(t_K) & ... & \bar{\phi}_L(t_{K}) \\
\end{bmatrix} \ ,
\end{equation}
which is in alignment with the measurement vectors $y_n$ in \VIZ{measure:time:course:def}. The (penalized) data fitting problem is defined as
\begin{equation}
\min_{C}   \sum_{n=1}^N \big\{||y_n - c_{n}^T \tilde{\phi}||_2^2
+ \lambda_C c_{n}^T \ddot{\bar{\Phi}} c_{n} \big\}
\label{cost:func:P2a}
\end{equation}
where $C$ is the $N\times L$ coefficient or weight matrix defined by 
\begin{equation}
C = [c_{1},...,c_{N}]^T \ ,
\end{equation}
while the second term is a smoothing penalty controlled by the non-negative parameter (i.e., weight) $\lambda_C$. The second term is designed  to penalize the ripples of the interpolated time-series by penalizing the squared $L_2$ norm of the derivatives of the time-series. This reads to $\int_{0}^{T} (\dot{x}_n(t))^2 dt = \langle\dot{x}_n, \dot{x}_n\rangle$ which can be rewritten as $ c_{n}^T \ddot{\bar{\Phi}} c_{j}$ where $\ddot{\bar{\Phi}}$ is a $L\times L$ symmetric matrix with elements $\ddot{\bar{\Phi}}_{l,k} = \langle \dot{\bar{\phi}}_l, \dot{\bar{\phi}}_k\rangle$. The derivation of the ripple-penalty is
\begin{equation*}
\begin{aligned}
\int_{0}^{T} (\dot{x}_n(t))^2 dt &=
\int_{0}^{T} \big(\sum_{l=1}^L c_{nl} \dot{\bar{\phi}}_l(t) \big) \big(\sum_{k=1}^L c_{nk} \dot{\bar{\phi}}_k(t) \big) dt \\
&= \sum_{l=1}^L \sum_{k=1}^L c_{nl} c_{nk} \langle \dot{\bar{\phi}}_l, \dot{\bar{\phi}}_k\rangle 
= c_{n}^T \ddot{\Phi} c_{n}
\end{aligned}
\end{equation*}

The optimization problem in \VIZ{cost:func:P2a} is a regularized Least Squares (RLS) problem. Additionally, it can be decoupled into $N$ independent optimization subproblems; one for each column of $C$. Thus, the solution for the $n$-th column (i.e., the $n$-th coefficient vector) is given by
\begin{equation}
\hat{c}_n = \left(\tilde{\phi}\tilde{\phi}^T  + \lambda_C \ddot{\Phi}\right)^{-1} \tilde{\phi} y_n^T \ ,
\end{equation}
with $n=1,...,N$. 

Cost functional \VIZ{cost:func:P2a} can be seen from a \emph{Bayesian perspective} as a sum of the log-likelihood and the log-prior distribution. The assumption incorporated in the prior distribution is that the trajectories are typically smooth functions of time. Both likelihood and prior distributions are Gaussian thus the data fitting problem is a linear optimization problem for the coefficient matrices which can be solved explicitly as shown above.
We would like to highlight also that a simple approach to create many trajectories is to repeatedly and randomly choose a subset of the measurements and solve \VIZ{cost:func:P2a}. A weight can be also assigned to each trajectory from the posterior distribution defined by \VIZ{cost:func:P2a}. The expectation when one feeds several trajectories to the learning algorithm is that the dynamics of the underlying process are more accurately sampled, thus, the performance is improved. Finally, there are cases where the distribution of the measurements are multi-modal or heteroscedastic. In such cases, the Gaussianity assumption is not valid resulting in sub-optimal solutions. Hence the cost functional should be enriched with multi-modal distributions such as the non-linear Gaussian Mixture Model. Alternatively, Monte Carlo methods which are more expensive computationally can be utilized for the sampling of the trajectories in a non-Gaussian case.

\section{Tackling Multiple Interventions}

Depending on the application at hand, there are various types of interventions that can be applied. Probably, the simplest intervention is to start from different initial values, then let the system evolve and measure it again. The wider exploration of the state space results in more informative measurements thus the estimation of the structure and the parameters of the dynamical system becomes in principle more identifiable. In systems biology for instance, such type of interventions are called activations.
Another type of intervention is (partial) inhibition where the effect of a state variable is neutralized or kept constant over time. For instance, if $n^*$ state variable is kept constant then we can impose the constraint $x_{n^*}(t)=c_0$ for all $t$ or equivalently $\dot{x}_{n^*}(t)=0$ which results to the constraint
$$\sum_{q=1}^Q a_{n^*q} \psi_q(x)=0 \ .$$
When an ODE system is augmented with constraints that do not contain time derivatives then the resulted set of equations is called a system of differential algebraic equations. More complex interventions and/or physical constraints (or laws) can be incorporated into the ODE system. The intervention type where $c_0=0$ is typically called inhibition.

However, experimentalists and scientists do not have full control on the effect of an intervention thus it is preferable to introduce a general way to model interventions other than hard constraints. Indeed, for ODE systems, the original equation, $\dot{x}_n = \sum_{q=1}^Q a_{nq} \psi_q(x)$, can be slightly extended to
\begin{equation}
\dot{x}_n = \sum_{q=1}^Q a_{nq} \psi_q(x) + b_n u_n\ , \ \ \ x_n(0)=x_{0n}\ , \ \ \ n=1,...,N \ ,
\label{ODE:intervent:eq2}
\end{equation}
where $b_n\in\mathbb R$ while $u_n=u_n(t)$ is a given function of time which can be thought as an input signal and constitutes the intervention signal. If for instance $n^*$ state variable is inhibited then this intervention is approximated by the reaction $X_{n^*}\to\emptyset$ with a large rate constant. In the modified ODE system \VIZ{ODE:intervent:eq2}, intervention of variable $n^*$ is modeled by setting $u_n=0$ for $n\ne n^*$ and $u_{n^*}=x_{n^*}$. 
Moreover, input signals can amount not only for inhibition but also for other type of interventions such as dosage level. Indeed, the form of the input signal for dosage intervention can be defined as $u_n(t) = u(t) = e^{-t/\tau}$ where $\tau$ controls the decay rate of the drug while $b_n$ reflects the strength of the effect of the drug to the $n$-th state variable.

Proceeding, assume that there are $R$ different experimental conditions thus there are $R$ different ODE systems with each one having different intervention input $u_n^{(r)}$, $r=1,...,R$. The weak formulation for the $r$-th ODE and the $n$-th variable is given by 
\begin{equation}
z^{(r)}_n = \Psi^{(r)} a_n + b^{(r)}_n v^{(r)}_n % + e^{(r)}_n
\end{equation}
where $z^{(r)}_n$ and $\Psi^{(r)}$ as in the main text while $v^{(r)}_n$ is the projection vector of the intervention input to the test functions with elements given by $v^{(r)}_{nm} = \langle u^{(r)}_n, \phi_m \rangle$.
The integrated ODE system can be written in a matrix form as
\begin{equation}
\begin{bmatrix} 
z^{(1)} \\ \vdots \\ z^{(R)}
\end{bmatrix}
=
\begin{bmatrix} 
\Psi^{(1)} & v^{(1)} & 0 & \cdots & 0 \\
\Psi^{(2)} & 0 & v^{(2)} & \cdots & 0 \\
\vdots & \vdots & \vdots & \vdots & \vdots \\
\Psi^{(R)} & 0 & 0 & \cdots & v^{(R)} \\
\end{bmatrix}
\begin{bmatrix} 
a \\ b^{(1)} \\ \vdots \\ b^{(R)}
\end{bmatrix}
\end{equation}

The derived system of equations fall again in the SSR category thus OMP can be utilized. Moreover, since we know a priori that there is driving input we start OMP with index set $\mathcal S = \{Q+1,\cdots,Q+R\}$. Finally, we remark that when $v^{(r)}=0$ for some $r$ then there is no input term. Hence, it becomes redundant in the above system of equations and the corresponding column of the measurement matrix as well the corresponding $b^{(r)}$ coefficient are removed.

\section{Sparse Signal Recovery}

After the application of the weak formulation, the transformed system of equations is written as
\begin{equation}
z = \Psi a + e
\end{equation}
where $z$, $\Psi$ and $a$ as defined in the main text while $e$ denotes the noise vector with $||e||_2 = \epsilon$ being the energy of the error/noise term. The SSR problem is defined as an $l_0$ quasi-norm minimization program
\begin{equation*}
\min_a ||a||_0 \ \ \text{subject to} \ \ ||z-\Psi a||_2 \le \epsilon \ ,
\label{l0:err:eq}
\tag{l0-error}
\end{equation*}
where $l_0$ quasi-norm counts the number of non-zero elements in its argument (i.e., $||a||_0 = \# \{q: a_q\ne 0 \}$). In order to solve this non-convex program, one must search through all possible solutions making this approach intractable for large systems since the search space is exponentially large \cite{Natarajan1995}. There is a wide spectrum of algorithms that try to approximate the solution of the above intractable program. We refer to \cite{Zhang2015} for a recent review. In the following, we present representative approximation algorithms for two different families that solve SSR. Moreover, the presented techniques are accompanied with theoretical guarantees which determine under which conditions the approximate solution is also the solution of \VIZ{l0:err:eq}.

Let us remark here that the columns of the measurement matrix are assumed to be normalized in the literature, however, $\Psi$ obtained by weak formulation has not normalized columns. In order to make the columns of $\Psi$ having $l_2$ norm equal to 1, the following computation is performed,
\begin{equation}
z=\Psi a + e = \sum_{q=1}^{Q} a_q \psi_q  + e = \sum_{q=1}^{Q} a_q ||\psi_q||_2 \frac{\psi_q}{||\psi_q||_2} + e = \sum_{q=1}^{Q} \bar{a}_q \bar{\psi}_q  + e = \bar{\Psi} \bar{a} + e \ .
\end{equation} 
Thus, the non-zero coefficients of $a$ are amplified by a factor which is proportional to the $l_2$-norm of the respective column of $\Psi$.

\subsection{Convex Relaxation Theory}

One approach to relax the above optimization problem and make it computationally feasible is to replace the $l_0$ quasi-norm with the $l_1$ norm and obtain the convex program
\begin{equation*}
\min_a ||a||_1 \ \ \text{subject to} \ \ ||z-\Psi a||_2 \le \delta \ ,
\label{l1:err:eq}
\tag{l1-error}
\end{equation*}
where $||a||_1=\sum_{q=1}^{Q} |a_q|$ is the $l_1$ norm while $\delta$ is the tolerance. It holds that $l_1$ norm is the closest convex function to $l_0$ quasi-norm \cite{Tropp2006}. Using standard linear programming software the \VIZ{l1:err:eq} minimization problem can be solved in polynomial time. Nevertheless, the solution of \VIZ{l1:err:eq} is just an approximation to the SSR problem and it is not guaranteed that it provides the same answer as the $l_0$ quasi-norm problem. Systematic investigations have been presented throughout the last fifteen years exploring when the solutions of the two problems are equal. For the sake of completeness, we provide a special case of a theorem by Tropp \cite{Tropp2006} which is based on ERC and determines under which conditions the solution of \VIZ{l1:err:eq} recovers the true support. In order to present the theorem, we have to define the following quantities. The $(p,q)$-matrix operator norm is defined for a matrix $B$ as
$$||B||_{p,q}:= \max_{x\ne0} \frac{||Bx||_q}{||x||_p} \ ,$$
while, for an index set $\mathcal S$, we define the restricted pseudo-inverse operator over $\mathcal S$ as
$$\bar{\Psi}_{\mathcal S}^\dagger := (\bar{\Psi}_{\mathcal S}^T\bar{\Psi}_{\mathcal S})^{-1} \bar{\Psi}_{\mathcal S}^T \ , $$
and the  best $l_2$  approximation of $z$ restricted on $\mathcal S$  as
$$\hat{z}_{\mathcal S} := \bar{\Psi}_{\mathcal S} \bar{\Psi}_{\mathcal S}^\dagger z \ . $$
Then, the following holds.

{\bf Theorem 1:} (Tropp '06) {\it Let $\mathcal T$ be the support of the true solution for which $ERC(\mathcal T) \ge 0$ and select tolerance to be
	$$\delta = \sqrt{\epsilon^2 + \left(\frac{||\bar{\Psi}(z-\hat{z}_{\mathcal T})||_\infty ||\bar{\Psi}_{\mathcal T}^\dagger ||_{2,1}}{ERC(\mathcal T)}  \right)^2} \ .$$
	If for all $q\in\mathcal T$ holds that 
	$$|a_q|>\frac{||\bar{\Psi}_{\mathcal T}^\dagger ||_{2,2}}{||\psi_q||_2} \delta$$
	then the support of the (unique) solution of \VIZ{l1:err:eq} convex program with tolerance $\delta$ is exactly $\mathcal T$.} \\

Another methodology to relax the $l_0$ quasi-norm optimization program is to consider the Lagrangian version of \VIZ{l1:err:eq}. It reads 
\begin{equation*}
\min_a \frac{1}{2}||z-\Psi a||_2^2 + \lambda ||a||_1 \ ,
\label{l1:pen:eq}
\tag{l1-penalty}
\end{equation*}
where $\lambda$ is a weight parameter that balances between complexity and sparsity. Indeed, increasing $\lambda$ results in sparser solutions. This optimization program is also known as LASSO \cite{Tibshirani1996} and there exist fast algorithms such as LARS \cite{Efron2004} which finds LASSO solutions.
As in the previous case, there are theoretical guarantees under which conditions the LASSO solution is also the solution of the SSR problem. We provide two theorems; one involving MIP and the other ERC which are special cases of theorems in \cite{Tropp2006}.

{\bf Theorem 2:} (Tropp '06) {\it Let $\mathcal T$ be the support of the true solution and $k:=|\mathcal T|$ its size. Suppose that $\mu\le \frac{1}{2k}$ as well as that
	$$||\bar{\Psi}_{\mathcal T} (z-\hat{z}_{\mathcal T})||_\infty \le \lambda \frac{1-(2k-1)\mu}{1-(k-1)\mu}$$
	for some $\lambda>0$ where $\hat{z}_{\mathcal T}$ is the best $l_2$ approximation over $\mathcal T$. If for all $q\in \mathcal T$ holds that
	$$|a_q|>\frac{\lambda}{||\psi_q||_2(1-(k-1)\mu)}$$
	then the support of the (unique) solution of the Lasso convex program with parameter $\lambda$ is exactly $\mathcal T$.}

It is noteworthy that for the noiseless case and given that $\mu\le \frac{1}{2k}$, there is always a small enough $\lambda$ which satisfies the condition on the non-zero coefficients of $a$. Moving forward, as we saw in the main text, the condition for MIP is rarely satisfied, mainly, due to collinearity between columns of $\Psi$ that might be irrelevant to the space spanned by the atoms of the actual solution. For this reason, we stated that MIP is rather conservative. The following theorem, again due to Tropp, is based on ERC and determines when Lasso solution recovers the true support.

{\bf Theorem 3:} (Tropp '06) {\it Let $\mathcal T$ be the support of the true solution for which $ERC(\mathcal T) \ge 0$. Suppose that 
	$$||\bar{\Psi}_{\mathcal T} (z-\hat{z}_{\mathcal T})||_\infty \le \lambda ERC(\mathcal T)$$
	for some $\lambda>0$. If for all $q\in \mathcal T$ holds that
	$$|a_q|>\frac{\lambda ||(\bar{\Psi}_{\mathcal T}^T\bar{\Psi}_{\mathcal T})^{-1}||_{\infty,\infty}}{||\psi_q||_2}$$
	then the support of the (unique) solution of the Lasso convex program with parameter $\lambda$ is exactly $\mathcal T$.}

The condition on ERC is harder to validate compared to MIP because the support of the true but unknown solution is required. Here we present the results when the true support is known, however, in Tropp \cite{Tropp2006}, the conditions are relaxed for arbitrary index set of linearly independent collection of dictionary atoms. Finally, we remark that a difficulty in Lasso formulation is that little or no information about the hyper-parameter $\lambda$ is provided and scientists usually apply information criteria in order to select $\lambda$. Nevertheless, guidance on the values of $\lambda$ can be provided from the above theorems since two counterbalancing conditions for $\lambda$ need to be satisfied.

\subsection{Orthogonal Matching Pursuit}
There are several greedy algorithms proposed in the literature trying to solve the SSR problem. Among them one of the fastest and surely the most popular is OMP. Apart from being very fast, OMP has the important property that its hyper-parameter is easy to interpret and approximate from the available data. For the sake of completeness, we present the basic OMP algorithm where we added an extra criterion on the maximum allowed number of non-zero elements which is denoted with $K$.

\begin{itemize}
	\item OMP Algorithm:
	\begin{enumerate}
		\item Initialize the residual $r_0=z$ and the set of selected indices $S=\emptyset$. Set counter to $i=1$.
		\item Find the next index as
		\begin{equation*}
		q' = \arg\max_q |\bar{\psi}_q^T r_{i-1}|
		\end{equation*}
		and add $q'$ to $S$. 
		\item Solve the Least Squares problem using only the selected indices
		\begin{equation*}
		\hat{a}_S = (\Psi_S^T\Psi_S)^{-1}\Psi_S^T z = \Psi_S^\dagger z
		\end{equation*}
		Update residual error $r_i=z-\Psi_S\hat{a}_S$.
		\item If $||r_i||_2<\epsilon$ or $|S|> K$, stop the algorithm. Otherwise, set $i=i+1$ and return to step 2.
	\end{enumerate}
\end{itemize}

The stopping criterion in OMP requires the knowledge of the $l_2$ norm of the true residual error which in principle is not available a priori. The energy of the true residual error is approximated as
$$\hat{\epsilon} = (1+\alpha) ||r_{LS}||_2 \ ,$$
where $r_{LS}$ is the residual when the complete dictionary is utilized given by $r_{LS} := z-\Psi\Psi^\dagger z$ while $\alpha$ is a small positive number in the range $10^{-1}-10^{-3}$. $\alpha$ is user-defined and corresponds to the potential over-fitting of the complete model representation. Notice that the stopping criterion can be rewritten as
$$ \frac{||r_i||_2-||r_{LS}||_2}{||r_{LS}||_2} < \alpha \ ,$$
implying that OMP stops when the relative residual energy becomes smaller than $\alpha$.

In the following, we present two theorems on perfect support recovery; one based on MIP and the other on ERC which were first proven in \cite{Cai2011}.

{\bf Theorem 4:} (Cai \& Wang '11) {\it Suppose $||e||_2\le \epsilon$ and $\mu<\frac{1}{2k-1}$ where $k$ is the number of nonzero elements of $a$. Then, OMP algorithm with stopping rule $||r_i||_2\le \epsilon$ recovers the true subset of correct dictionary atoms indexed by $\mathcal T$ if for all $q\in \mathcal T$ holds that
	$$|a_q|\ge \frac{2}{SNR(q) (1-(2k-1)\mu)} \ .$$}

The next theorem collects the pieces that has been presented in the main text.

{\bf Theorem 5:} (Cai \& Wang '11) {\it Suppose $||e||_2\le \epsilon$ and $ERC(\mathcal T)>0$ where $\mathcal T$ is the support of the true solution. Then,  OMP algorithm with stopping rule $||r_i||_2\le \epsilon$ recovers the true subset of correct dictionary atoms, $\mathcal T$, if for all $q\in \mathcal T$ holds that
	$$|a_q|\ge \frac{2}{SNR(q) ERC(\mathcal T) \lambda_{\min}(\mathcal T)} \ .$$}

Additional theoretical results can be found in \cite{Cai2011}. For instance, by varying the stopping criterion, one can guarantee that the strongest components of the solution will be obtained showing that the strongest components are selected first.

\subsubsection{Adding prior Knowledge}
In some applications, user may partially know the driving forces of a dynamical system or she has an instrument that intervene the system in a systematic and known way. Thus, being able to add prior knowledge to the inference algorithm is very important. Another convenient feature of OMP is that adding prior knowledge is straightforward. Indeed, instead of starting with an empty index set, $S$ in step 1 of OMP will be initialized with the provided indices that encode the prior knowledge. Accordingly, the initial residual error will be evaluated after the subtraction of the known non-zero components.

\section{Further Demonstrations}
We provide here several details on the production of the figures in the main text. Additional experiments are also presented for each of the demonstration examples.

\subsection{Protein interaction network}
%\label{gene:reg:sec}

This demonstration example emulates the output of a single-cell mass cytometry machine with a prototypical protein interaction network where protein $P_1$ interacts with protein $P_3$ through protein $P_2$. The complete biochemical reaction network is given in Table~\ref{gene:reg:reactions} and the corresponding ODE system has been simulated with standard ODE solver. 
It consists of six reactions which follow the law of mass action kinetics. $R_1$ and $R_2$ produce $P_2$ and $P_3$, respectively, while $R_3$ and $R_4$ correspond to the binding and unbinding of the $P_3$ to $P_1$. $R_5$ and $R_6$ are simple degradation mechanics. The initial concentration of the state variables (i.e., the species) are 1 for $P_1$ and 0 for the rest species. Furthermore, we perform four interventions. The first intervention was to change the initial concentration of $P_2$ from 0 to 5 while the second intervention was to change the initial concentration of $P_3$ from 0 to 15. The third intervention was the inhibition of $P_2$ while the final one was the inhibition of $P_3$. Inhibition was implemented by augmenting a new and fast degradation reaction with rate constant equal to 100.

\begin{table}[!htb]
	\begin{center}
		\caption{The reaction table with $P_1-P_3$ corresponding to the 3 proteins while $[P_1P_3]$
			corresponds to the Protein-Protein complex. The state of the reaction model is defined as $x=[P_1,P_2,P_3,[P_1P_3]]^T$.}
		\begin{tabular}{|c|l|l|l|} \hline
			Event & Reaction & Rate & Rate constant \\ \hline \hline
			$R_1$ & $P_1 \rightarrow P_1 + P_2$ & $a_1({ x}) = k_1 P_1$ & $k_1=4$ \\ \hline
			$R_2$ & $P_2 \rightarrow P_2 + P_3$ & $a_2({ x}) = k_2 P_2$ & $k_2=0.05$ \\ \hline
			$R_3$ & $P_1 + P_3 \rightarrow [P_1P_3]$ & $a_3({ x}) = k_b P_1 \cdot P_3$ & $k_b=0.01$ \\ \hline
			$R_4$ & $[P_1P_3] \rightarrow P_1 + P_3$ & $a_4({ x}) = k_u [P_1P_3]$ & $k_u=0.001$ \\ \hline
			$R_5$ & $P_2 \rightarrow \emptyset$ & $a_5({ x}) = k_{-2} P_2$ & $k_{-2}=0.4$ \\ \hline
			$R_6$ & $P_3 \rightarrow \emptyset$ & $a_6({ x}) = k_{-3} P_3$ & $k_{-3}=0.2$ \\ \hline
		\end{tabular}
		\label{gene:reg:reactions}
	\end{center}
\end{table}

The resulted dynamical system has a subtle property. Due to the fact that $dP_1/dt + d[P_1P_3]/dt = 0$, the following conservation law holds
$$P_1+[P_1P_3]=1 \ .$$
Eventually, the dynamical system consists of three equations with four unknowns. Therefore, the complete reaction network is inherently unidentifiable when the conservation law is not taken into account making it an inappropriate example for testing the USDL algorithm. This is the main reason why we choose a coarser and thus simpler dynamical model for the learning algorithms. Moreover, the property of unidentifiability is very typical in biochemical reaction networks and in conjunction with the sloppiness of the parameters \cite{Gutenkunst2007}, the structure inference of the complete biochemical network of interactions is impossible most of the times. % (``Universally Sloppy Parameter Sensitivities in Systems Biology Models'', Gutenkunst et al., PLOS Comp Bio, 2007)

\begin{figure}[!htb]
	\centering
	\subfigure[F1 scores]{
		\includegraphics[width=.47\textwidth]{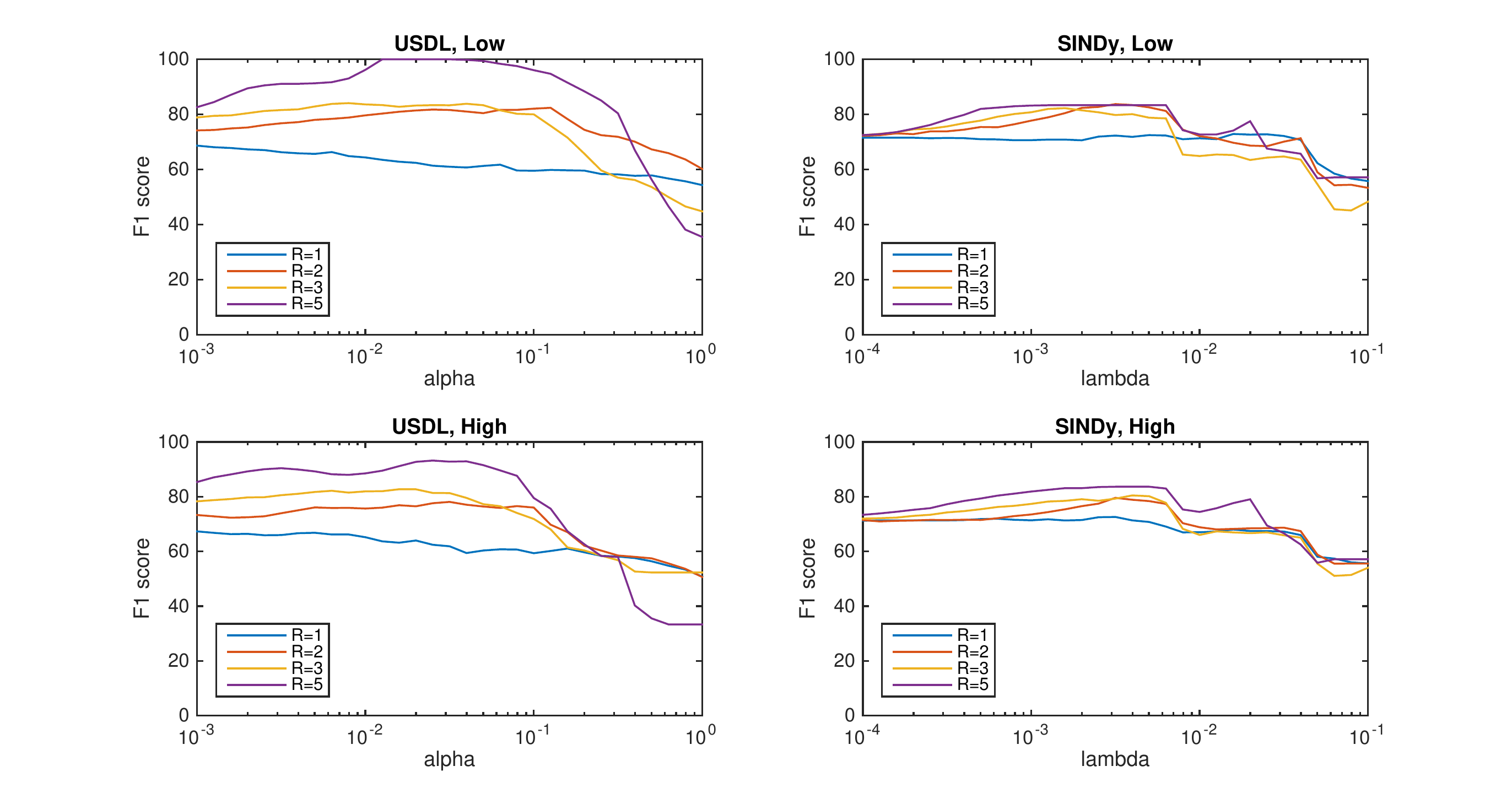}
		\label{F1score:gene:reg:basic}}
	\subfigure[Time-series forecasts]{%
		\includegraphics[width=.5\textwidth,height=.19\textheight]{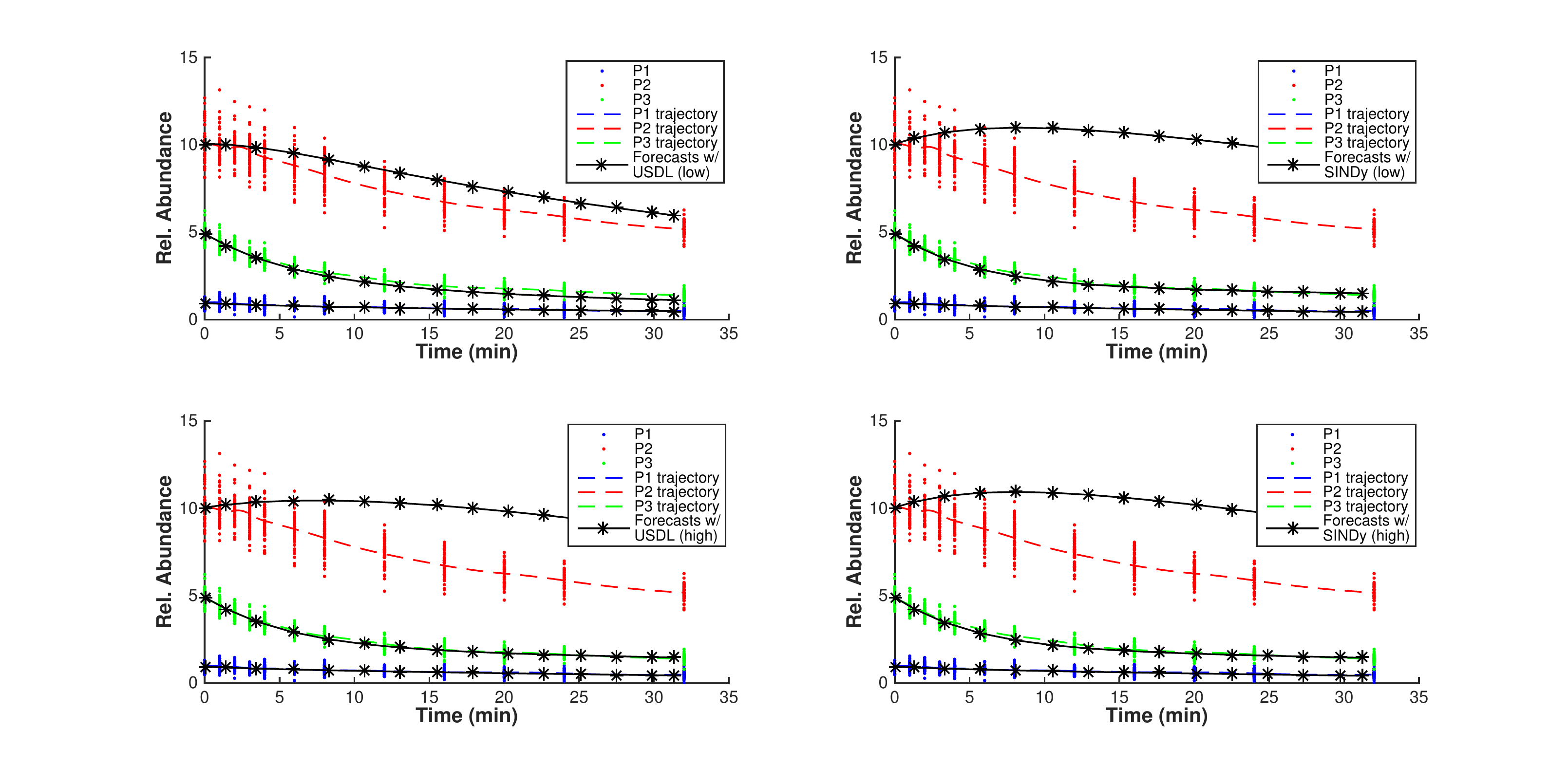}
		\label{forecasts:gene:reg:basic}}
	\caption{(a) The F1 score as a function of the hyperparameter for both USDL (left column) and SINDy (right column) and various number of experiments and noise levels. (b) Time-series forecasts (black curves) based on the estimated parameter values for both USDL (left column) and SINDy (right column).}
	\label{gene:reg:basic:fig}
\end{figure}

Figure~\ref{F1score:gene:reg:basic} presents the F1 score (i.e., the harmonic mean between precision and recall) for both USDL (left column) and SINDy (right column) and various number of experiments and noise levels. Perfect reconstruction which is implied by an F1 score at 100\% is achieved only by the USDL algorithm when the noise has low variance and data from all five experiments are used. As already stated in the main text, we choose the hyperparameter value that maximizes the F1 score. We also observe that utilizing up to $R=3$ experiments results has similar F1 score performance for both approaches however the USDL algorithm is superior when all available data are exploited.

Figure~\ref{forecasts:gene:reg:basic} presents the time-series forecasts (black curves) based on the estimated parameter values for both USDL (left column) and SINDy (right column). Our simulation is performed on a new experiment which is different from the five experiments used for the inference. Indeed, the initial concentration is 10 for $P_2$ and 5 for $P_3$. Let us remark also that we use as initial values to the ode solver the mean of variable's concentration at time point 0. In accordance with the ERC metric, the problematic variable with the highest error is $P_2$.

\begin{figure}[!htb]
	\centering
	\subfigure[Time-course measurements and time-series interpolation]{%
		\includegraphics[width=.38\textwidth]{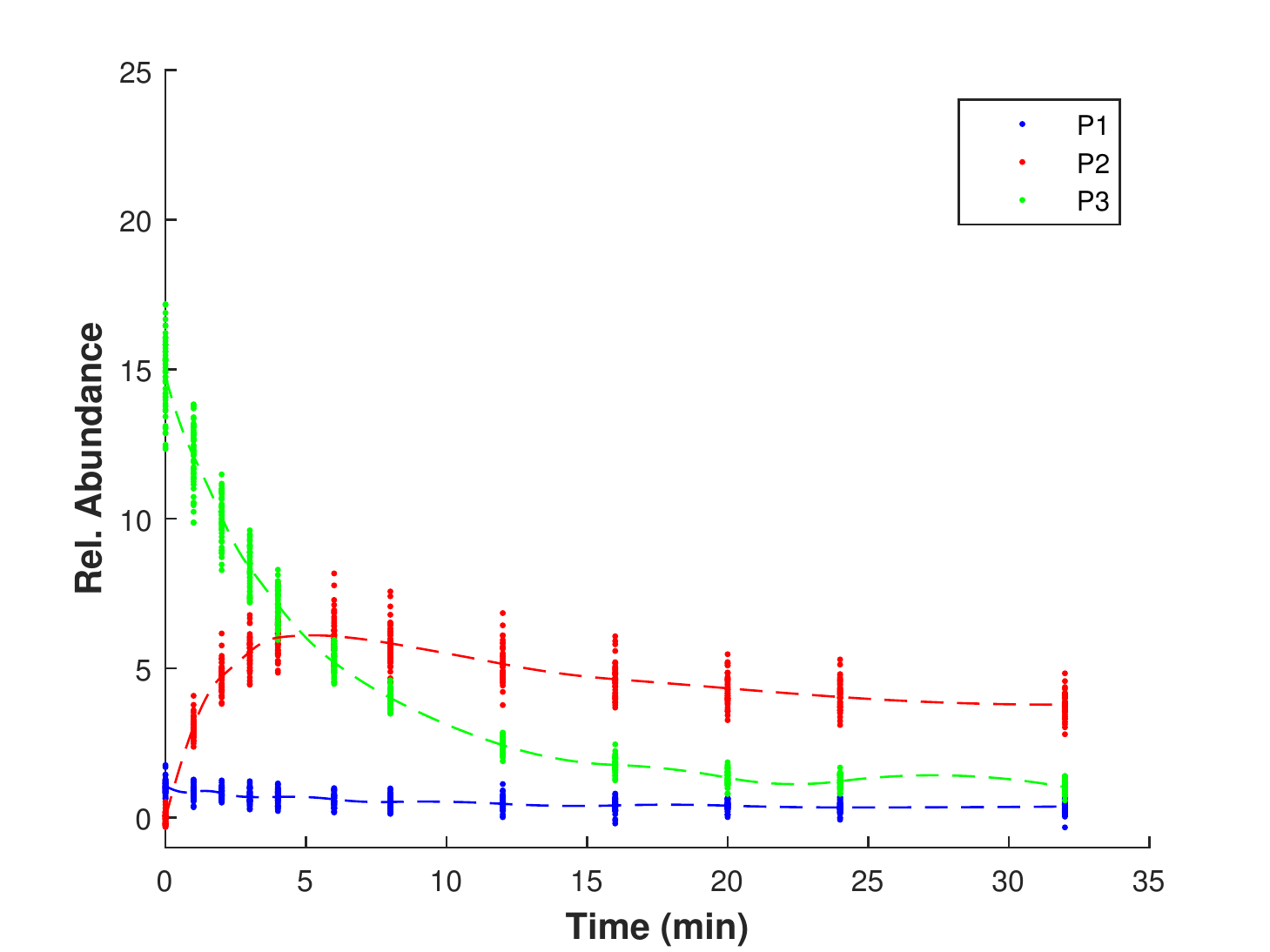}
		\label{tc:tp:gene:reg:lC:1}}
	\subfigure[MIP and ERC]{%
		\includegraphics[width=.38\textwidth]{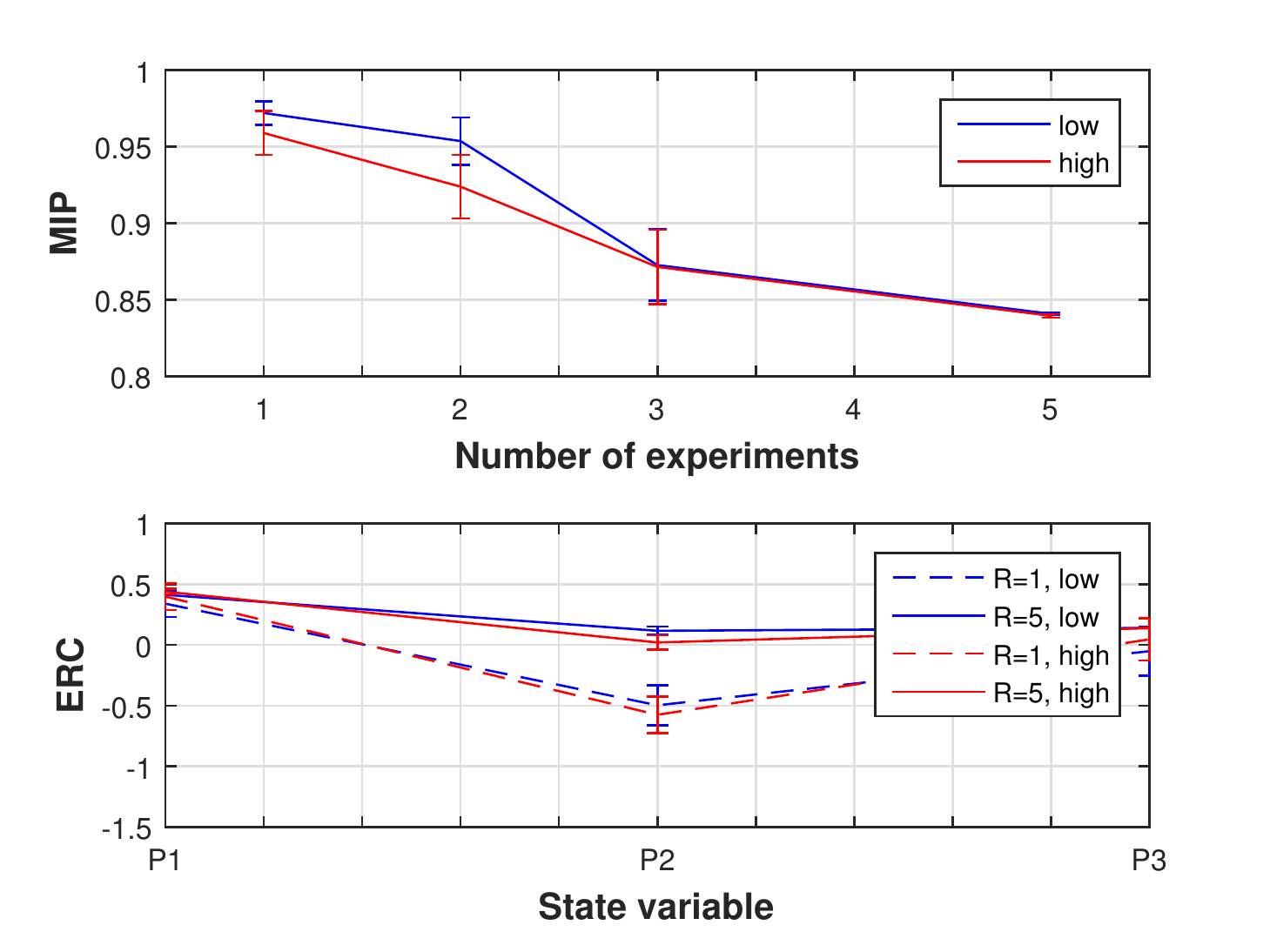}
		\label{mip:erc:gene:reg:lC:1}}
	\quad
	\subfigure[Precision and recall]{
		\includegraphics[width=.38\textwidth]{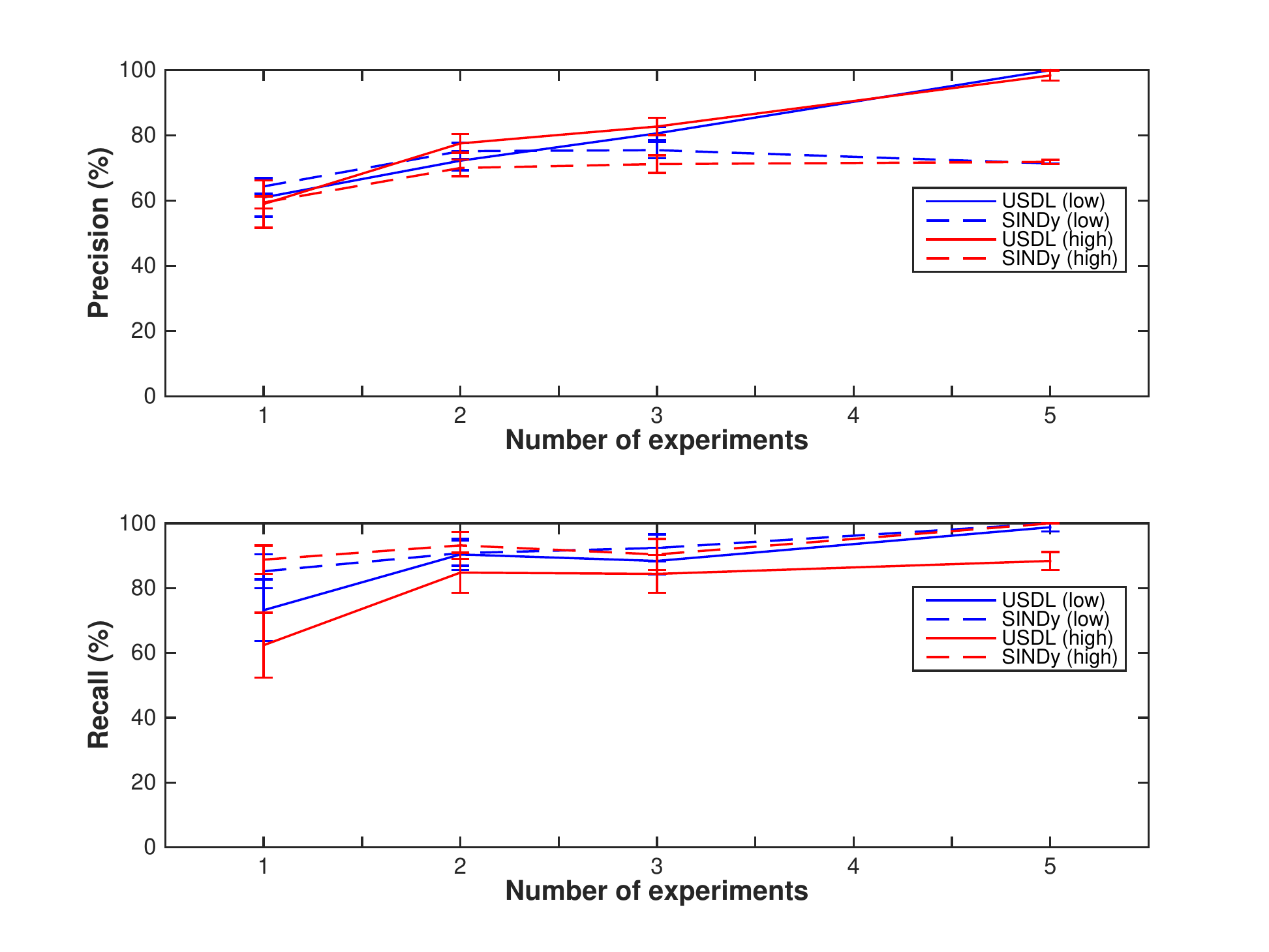}
		\label{pre:rec:gene:reg:lC:1}}
	\quad
	\subfigure[Time-series forecasts]{
		\includegraphics[width=.48\textwidth,height=.22\textheight]{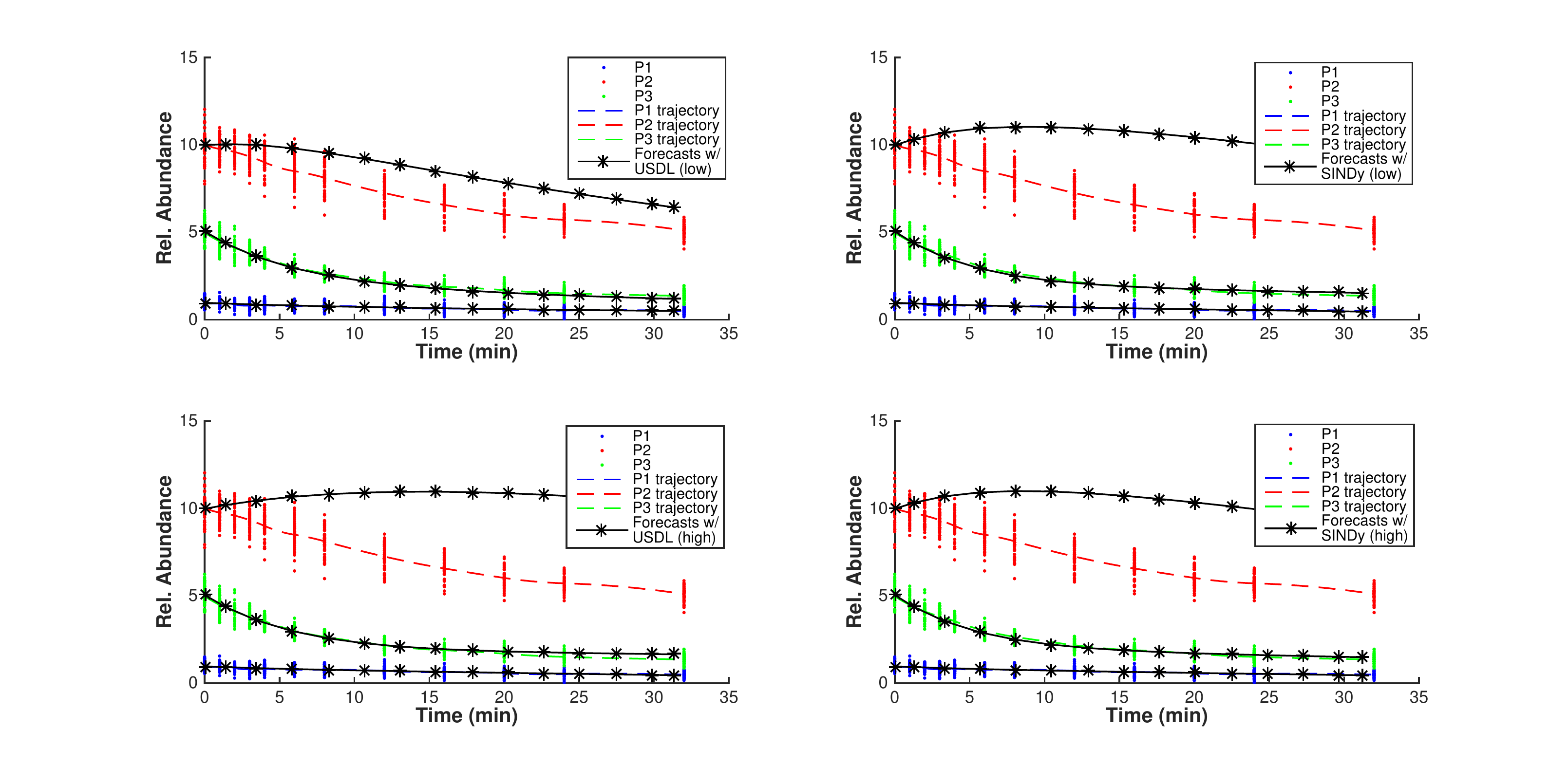}
		\label{forecast:gene:reg:lC:1}}
	\caption{(a) Time-series interpolation for the protein interaction network when the smoothing weight parameter is $\lambda_C=1$. Trajectories are not smooth introducing further noise, nevertheless, the dynamics of the measurements are correctly captured. (b) MIP (upper panel) and ERC per state variable (lower panel) for two noise levels; low (blue) and high (red). ERC is negative when $R=1$ experimental condition (dashed lines) is considered while it is positive when $R=5$ experimental conditions (solid lines) are considered. (c)  Precision (upper panel) and recall (lower panel) curves as a function of $R$ for both USDL (solid lines) and SINDy (dashed lines) algorithms. Perfect reconstruction is achieved for the low noise level and when $R=5$ interventions are provided.  (d) Time-series forecasts (black curves) based on the estimated parameter values for both USDL (left column) and SINDy (right column). Forecasts are similar to the case where $\lambda_C=100$.}
	\label{gene:reg:fig:lC:1}
\end{figure}

\subsection{Further exploration}

\begin{figure}[!htb]
	\centering
	\subfigure[Time-course measurements and time-series interpolation]{%
		\includegraphics[width=.4\textwidth]{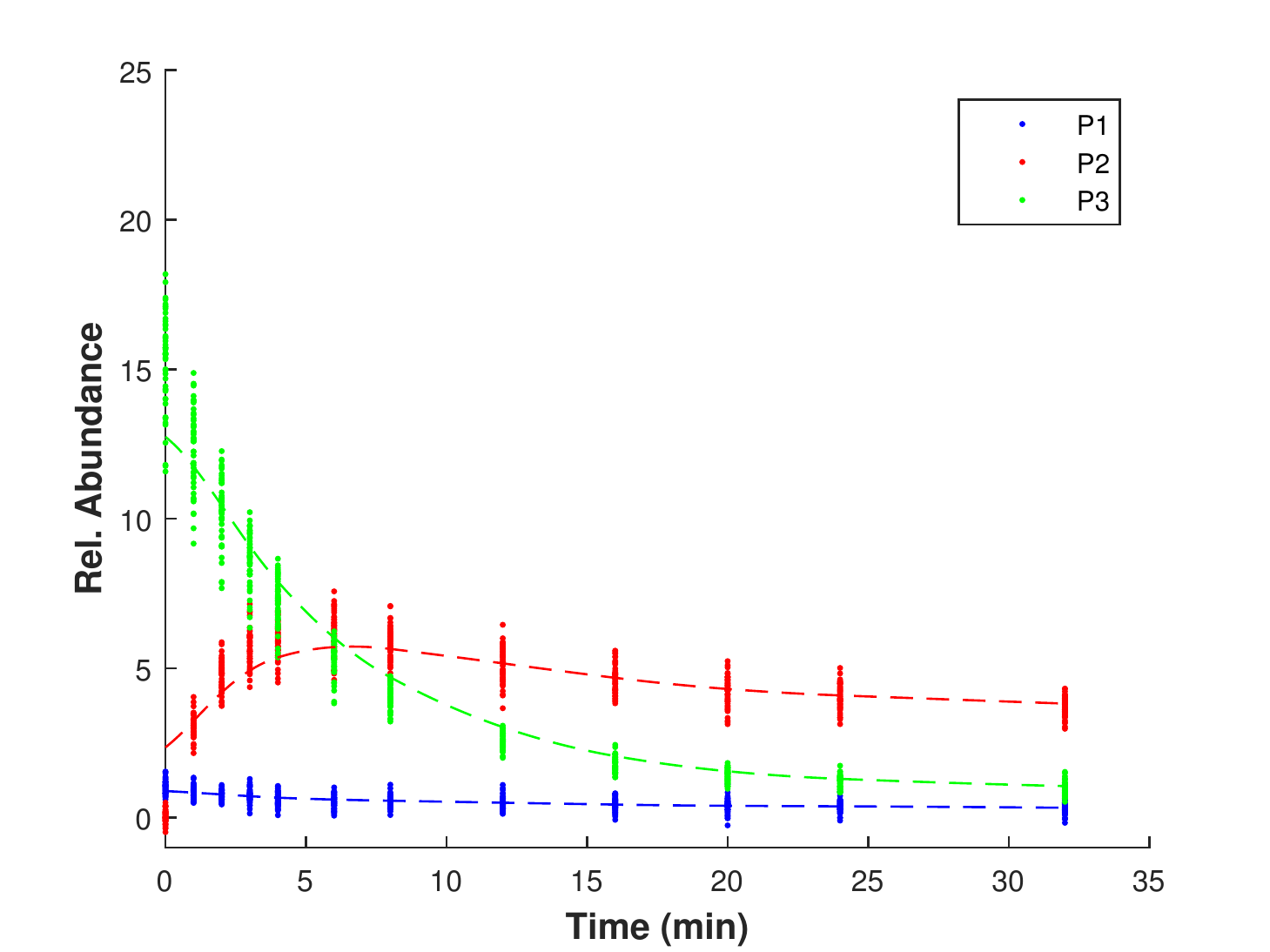}
		\label{tc:tp:gene:reg:lC:10000}}
	\subfigure[MIP and ERC]{%
		\includegraphics[width=.4\textwidth]{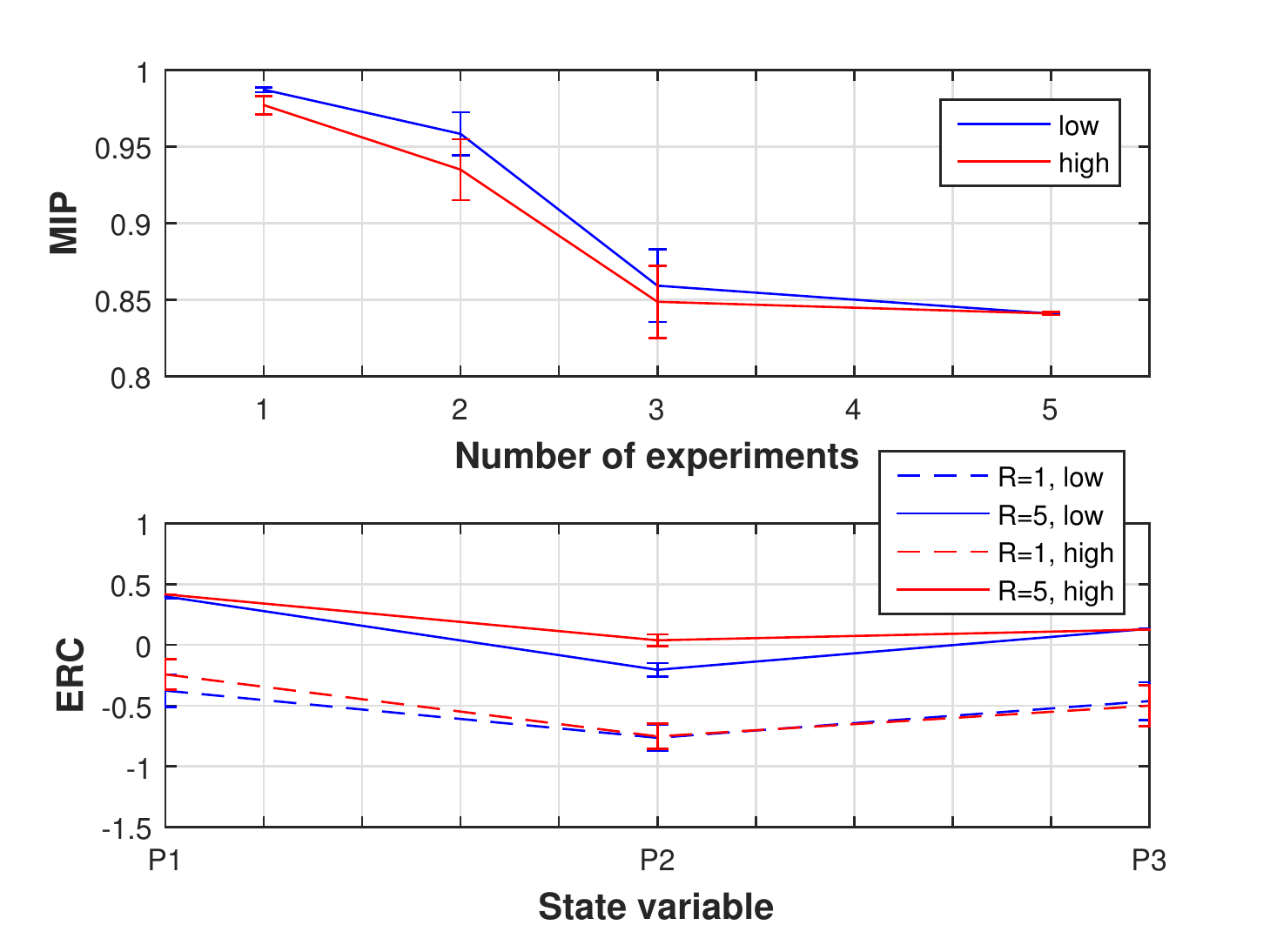}
		\label{mip:erc:gene:reg:lC:10000}}
	\quad
	\subfigure[Precision and recall]{
		\includegraphics[width=.4\textwidth]{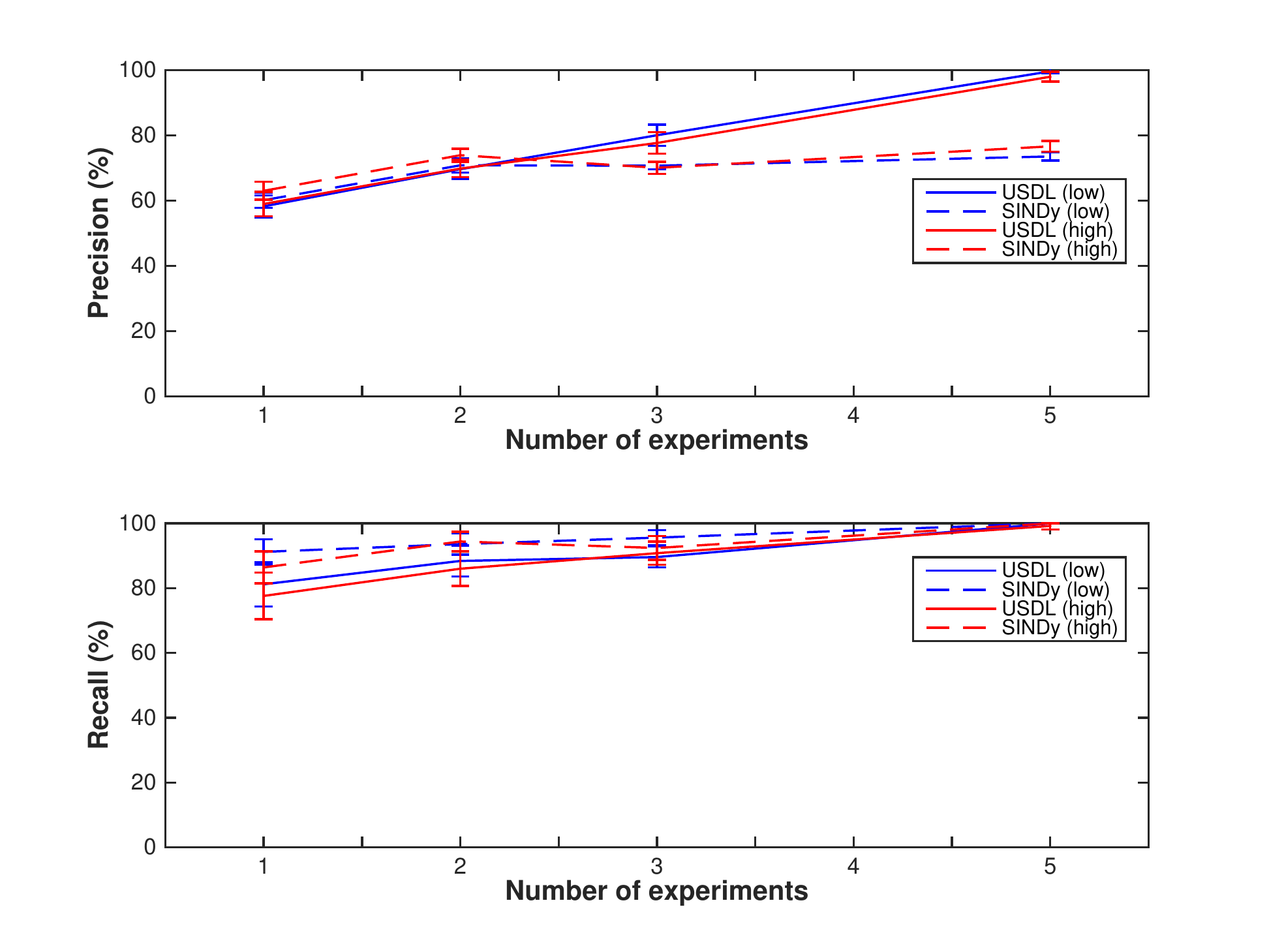}
		\label{pre:rec:gene:reg:lC:10000}}
	\quad
	\subfigure[Time-series forecasts]{
		\includegraphics[width=.5\textwidth,height=.22\textheight]{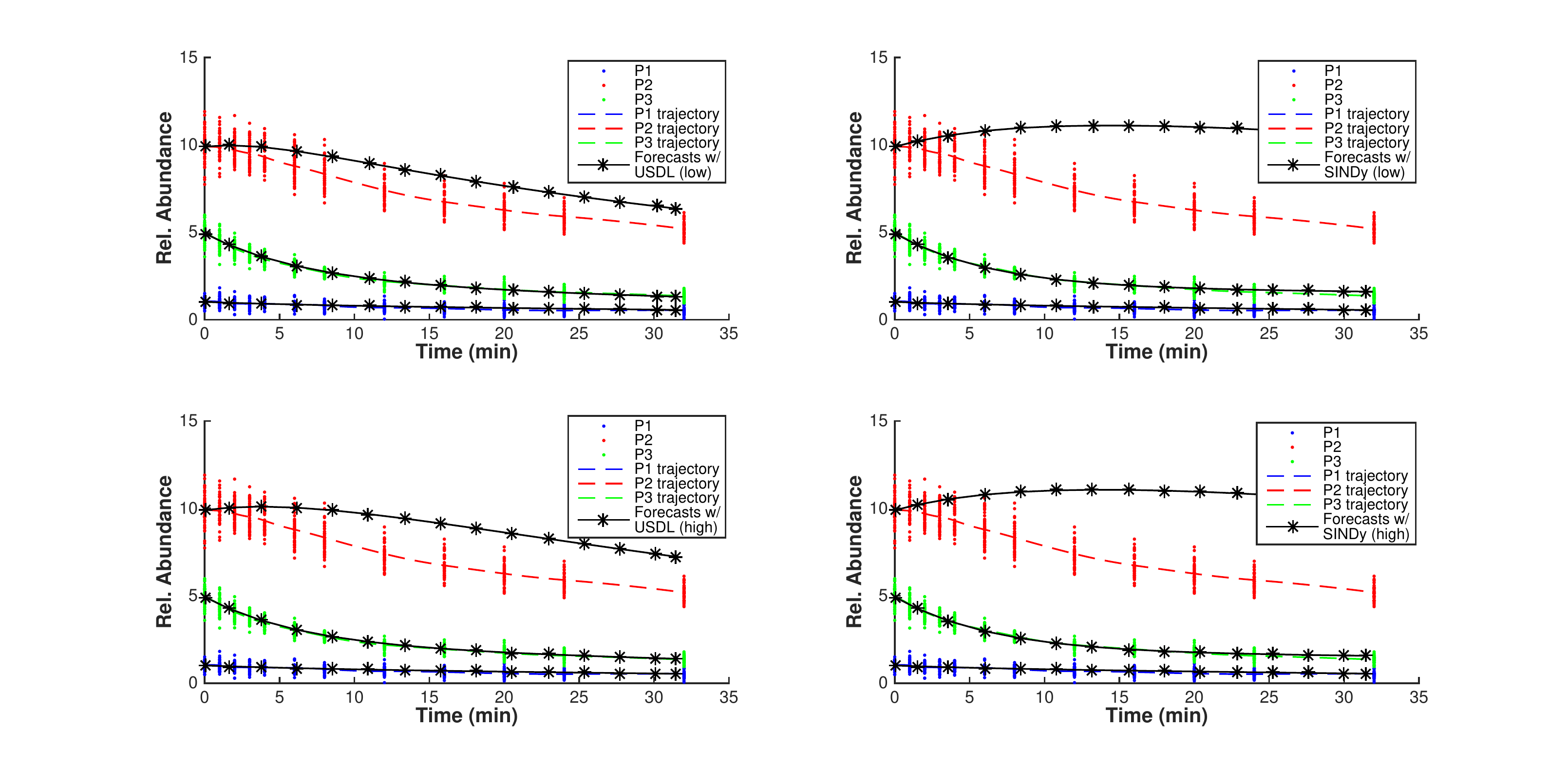}
		\label{forecast:gene:reg:lC:10000}}
	\caption{(a) Time-series interpolation for the protein interaction network when the smoothing weight parameter is $\lambda_C=10^4$. Trajectories are excessively smoothed resulting in incapability of  capturing correctly the dynamics. (b)  MIP (upper panel) and ERC per state variable (lower panel) for two noise levels; low (blue) and high (red). (c)  Precision (upper panel) and recall (lower panel) curves as a function of $R$ for both USDL (solid lines) and SINDy (dashed lines) algorithms. Perfect reconstruction is achieved for USDL algorithm at both low and high noise levels.  (d) Time-series forecasts (black curves) based on the estimated parameter values for both USDL (left column) and SINDy (right column).}
	\label{gene:reg:fig:lC:10000}
\end{figure}

We proceed by showing the results on the performance of the proposed approach under various experimental conditions. We test the performance when the number of sampling points is reduced as well as when different weights for the smoothing penalty in the collocation method are shown next. In all experiments of this subsection, we set the hyperparameter of both USDL and SINDy algorithms based on the optimal F1 score value. We do not show these plots since they are very similar to Figure~\ref{F1score:gene:reg:basic}.
Figure~\ref{gene:reg:fig:lC:1} presents the interpolated time-series of each species when $\lambda_C=1$. It is evident with naked eye that the bending of the trajectories are harsher, especially, near the sampling points. Nevertheless, positive values of ERC for all species when all five interventions are taken into account indicate the perfect reconstruction of the protein interaction network. Indeed, the performance of the USDL algorithm measured by precision/recall metrics and forecast capacity on new experiment is only slightly affected as shown in Figures~\ref{pre:rec:gene:reg:lC:1} \& (d), respectively. SINDy algorithm, as in the main text, does not exploit the information from the multiple interventions. Thus, SINDy's performance is suboptimal.
On the opposite direction where smoothing weight is large ($\lambda_C=10^4$), the constructed trajectories are fairly smoother as Figure~\ref{gene:reg:fig:lC:10000} shows. Despite the error on the dynamics, the performance of the USDL algorithm is not negatively affected as precision and recall plots reveal. Evidently, the accuracy of both algorithms is slightly improved for the high level noise with USDL achieving perfect reconstruction when fed with data from five interventions. It is noteworthy that ERC indicates this behavior since it is larger under high noise (red solid line in Figure~\ref{gene:reg:fig:lC:10000}(b)). In both cases the inference methods where robust against the hyperparameter of the collocation method.

\begin{figure}[!htb]
	\centering
	\subfigure[Time-course measurements and time-series interpolation]{%
		\includegraphics[width=.4\textwidth]{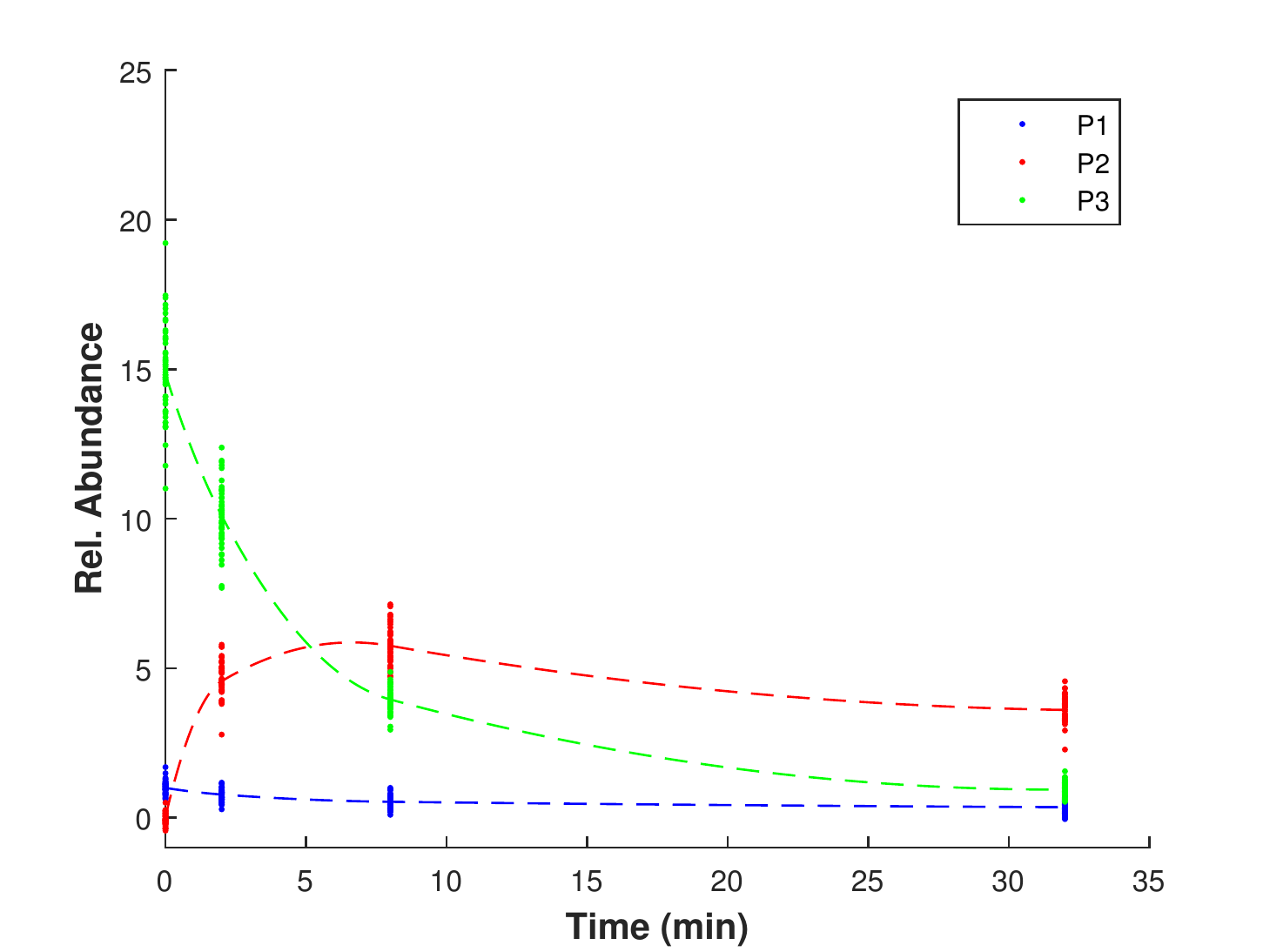}
		\label{tc:tp:gene:reg:less:sampl}}
	\subfigure[MIP and ERC]{%
		\includegraphics[width=.4\textwidth]{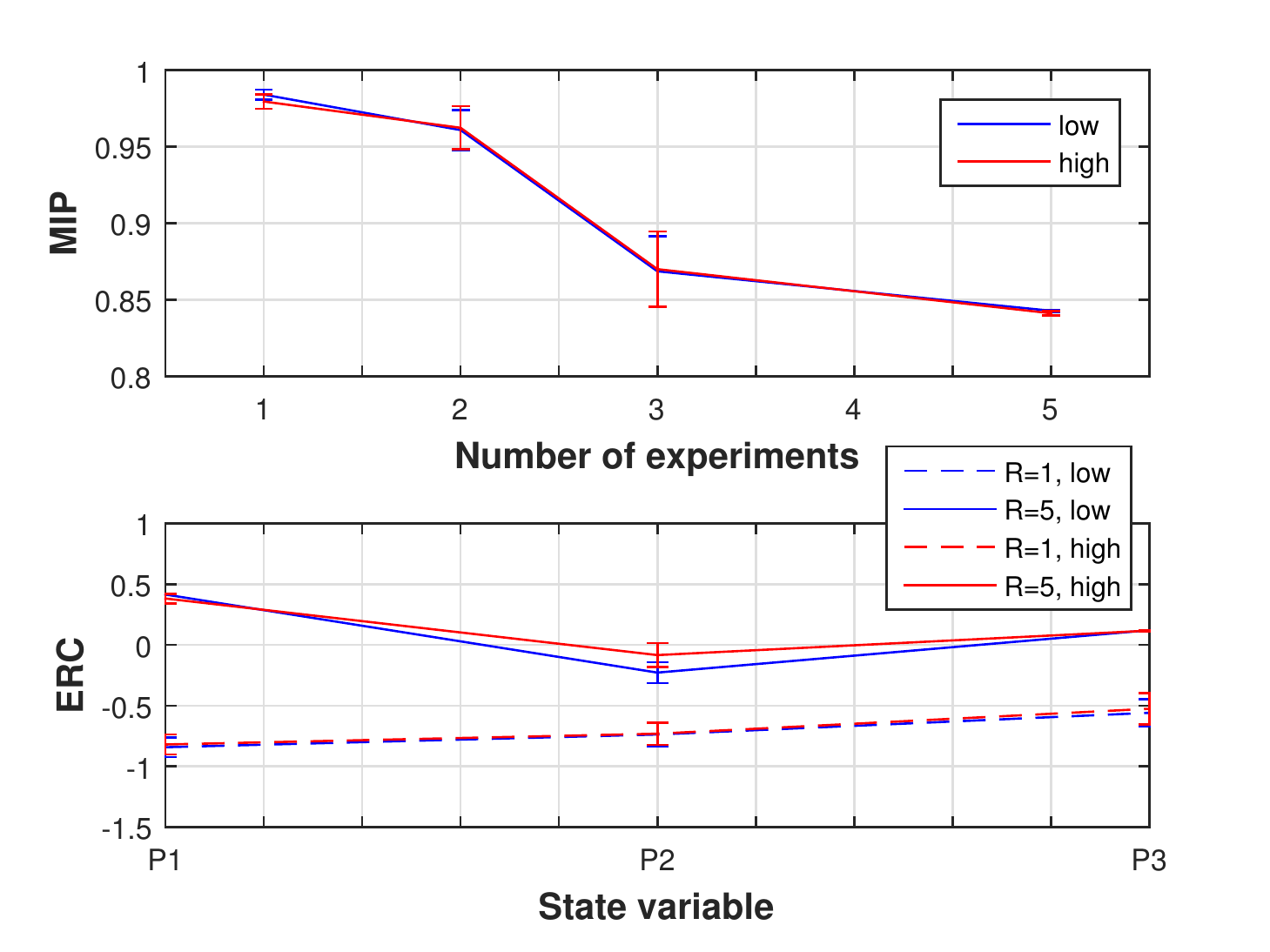}
		\label{mip:erc:gene:reg:less:sampl}}
	\quad
	\subfigure[Precision and recall]{
		\includegraphics[width=.4\textwidth]{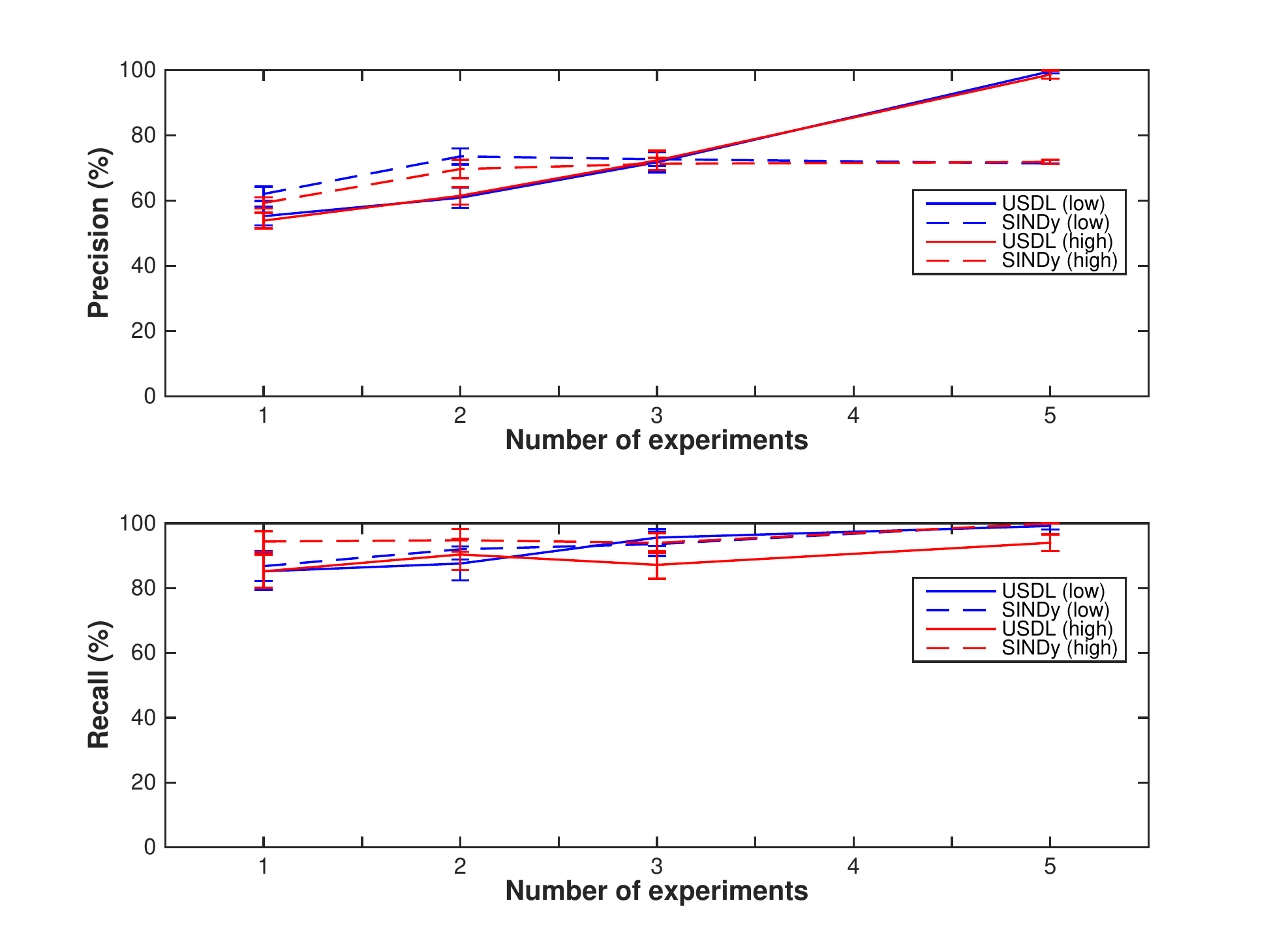}
		\label{pre:rec:gene:reg:less:sampl}}
	\quad
	\subfigure[Time-series forecasts]{
		\includegraphics[width=.5\textwidth,height=.22\textheight]{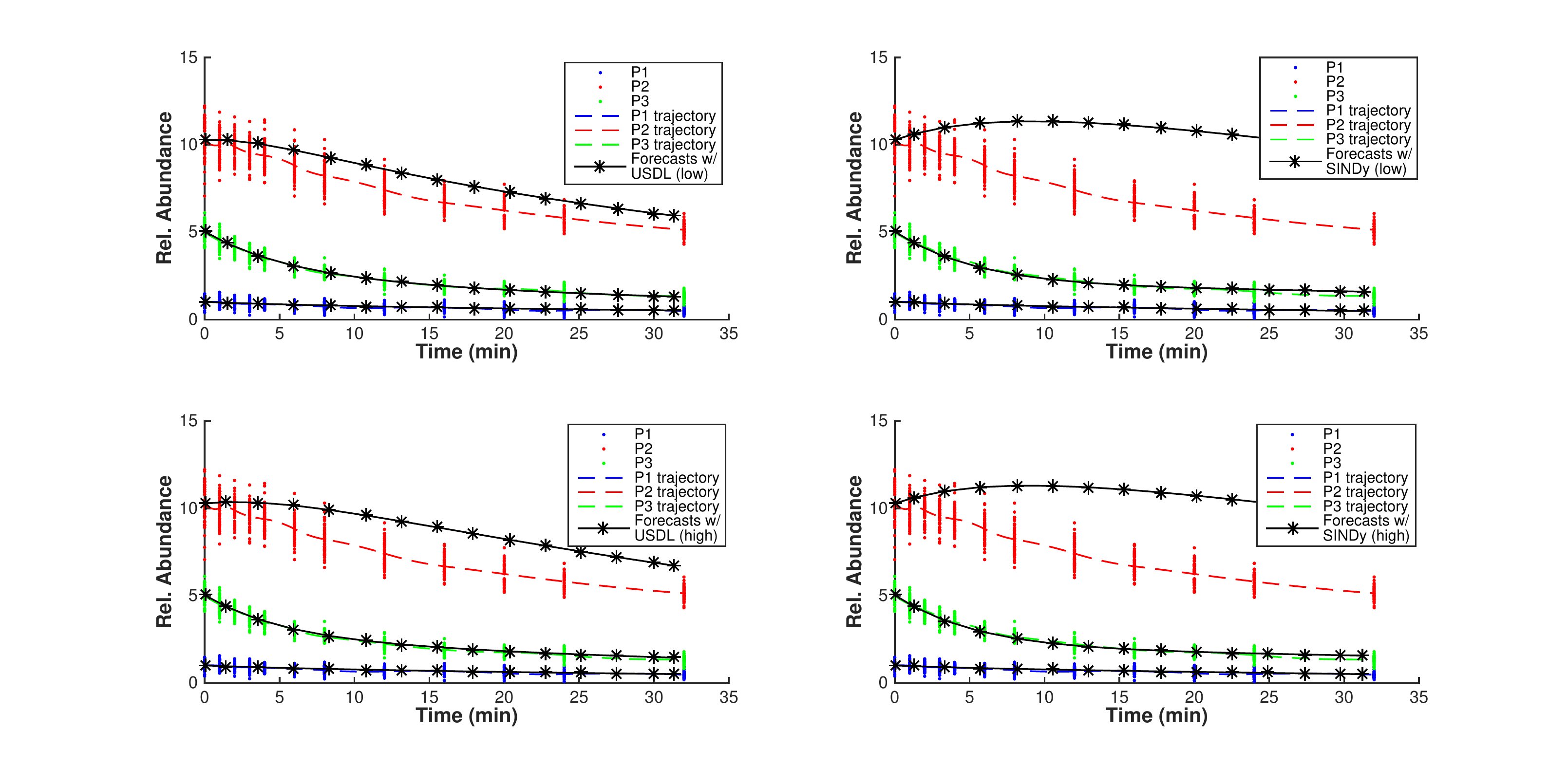}
		\label{forecast:gene:reg:less:sampl}}
	\caption{(a) Time-series interpolation for the protein interaction network when less sampling points are measured. (b)  MIP (upper panel) and ERC per state variable (lower panel) for two noise levels; low (blue) and high (red). ERC is positive when $R=5$ interventions are considered (solid lines) while it is negative when $R=1$ intervention is considered (dashed lines). (c)  Precision (upper panel) and recall (lower panel) curves as a function of $R$ for both USDL (solid lines) and SINDy (dashed lines) algorithms. Overall, the right positioning of the sampling instants resulted in robust inference of the network of interactions.  (d) Time-series forecasts (black curves) based on the estimated parameter values for both USDL (left column) and SINDy (right column).}
	\label{gene:reg:fig:less:sampl}
\end{figure}

The next experiment assesses the performance of the proposed algorithm when measurements from fewer time-points are obtained. Since dense sampling is more expensive both in time and in money than sparse sampling, it is desirable the learning methods be robust against less sampling points. Figure~\ref{gene:reg:fig:less:sampl} presents the results when we reduce the number of sampling times by a factor of 3 (from 12 to 4) and keeping the time instants at 0,2, 8 and 32. Interestingly, the performance of the USDL algorithm remained the same with perfect reconstruction being achieved when all five interventions are considered under low noise. This implies that the collocation method was able to correctly estimate the trajectories from these four time points. Perfect reconstruction was also predicted from the positivity of the ERC for all variables for the case of five interventions. We would like to remark here that ERC can be utilized as an experimental design metric for the determination of the optimal sampling points. Indeed, using prior information about the dynamics of the system under study, the sampling points space could be explored. Optimal conditions may be achieved by making ERC as large as possible for all variables. Additionally, the number as well as the type of interventions would be chosen based on the ERC, however, this is a different study which we leave it as future work.

\begin{figure}[!htb]
	\centering
	\subfigure[Time-course measurements and time-series interpolation]{%
		\includegraphics[width=.4\textwidth]{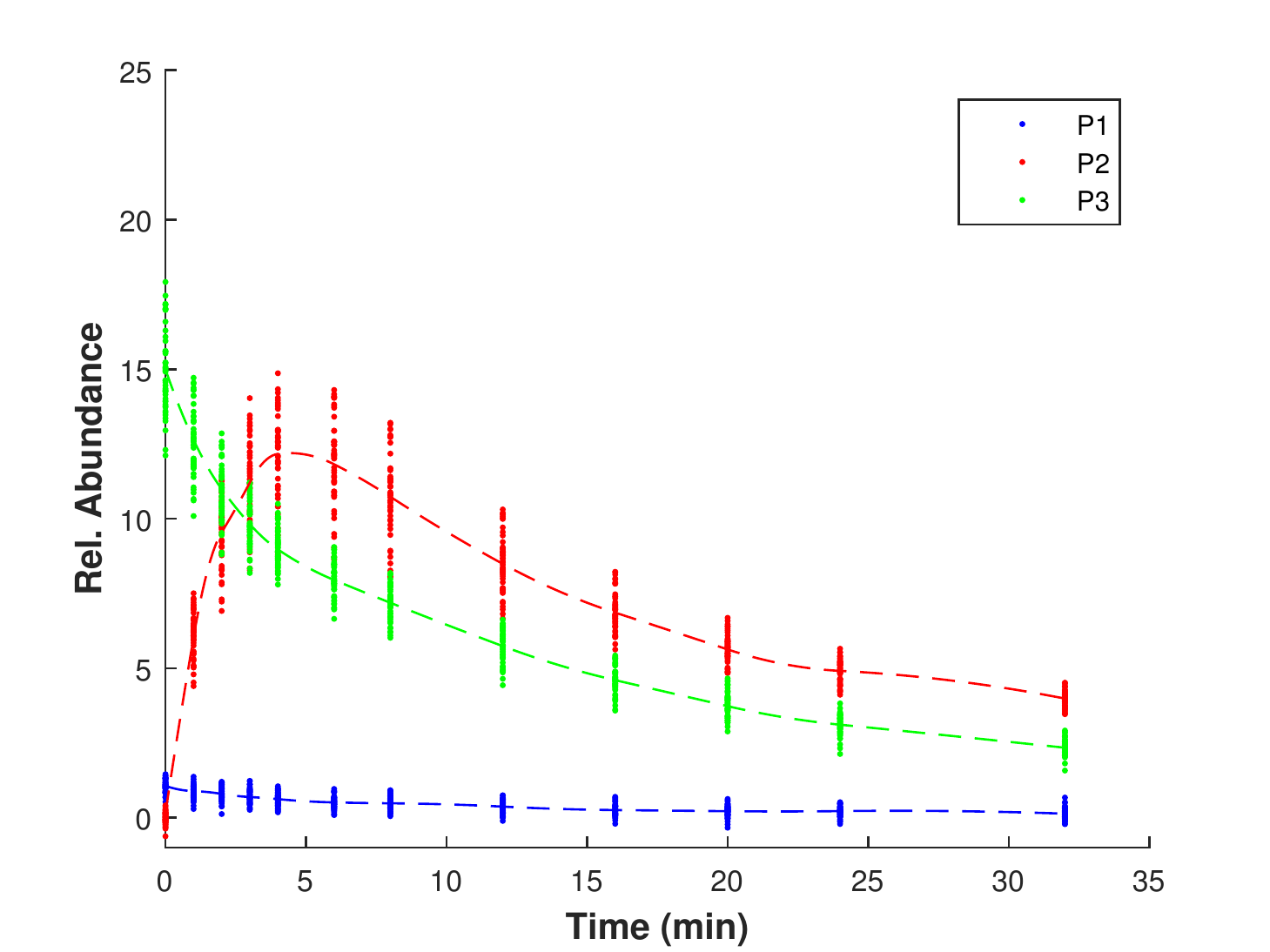}
		\label{tc:tp:gene:reg:diff:rxn:net}}
	\subfigure[MIP and ERC]{%
		\includegraphics[width=.4\textwidth]{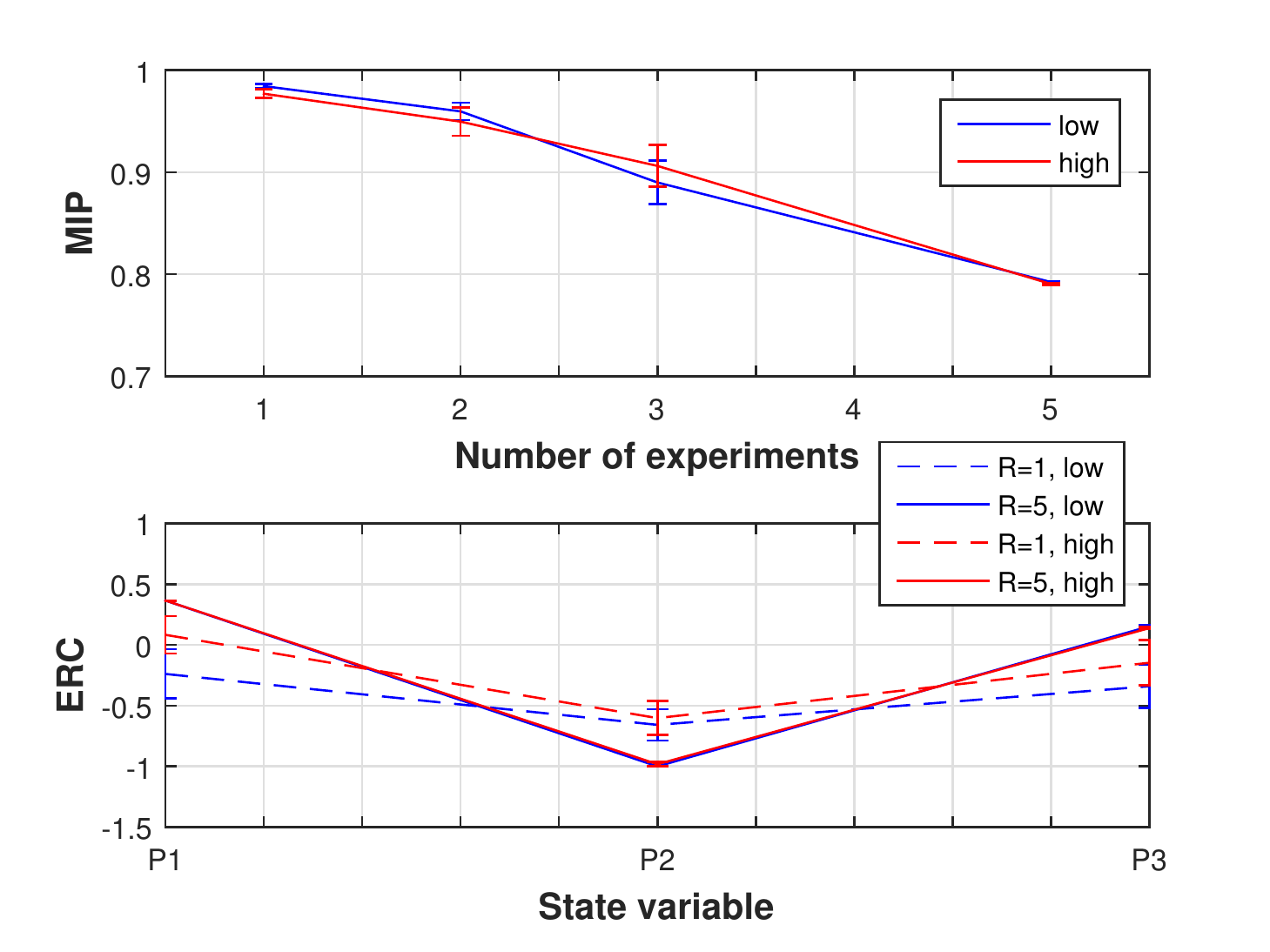}
		\label{mip:erc:gene:reg:diff:rxn:net}}
	\quad
	\subfigure[Precision and recall]{
		\includegraphics[width=.4\textwidth]{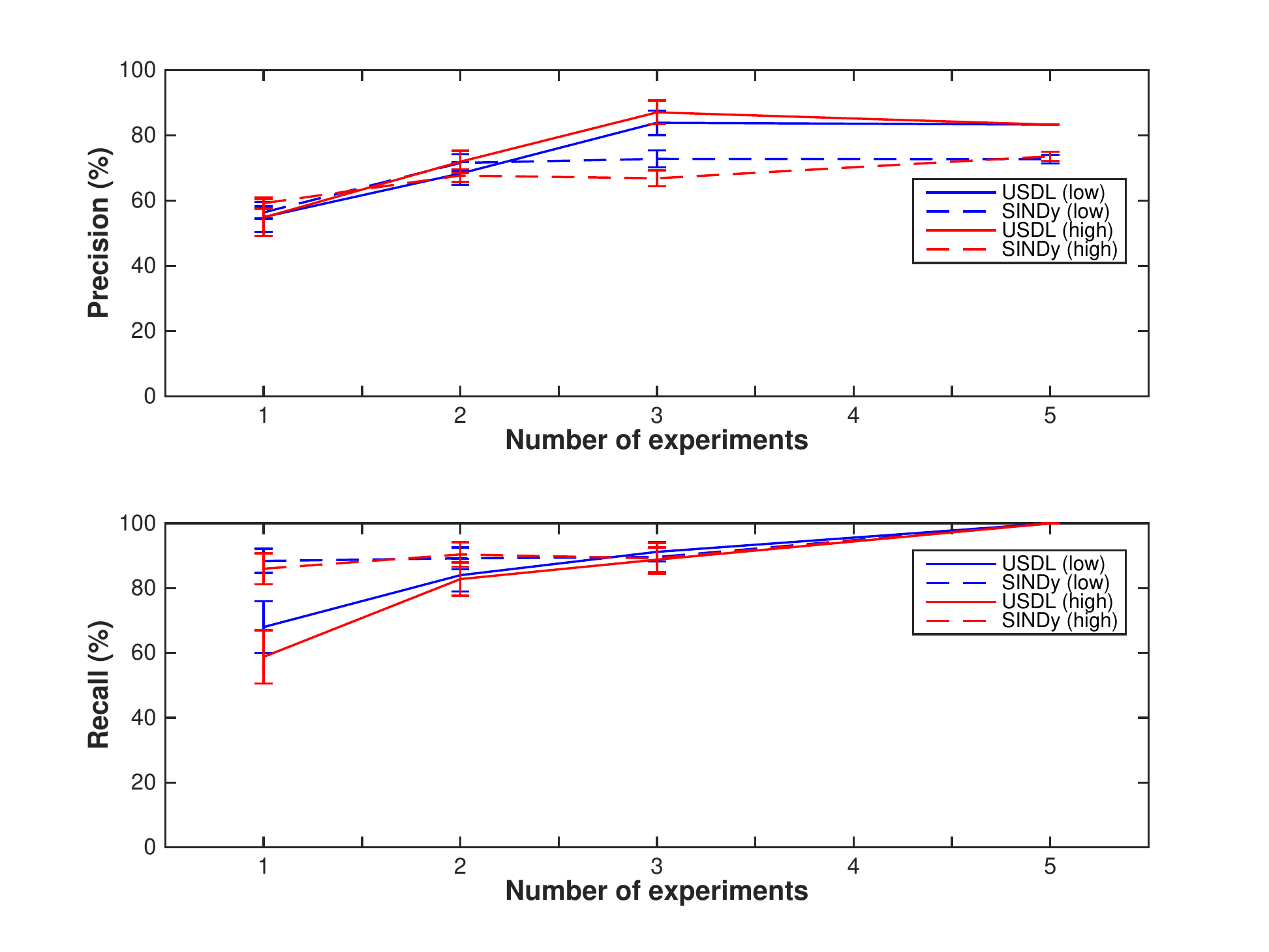}
		\label{pre:rec:gene:reg:diff:rxn:net}}
	\quad
	\subfigure[Time-series forecasts]{
		\includegraphics[width=.5\textwidth,height=.22\textheight]{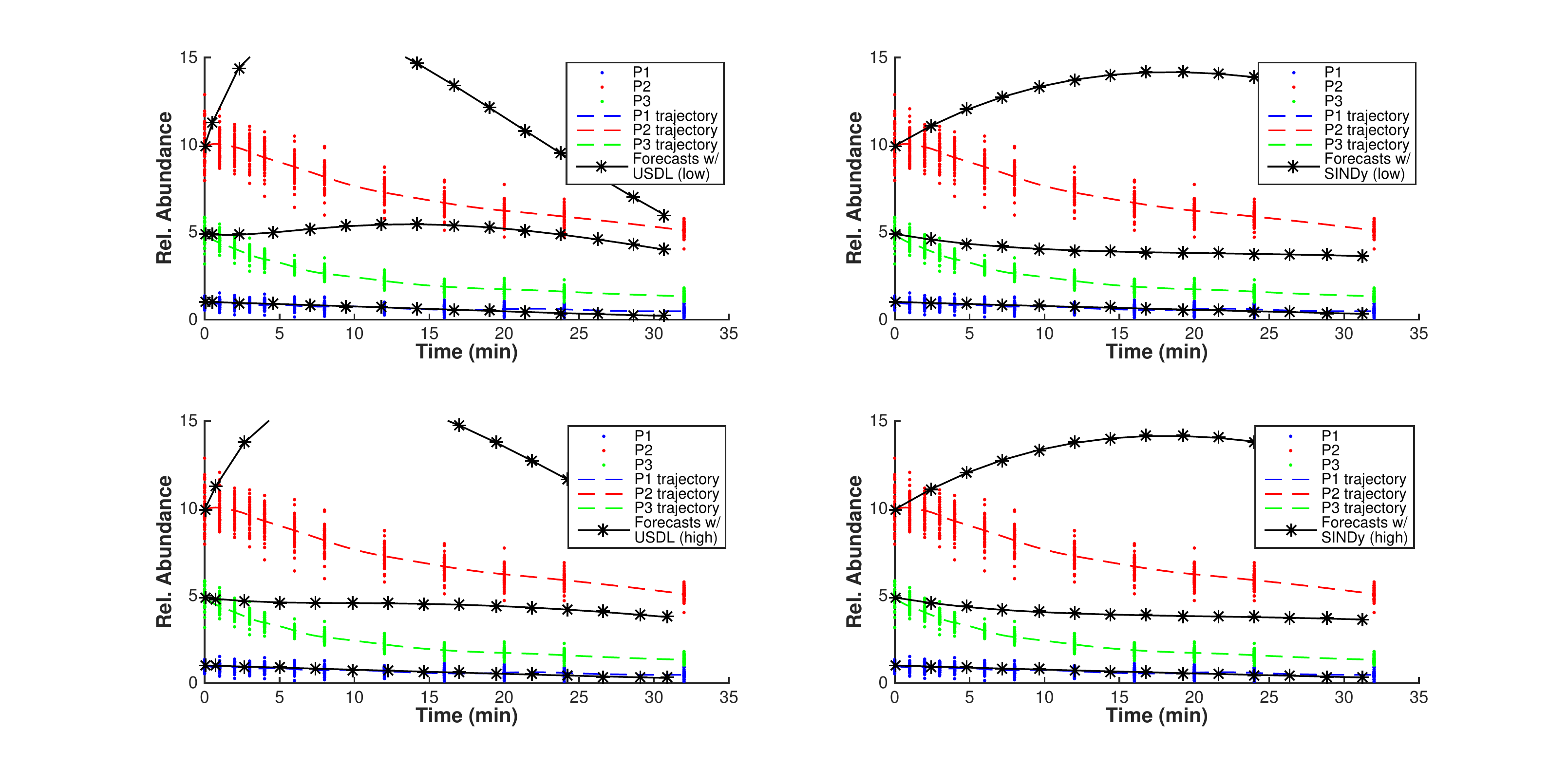}
		\label{forecast:gene:reg:diff:rxn:net}}
	\caption{(a) Time-series interpolation for the protein interaction network when different reaction constants are assumed without changing the topology of the network. (b)  MIP (upper panel) and ERC per state variable (lower panel) for two noise levels; low (blue) and high (red). ERC is negative for $P_2$ even when $R=5$ interventions are considered (solid lines). (c) Precision (upper panel) and recall (lower panel) curves as a function of $R$ for both USDL (solid lines) and SINDy (dashed lines) algorithms. Performance results indicate that perfect reconstruction is not achieved at any case which was again expected from the negativity of ERC for $P_2$. Perfect network reconstruction requires more interventions. (d) Time-series forecasts (black curves) based on the estimated parameter values for both USDL (left column) and SINDy (right column). The forecast error for $P_2$ is larger for this case confirming once again that  ERC nicely correlates with the forecast outcome.}
	\label{gene:reg:fig:diff:rxn:net}
\end{figure}

The final experiment assesses the performance of the proposed algorithm when the protein interaction network has different reaction constants. Nature rarely has the favourite constant values for an algorithmic approach and it is important to test it under different conditions. Figure~\ref{gene:reg:fig:diff:rxn:net} presents the results when the reaction constants of the first two reactions are doubled (i.e., $k_1=8$ and $k_2=0.1$). Even though the reduction of MIP, the perfect reconstruction of the network is lost and possibly more than five interventions are required in order to achieve true recovery of the protein interaction network. Interestingly, ERC is not positive for all variables with the problematic species being $P_2$ whose ERC is negative. For the case where all five interventions are taken into account, the inspection of the estimated connectivity matrix (i.e., of $A$) revealed that the row that describes the dynamics of $P_2$ was most of the times wrong (more than 95\% wrong) while the other two rows that describe the dynamics of $P_1$ and $P_3$ were most of the times correct (more than 95\% correct). Nevertheless, the total error in the structure inference significantly affect the forecast accuracy of all variables as Figure~\ref{forecast:gene:reg:diff:rxn:net} demonstrates.

Overall, this example demonstrates that when ERC is positive for all variables then with high probability the protein interaction network will be perfectly reconstructed. In contrast, when there are variables whose ERC is negative then the probability of perfect reconstruction becomes very low and more experiments are required in order to make ERC positive and thus recover the true network of interactions. Since ERC is a variable-dependent quantity, we can be confident that the found interactions between variables with positive ERC are true.
Finally, we remark that for small systems of interactions like this one, a brute force alternative is feasible. A complete search of all possible solutions when the non-zero components are less than ten is computationally tractable for dictionary size up to twenty atoms. However, such an approach will provide little or no information on how to design a new experiment or a new data acquisition policy compared with greedy algorithms or convex relaxation methods where metrics such as MIP and ERC can guide the experimental designer.

%%%%%%%%%%%%%%%%%%%%%%%%%%%%%%%%%%%%%%%%%%%%%%%%%%%%%%%%%%%%%%%%%
\subsection{Mass Cytometry}

\begin{figure}[!htb]
	\centering
	\includegraphics[width=.7\textwidth]{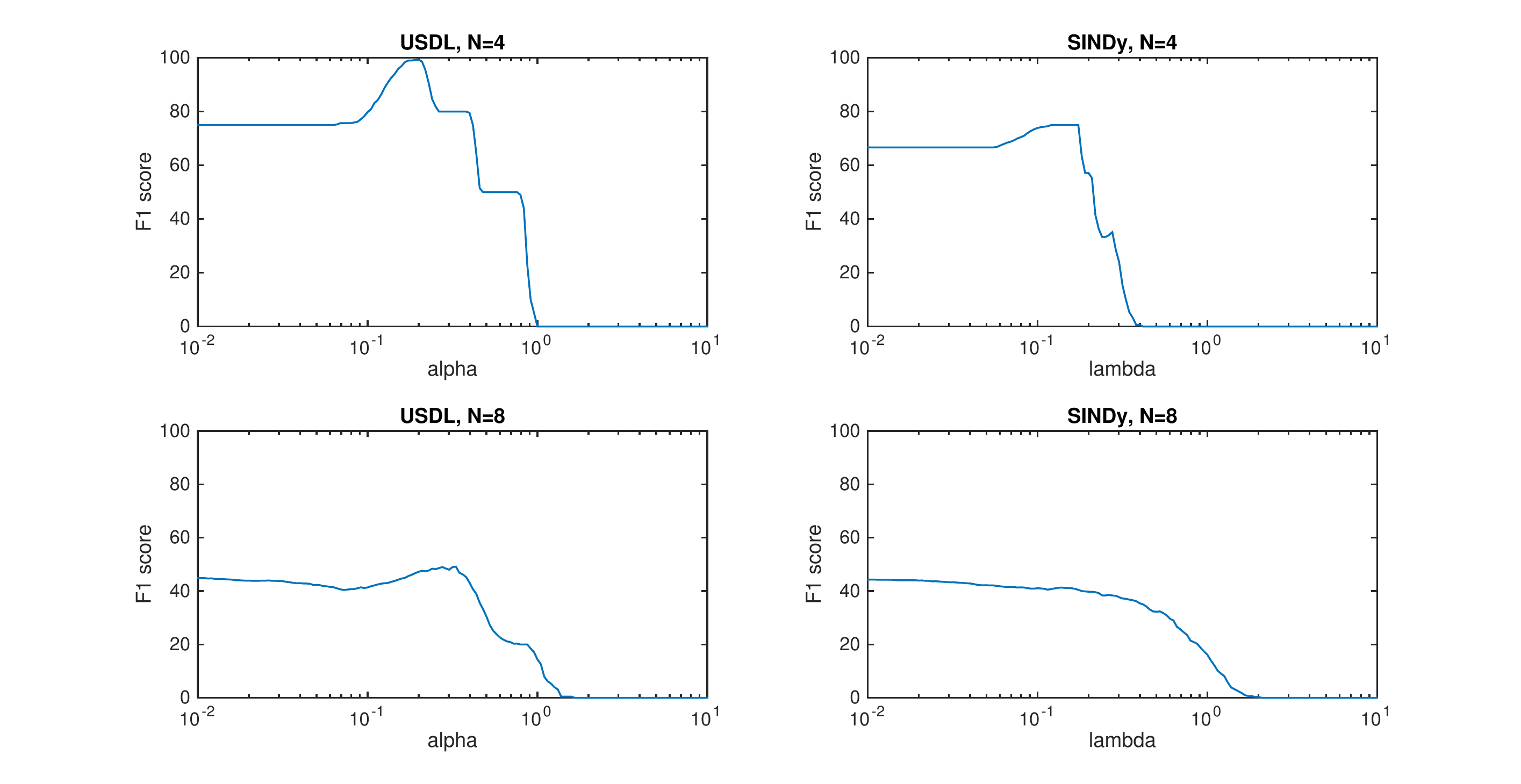}
	\caption{The F1 score as a function of the hyperparameter for both USDL (left column) and SINDy (right column). The first row shows the F1 score for the four protein subnetwork while the second row for the eight protein subnetwork. Evidently, F1 score decreases as the number of proteins is increased for both approaches.}
	\label{F1score:mass:cyto:fig}
\end{figure}

The protein interactions found in the literature are reported on Table~\ref{true:bio:net:table}. In the right most column, the scientific source of the corresponding interaction is provided. The smoothing weight in collocation method is $\lambda_C=10^4$. In order to make the results robust, we repeat the inference 100 times by using half of the points in each iteration. An edge is added to the network when it is found at least 80\% of the times. The hyperparameter tuning is based on the optimal F1 scores shown in Figure~\ref{F1score:mass:cyto:fig}.

\begin{table}[!htb]
	\begin{center}
		\caption{Protein interactions found in the literature.}
		\begin{tabular}{|c|c|} \hline
			Interaction & Literature \\ \hline \hline
			CD3z $\rightarrow$ Slp76 & Krishnaswamy et al. \cite{Krishnaswamy2014} \\ \hline
			Slp76 $\rightarrow$ Erk & Krishnaswamy et al. \cite{Krishnaswamy2014} \\ \hline
			Erk $\rightarrow$ S6 & Krishnaswamy et al. \cite{Krishnaswamy2014} \\ \hline
			Erk $\rightarrow$ Creb & Krishnaswamy et al. \cite{Krishnaswamy2014} \\ \hline
			CD3z $\rightarrow$ MAPKAPKII & Krishnaswamy et al. \cite{Krishnaswamy2014} \\ \hline
			MAPKAPKII $\rightarrow$ Creb & Krishnaswamy et al. \cite{Krishnaswamy2014} \\ \hline
			Akt $\rightarrow$ Rb & Krishnaswamy et al. \cite{Krishnaswamy2014} \\ \hline
			Akt $\rightarrow$ S6 & Krishnaswamy et al. \cite{Krishnaswamy2014} \\ \hline
			Akt $\rightarrow$ Creb & PI3K-AKT Signaling (KEGG) \\ \hline
%			Akt $\rightarrow$ Erk & PI3K-AKT Signaling (KEGG) \\ \hline
		\end{tabular}
		\label{true:bio:net:table}
	\end{center}
\end{table}

Table~\ref{mass:cyto:mip:erc} reports the value of MIP and ERC per variable for the two subnetworks presented in the main text. MIP is almost 1 showing that there is strong collinearity between the columns of dictionary matrix, $\Psi$. ERC values for the small subnetwork are positive increasing the confidence that the inferred network is correct. On the contrary, some variables have negative ERC for the larger subnetwork casting doubt on the correctness of the inferred network.

\begin{table}[!htb]
	\begin{center}
		\caption{MIP and ERC values for the two subnetworks of proteins. Positive ERC adds confidence to the inference results, however, it is just an estimate since the ground truth is unknown.}
		\begin{tabular}{|c||c||c|c|c|c|c|c|c|c|} \hline
			& MIP & \multicolumn{8}{c|}{ERC} \\ \hline
			& -- & CD3z & SLP76 & Erk & S6  & MAPKAPKII & Creb & Akt & Rb \\ \hline 
			Subnet 1 & 0.967 & 0.032 & 0.114 & 0.020  & 0.065 & -- & -- & -- & -- \\ \hline \hline
			Subnet 2 & 0.980 & 0.031 & 0.119 & -0.314  & -0.256  & -0.163 & 0.031 & 0.119 & -2.545 \\ \hline
		\end{tabular}
		\label{mass:cyto:mip:erc}
	\end{center}
\end{table}

\begin{figure}[!htb]
	\centering
	\subfigure[USDL]{%
		\includegraphics[width=.13\textwidth]{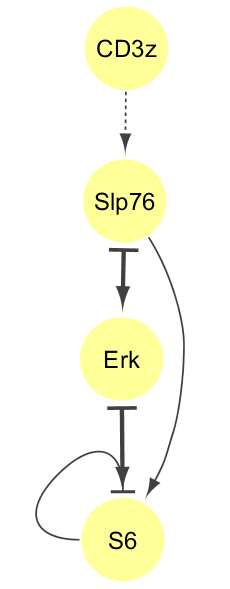}
		\label{prot4:less:sampl}}
	\quad
	\subfigure[SINDy]{%
		\includegraphics[width=.14\textwidth]{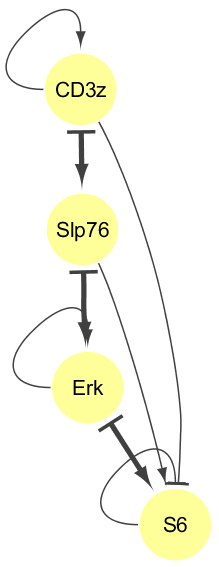}
		\label{prot4:less:sampl:sindy}}
	\quad
	\subfigure[USDL]{%
		\includegraphics[width=.31\textwidth]{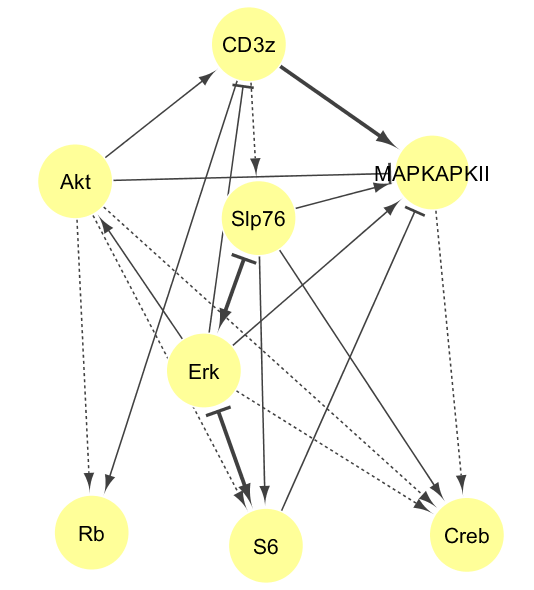}
		\label{prot8:less:sampl}}
	\quad
	\subfigure[SINDy]{%
		\includegraphics[width=.29\textwidth]{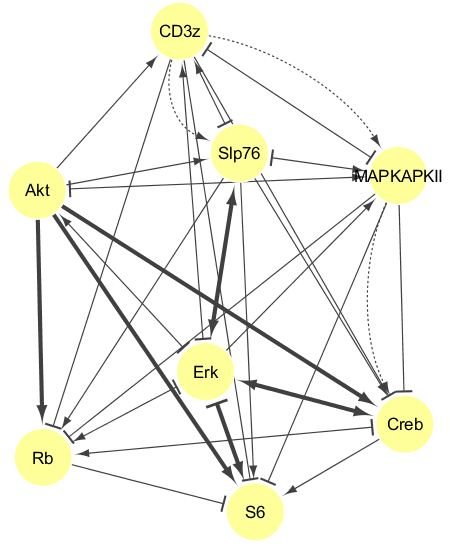}
		\label{prot8:less:sampl:sindy}}
	\caption{Reconstructed protein signalling networks when half of the sampling points are provided.}
	\label{prot4:dremi:fig1}
\end{figure}

Next, we examine the inference capabilities of the proposed approach under various experimental conditions. Figure~\ref{prot4:dremi:fig1} presents the first experiment where we eliminate half of the sampling points. The new sampling points are at 0, 2, 5, 8, 20 minutes. As Figure~\ref{prot4:less:sampl} asserts the network inference with USDL deteriorates since the interaction CD3z $\rightarrow$ Slp76 is missed while an additional interaction that Slp76 produces S6 is found. SINDy is minimally affected by the elimination of the half measurements inferring though an almost fully-connected graph. For the eight-protein network, the inference deteriorates when less sampling points are present for both USDL and SINDy (see Figure~\ref{prot8:less:sampl} \& (d), respectively).

\begin{figure}[!htb]
	\centering
	\subfigure[USDL]{%
		\includegraphics[width=.145\textwidth]{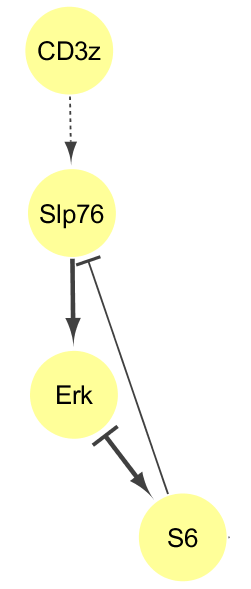}
		\label{prot4:all:act}}
	\quad
	\subfigure[SINDy]{%
		\includegraphics[width=.09\textwidth]{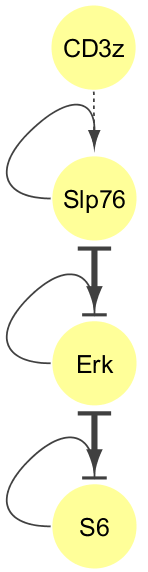}
		\label{prot4:all:act:sindy}}
	\quad
	\subfigure[USDL]{%
		\includegraphics[width=.33\textwidth]{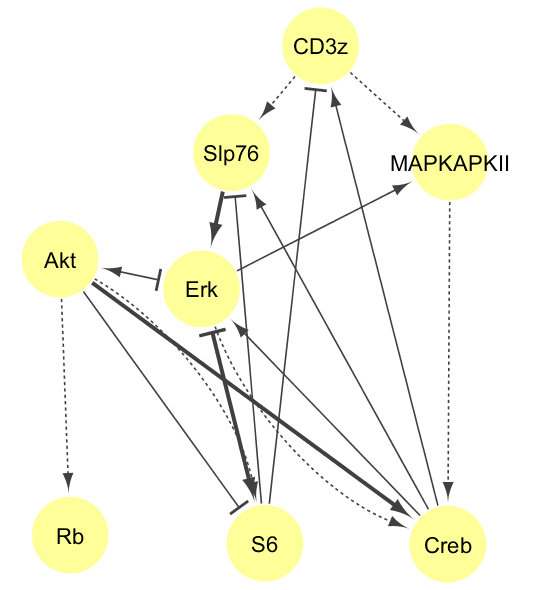}
		\label{prot8:all:act}}
	\quad
	\subfigure[SINDy]{%
		\includegraphics[width=.28\textwidth]{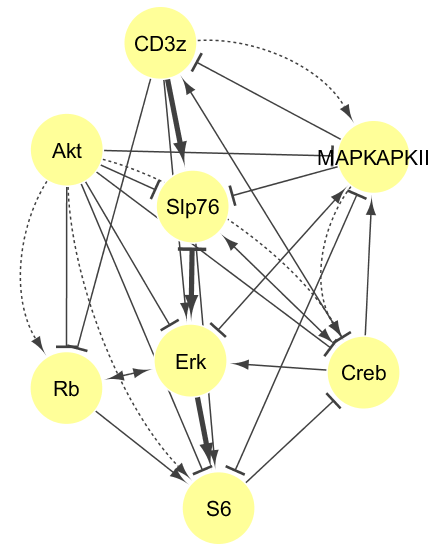}
		\label{prot8:all:act:sindy}}
	\caption{Reconstructed protein signalling network when both activation cocktails are provided.}
	\label{prot4:dremi:fig2}
\end{figure}

\begin{figure}[!htb]
	\centering
	\subfigure[USDL]{%
		\includegraphics[width=.13\textwidth]{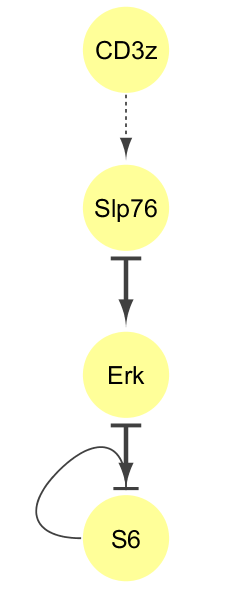}
		\label{prot4:all:pop}}
	\quad
	\subfigure[SINDy]{%
		\includegraphics[width=.15\textwidth]{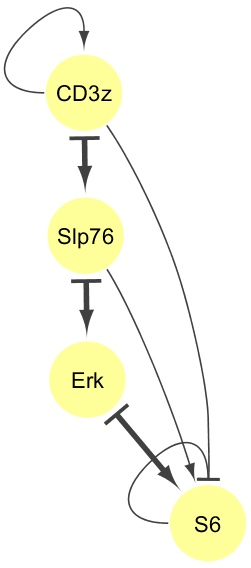}
		\label{prot4:all:pop:sindy}}
	\quad
	\subfigure[USDL]{%
		\includegraphics[width=.31\textwidth]{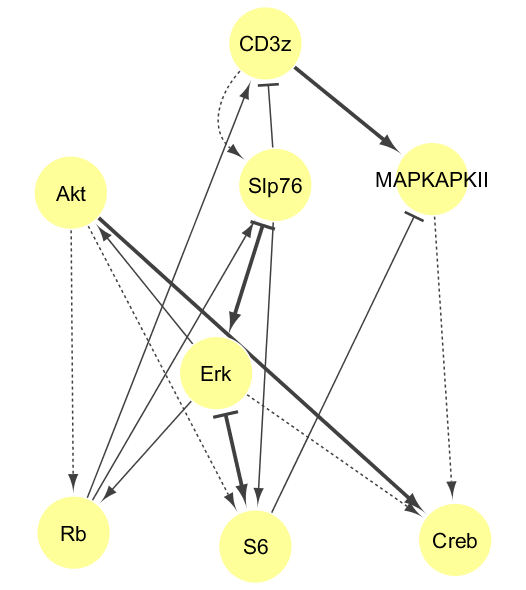}
		\label{prot8:all:pop}}
	\quad
	\subfigure[SINDy]{%
		\includegraphics[width=.28\textwidth]{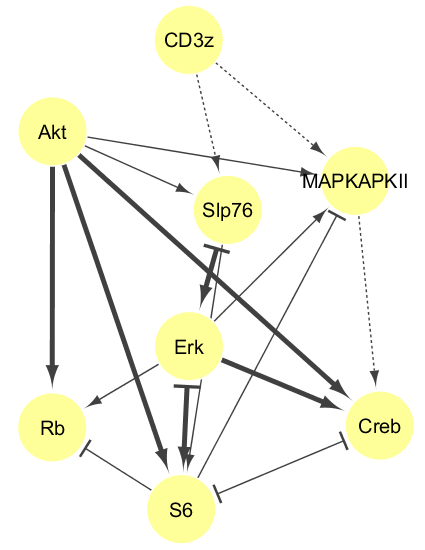}
		\label{prot8:all:pop:sindy}}
	\caption{Reconstructed protein signalling network when three dinstict populations are provided.}
	\label{prot4:dremi:fig3}
\end{figure}

Figure~\ref{prot4:dremi:fig2} demonstrates the reconstructed networks when both activation cocktails are taken into consideration. Again, network inference deteriorates for both algorithms even though we provide additional data from another intervention. A probable explanation is that there exist interference between measured and unmeasured proteins with the unmeasured ones serve as latent variables. These extra sources of noise confuse the inference algorithm resulting in inferior performance. 
We also tested the performance when all three available populations are taken into consideration and fed to the inference algorithms and the results are presented in Figure~\ref{prot4:dremi:fig3}. One activation cocktail (CD3/CD28) is considered. For the USDL algorithm there are few edges that are missing. On the contrary, SINDy algorithm especially for the eight-protein network is able to reconstruct most of the interactions with a minimum number of false positives. This is a rather unexpected result since different populations have different variable connectivity.

The final experiment is to add all the available measurements with all three populations and both activation cocktails (Figure~\ref{prot4:dremi:fig3}). The results as quantified by the reconstructed networks are similar to the case where both activation cocktails and one population (i.e., CD4 naive) is considered. Overall, we conclude that more data does not always produce superior results and caution is necessary in the experimental design in order to reveal the actual protein interactions. Interestingly, the worst performance results for both inference algorithms are obtained when all activators are considered . In this case, only two protein interactions that have been reported in the literature are found.
It is also noteworthy that there are two interactions that are present in all settings of the experiments. These interactions form the cascade Slp76 $\rightarrow$ Erk $\rightarrow$ S6. Moreover, in all experiments with USDL, the edge between Erk and Akt is always inferred which can be either a true interaction not reported or a way to up-regulate Akt since in the T-cell signaling pathway Akt is phosphorylated by CD28 which is not measured.

\begin{figure}[!htb]
	\centering
	\subfigure[USDL]{%
		\includegraphics[width=.165\textwidth]{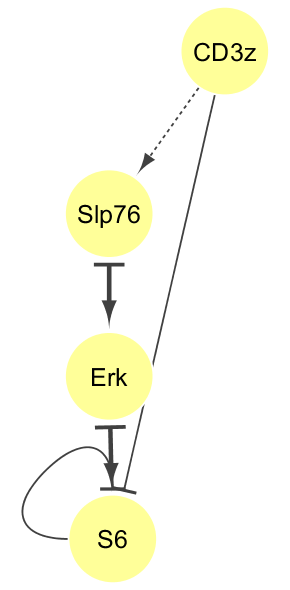}
		\label{prot4:all:act:all:pop}}
	\quad
	\subfigure[SINDy]{%
		\includegraphics[width=.095\textwidth]{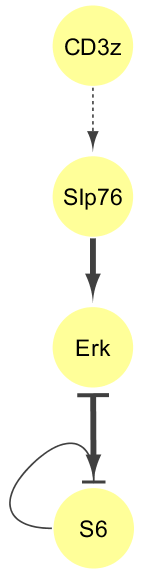}
		\label{prot4:all:act:all:pop:sindy}}
	\quad
	\subfigure[USDL]{%
		\includegraphics[width=.32\textwidth]{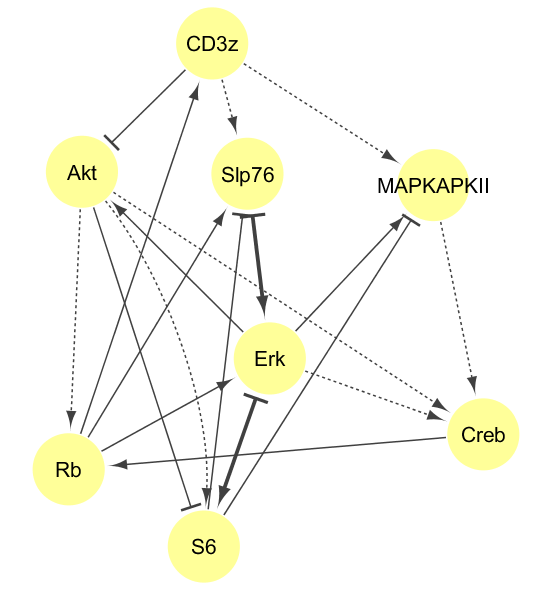}
		\label{prot8:all:act:all:pop}}
	\quad
	\subfigure[SINDy]{%
		\includegraphics[width=.295\textwidth]{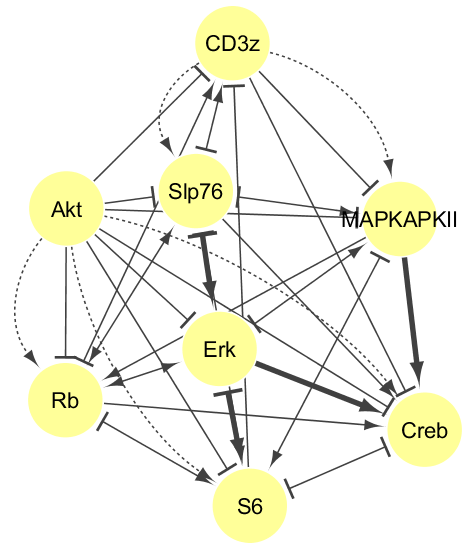}
		\label{prot8:all:act:all:pop:sindy}}
	\caption{Reconstructed protein signalling network when all available data are provided.}
	\label{prot4:dremi:fig4}
\end{figure}

\begin{figure}[!htb]
	\centering
	\includegraphics[width=.48\textwidth]{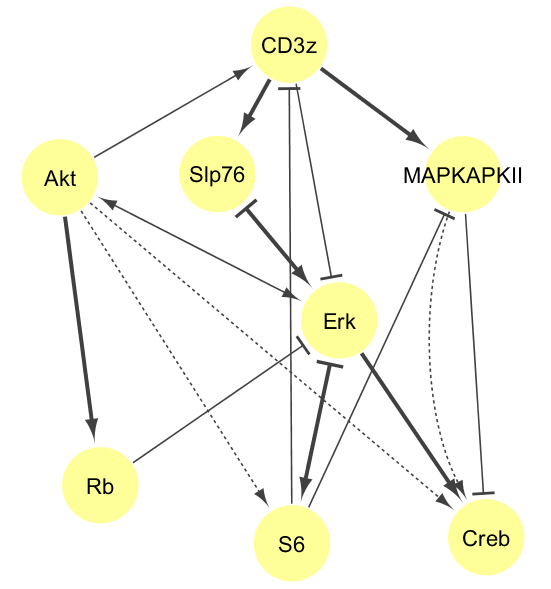}
	\caption{Reconstructed eight-protein subnetwork when prior knowledge is added. Measurements are from CD4+ naive cells with CD3/CD28 activation cocktail.}
	\label{prot8:dremi:prior:fig}
\end{figure}

Finally, we present an experiment where prior knowledge is provided to the inference algorithm. It is legitimate to use any available prior information to guide the inference algorithm especially when a scientist is searching for new knowledge. Our algorithm and in particular OMP can easily incorporate prior knowledge as discussed earlier. Thus, one can provide the known interactions and look for new ones. In Figure~\ref{prot8:dremi:prior:fig} we present the reconstructed network when the following interactions are a priori provided: CD3z $\rightarrow$ Slp76, CD3z $\rightarrow$ MAPKAPKII, MAPKAPKI I $\rightarrow$ Creb and Erk $\rightarrow$ Creb. There are two immediate observations related with Creb that can be made. First, the interaction Akt $\rightarrow$ Creb is lost probably because it has been replaced by Erk $\rightarrow$ Creb and, second, instead of up-regulating Creb, MAPKAPKII down-regulates it questioning the existence of this interaction.

\newpage

%%%%%%%%%%%%%%%%%%%%%%%%%%%%%%%%%%%%%%%%%%%%%%%%%%%%%%%%%%%%%%%%%%%%%%%%%%%%%%%%%%
\subsection{Multidimensional Ornstein-Uhlenbeck process}
The multidimensional Ornstein-Uhlenbeck (OU) process is defined in the more standard notation as
\begin{equation}
d x_t = -A x_t dt + \sigma dB_t
\end{equation}
where $A$ is the connectivity matrix, $B_t$ is an $N$-dimensional standard Brownian motion while $\sigma$ is the noise level which is set to 0.25. The OU process satisfies the detailed balance condition \cite{Gardiner2004} thus, at the stationary regime, it is time-reversible meaning that the reflected time-series have the same path distribution. Moreover, the stationary distribution, $\mu(x)$, is proportional to
\begin{equation}
\mu(x) \propto e^{-\frac{1}{\sigma^2} x^T\Sigma^{-1}x} \ .
\end{equation}
Obviously, it is a zero-mean Gaussian. The covariance matrix $\Sigma$ is found as the solution of the Lyapunov equation $A\Sigma+\Sigma A^T + I = 0$. Concentration matrix which is defined as the inverse of the covariance matrix is not necessarily sparse even when $A$ is sparse. This implies that if the measurements are obtained as i.i.d. samples from the stationary distribution then it is impossible to infer the causal relationships between the state variables since they have been lost. On the contrary, using the richer information that is contained by the  dynamics, primarily time correlations, the true connectivity matrix is estimated and the causal relations can be correctly inferred.

\begin{figure}[!htb]
	\centering
	\subfigure[Time-series for both regimes]{
		\includegraphics[width=.47\textwidth]{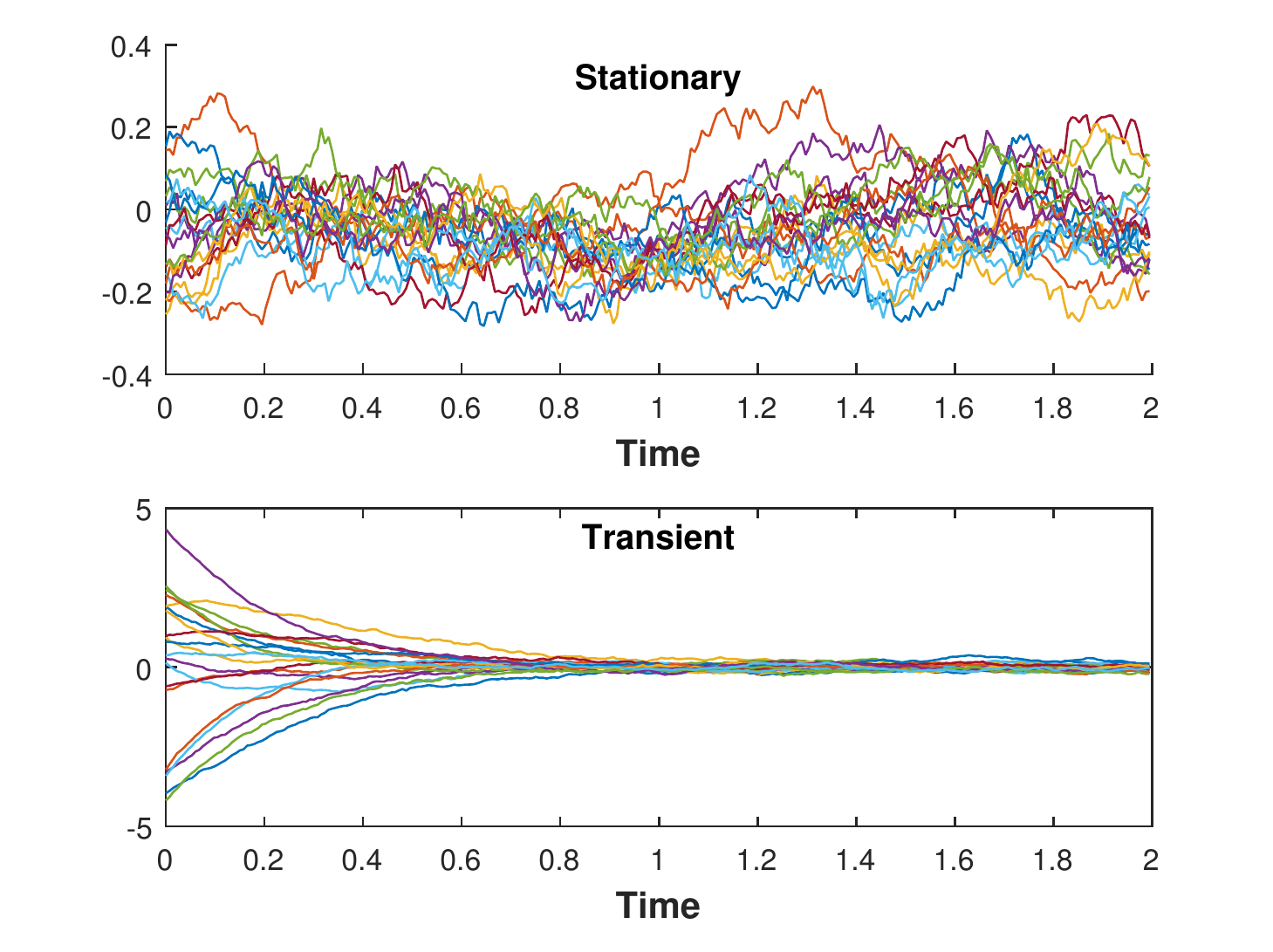}
		\label{OUprocess:timeseries:fig}}
	\quad
	\subfigure[ERC per state variable]{
		\includegraphics[width=.47\textwidth]{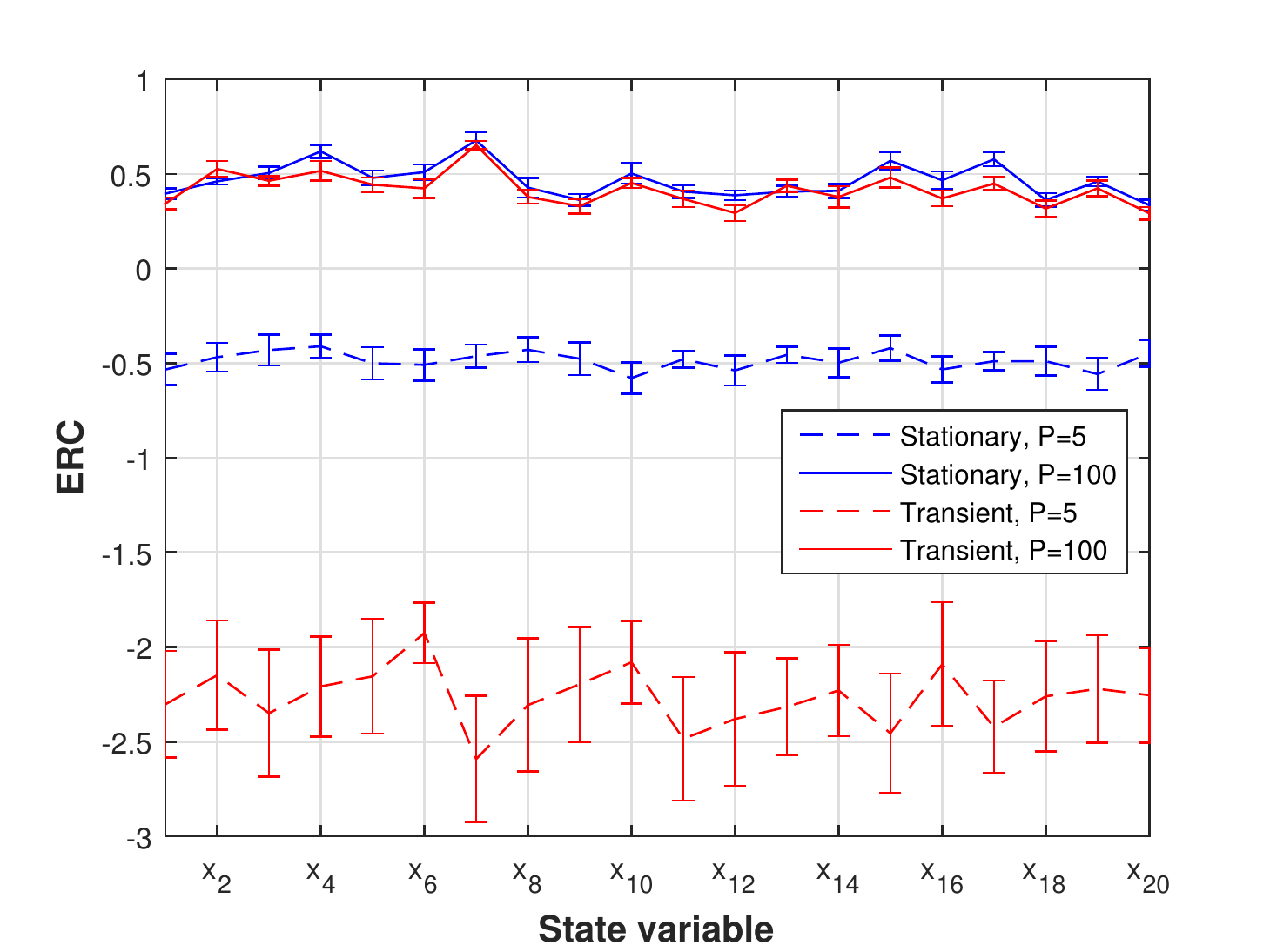}
		\label{OUprocess:ERC:fig}}
	\caption{(a) Time-series of a multi-dimensional OU process at stationary regime (upper plot) as well at transient regime (lower plot). (b) ERC for each variable of the OU process at both regimes and two different number of trajectories. ERC is positive when $P=200$ time-series are given (solid lines).}
\end{figure}

\begin{figure}[!htb]
	\centering
	\subfigure[F1 score]{
		\includegraphics[width=.5\textwidth,height=0.21\textheight]{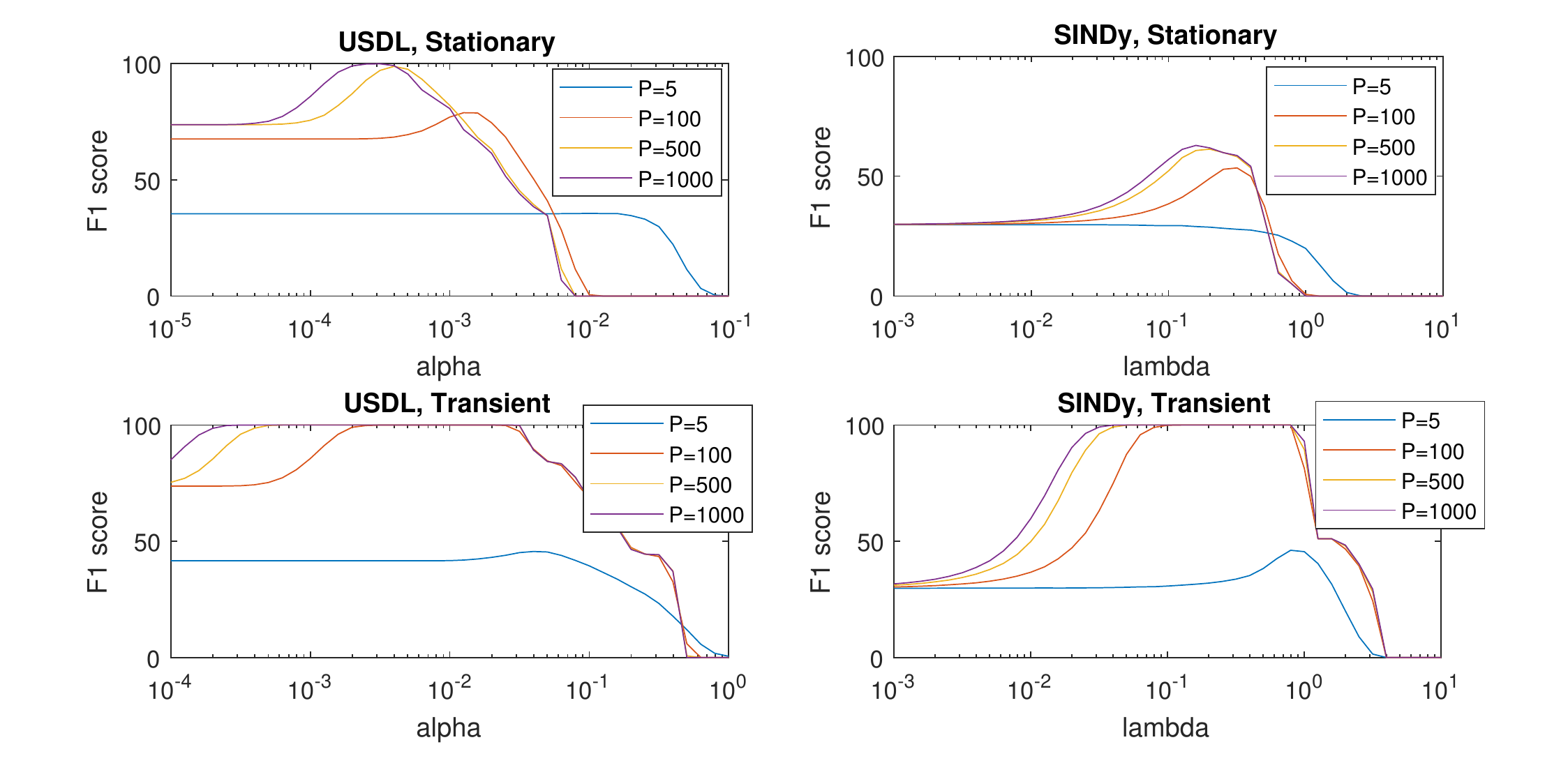}
		\label{F1score:OUprocess:peakyFourier}}
	\quad
	\subfigure[RMSE]{
		\includegraphics[width=.4\textwidth]{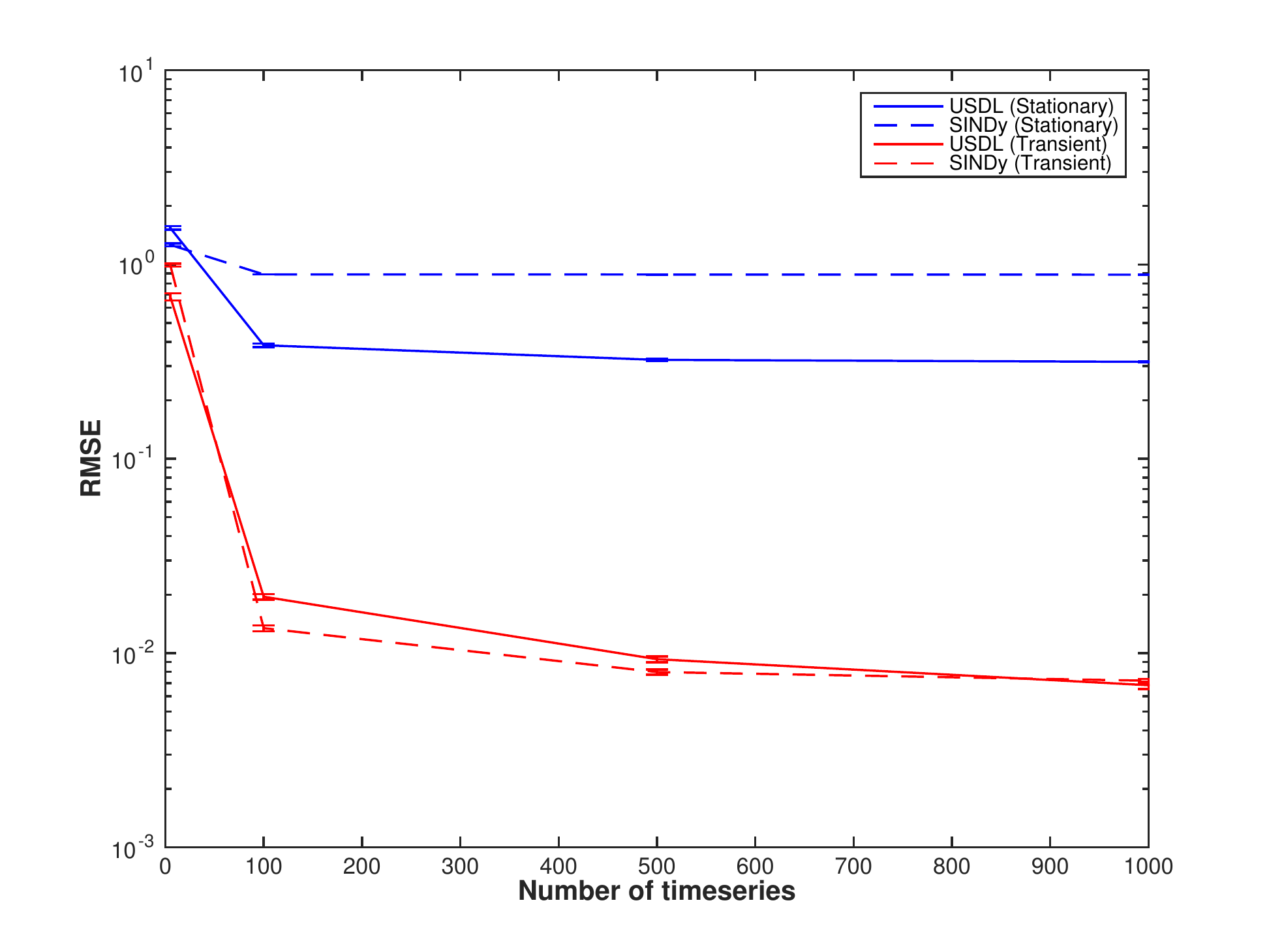}
		\label{rmse:OUprocess:peakyFourier}}
	\caption{(a) The F1 score as a function of the hyperparameter for both USDL (left column) and SINDy (right column) and various number of time-series using the peaky Fourier modes as test functions. (b) The respective RMSE for both USDL (solid lines) and SINDy (dashed lines). The parameter estimation is more accurate at the transient regime (almost two orders of magnitude) due to the higher signal-to-noise ratio compared to the stationary regime.}
	\label{F1score:OUprocess:fig}
\end{figure}

Proceeding, the time-series of an $N=20$-dimensional Ornstein-Uhlenbeck (OU) process are presented in Figure~\ref{OUprocess:timeseries:fig}. The upper panel shows the time-series at the stationary regime while the lower panel shows the time-series at the transient regime. At the transient regime, the initial values of the process were independently sampled from a Gaussian distribution with variance one which obviously result to starting the process out of equilibrium. The OU process converges to equilibrium after one time unit. The numerical integration is performed using Euler-Maruyama scheme which is a first-order finite difference scheme with time step of $\Delta t=0.001$. 
The results of Fig. 3 in the main text are produced by setting the maximum allowed non-zero elements in USDL algorithm to be $K=6$ while the threshold is set according to the maximum value of the F1 score which is shown in Figure~\ref{F1score:OUprocess:peakyFourier}. SINDy algorithm's hyperparameter is similarly set. Figure~\ref{F1score:OUprocess:peakyFourier} also reveals that there is a large region of hyperparameter values that produce perfect reconstruction at the transient regime. The region of optimal values is much smaller at the stationary regime but as we increase the number of time-series it tends to increase at least for the USDL algorithm. Additionally, this figure highlights the significant superiority of the USDL algorithm over SINDy algorithm when the noise is prominent (i.e., at the stationary regime) showing the generality of the proposed methodology.

Figure~\ref{OUprocess:ERC:fig} shows ERC values per variable. It is evident that when $P=5$ time-series are fed to USDL algorithm (dashed lines), ERC is negative in both regimes for all variables. Moreover, ERC for the transient regime is worse that the stationary regime which might be caused due to different threshold values.
When $P=100$ time-series are fed to the inference algorithm then ERC at both regimes is positive for all variables and ERC take slightly larger values in the transient regime. Nevertheless, as it is evident from the precision-recall curves (Fig. 2(b) in the main text where precision is slightly above 50\%), perfect reconstruction is not achieved with stationary time-series. 
This is a consequence of the fact that the strength of the interaction coefficients (i.e., non-zero elements of $A$) compared to the noise level is low. Indeed, noise is the primal driving force of the dynamics in the stationary regime thus it is harder to infer accurately the connectivity matrix. Only when $P=1000$ the signal-to-noise ratio is high enough for perfect reconstruction of the dynamical system.

Figure~\ref{rmse:OUprocess:peakyFourier} presents the RMSE results for the parameter estimation of the connectivity matrix $A$. RMSE results are qualitatively similar to the precision-recall results since RMSE at the transient regime is two orders of magnitude less that the RMSE at the stationary regime. Furthermore, USDL performs better at the stationary regime than SINDy revealing that a well-educated selection of the family of test functions could also improve the parameter estimation outcomes.

\subsubsection{Selection of Test Functions}
\begin{figure}[!htb]
	\centering
	\subfigure[MIP and ERC]{%
		\includegraphics[width=.47\textwidth]{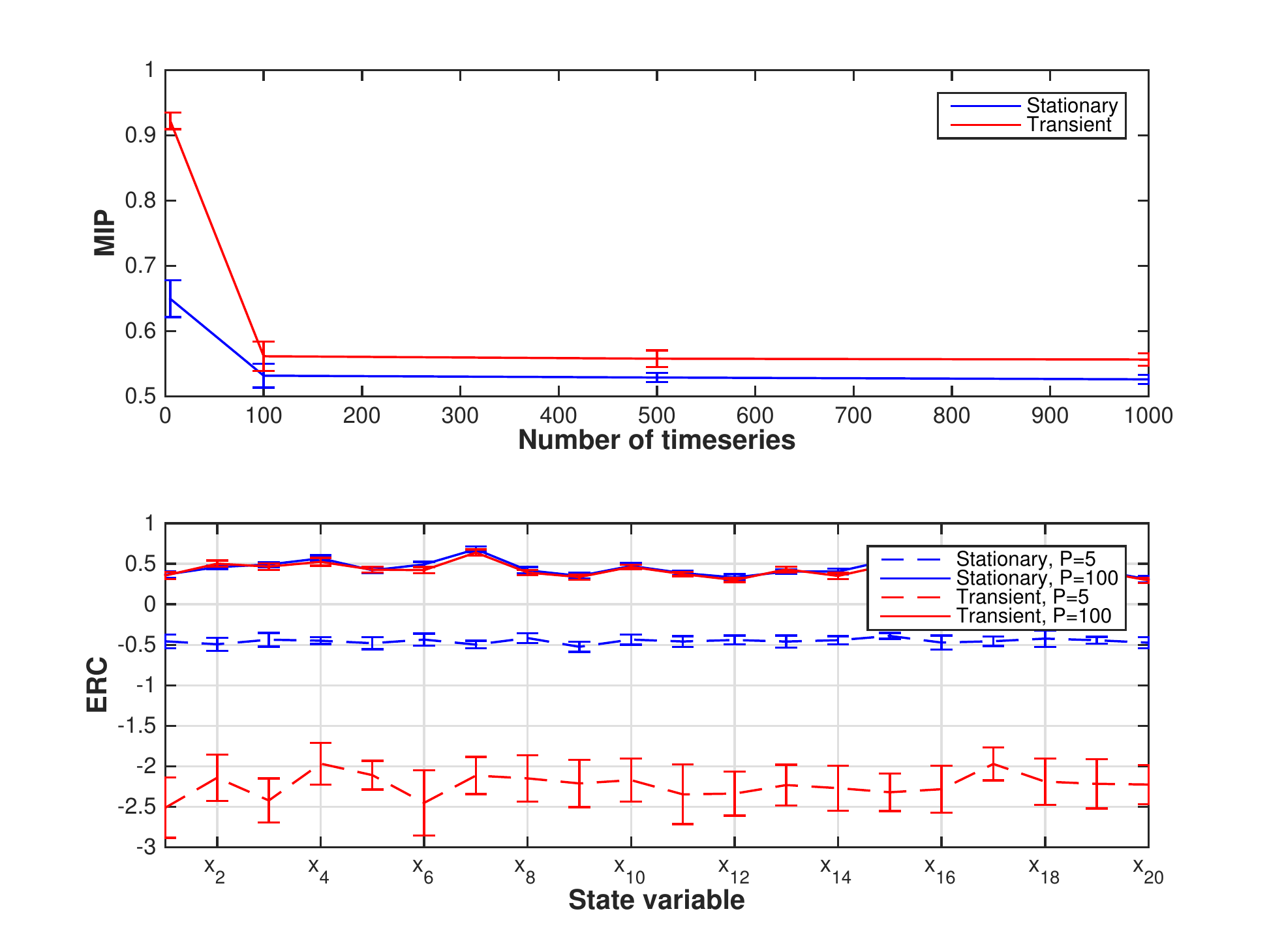}
		\label{mip:erc:OUprocess:more:tf}}
	\quad
	\subfigure[Precision and recall]{
		\includegraphics[width=.47\textwidth]{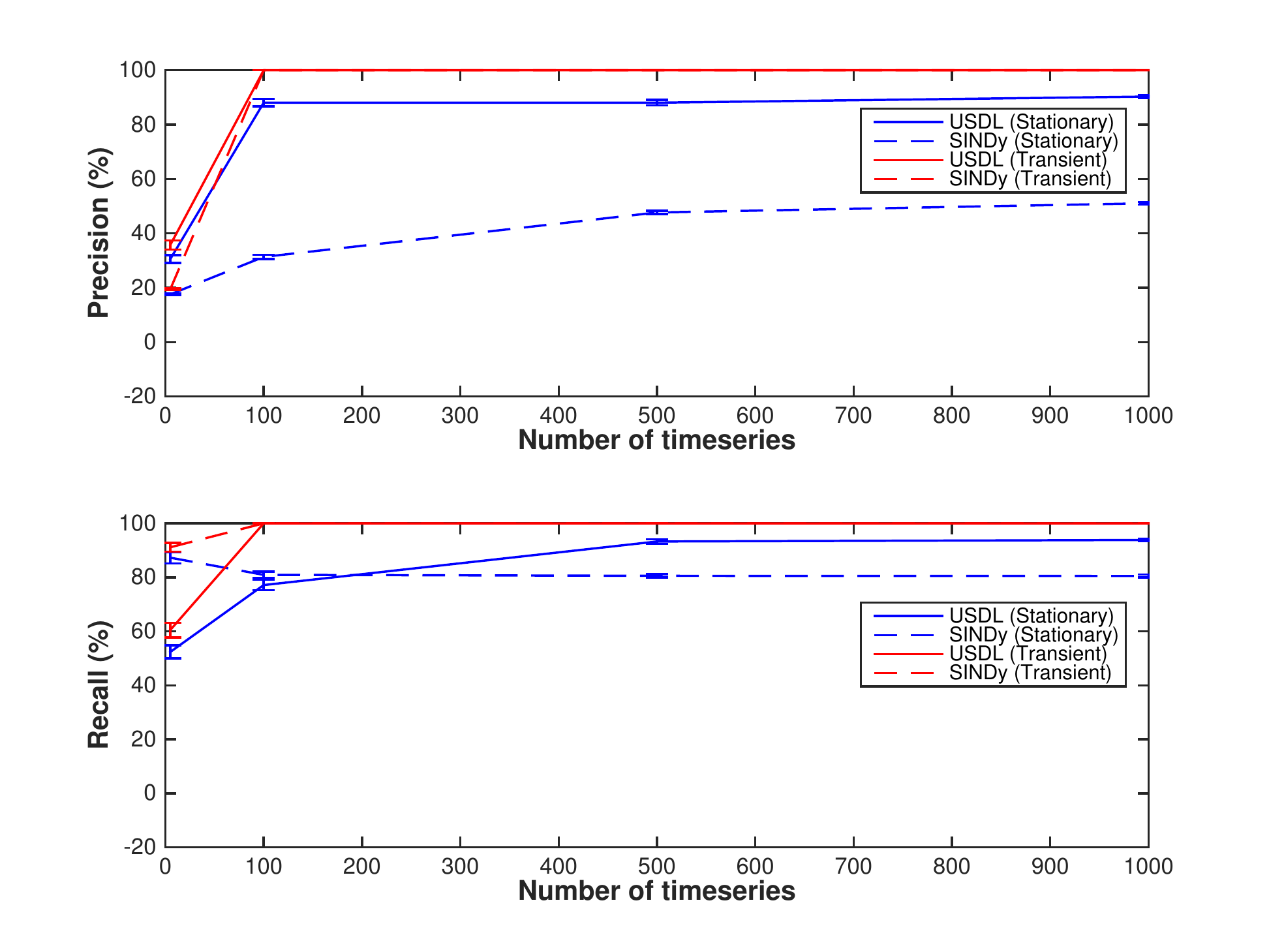}
		\label{pre:rec:OUprocess:more:tf}}
	\caption{(a) MIP (upper panel) and ERC per state variable (lower panel) for the OU process when more test functions are considered. The number of Fourier modes is $M=81$. (b) Precision (upper panel) and recall (lower panel) curves as a function of $P$, i.e., the number of trajectories for both USDL and SINDy algorithms. However, perfect reconstruction is not achieved.}
	\label{OUprocess:fig:more:tf}
\end{figure}

\begin{figure}[!htb]
	\centering
		\subfigure[F1 score]{%
		\includegraphics[width=.5\textwidth,height=0.21\textheight]{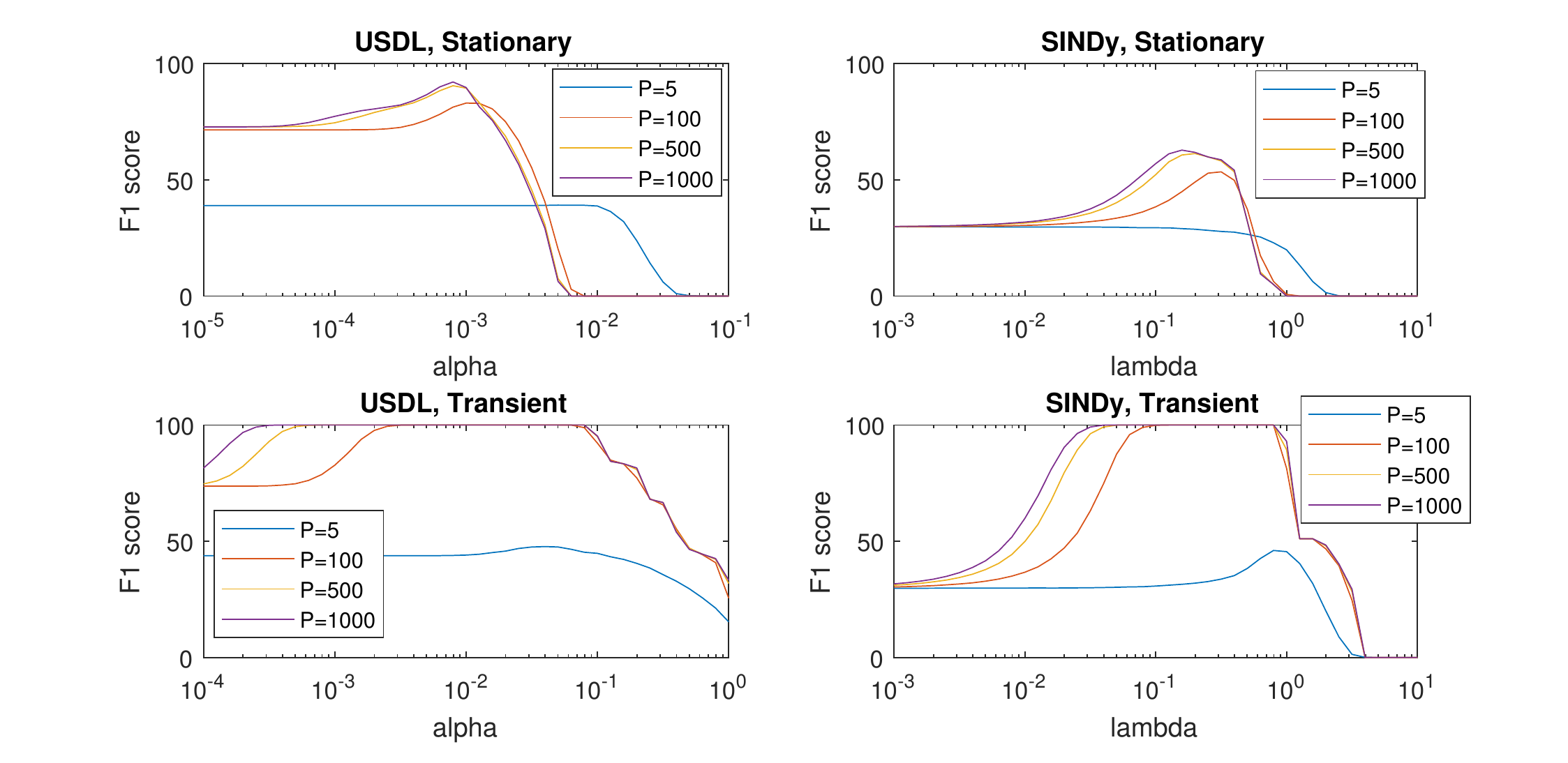}
		\label{F1score:OUprocess:Fourier}}
	\quad
	\subfigure[RMSE]{%
		\includegraphics[width=.4\textwidth]{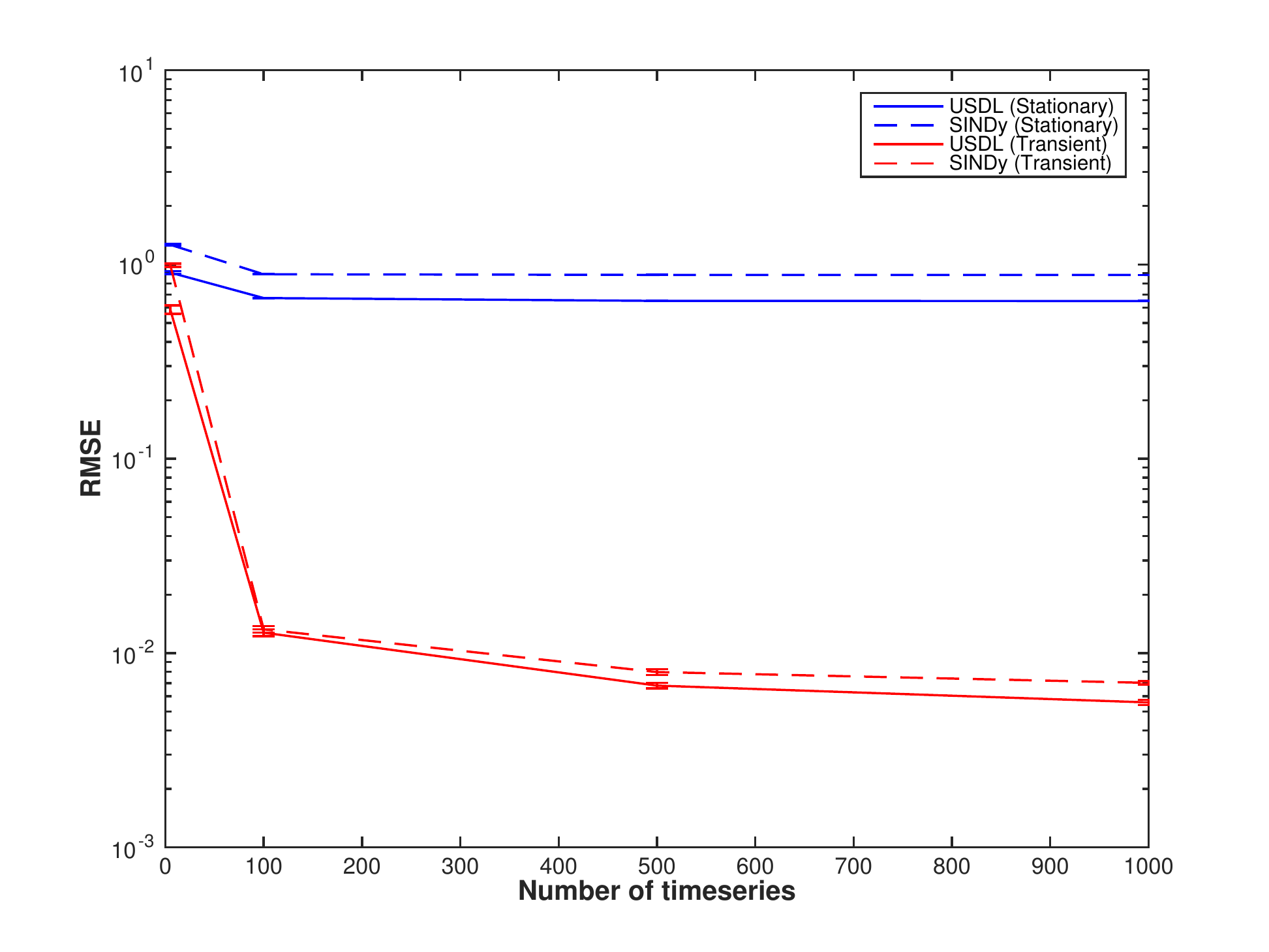}
		\label{rmse:OUprocess:Fourier}}
	\caption{(a) The F1 score as a function of the hyperparameter for both USDL (left column) and SINDy (right column) and various number of time-series using the standard Fourier modes. (b) The respective RMSE for both USDL (solid lines) and SINDy (dashed lines).}
	\label{rmse:OUprocess:fig}
\end{figure}

\begin{figure}[!htb]
	\centering
	\includegraphics[width=\linewidth]{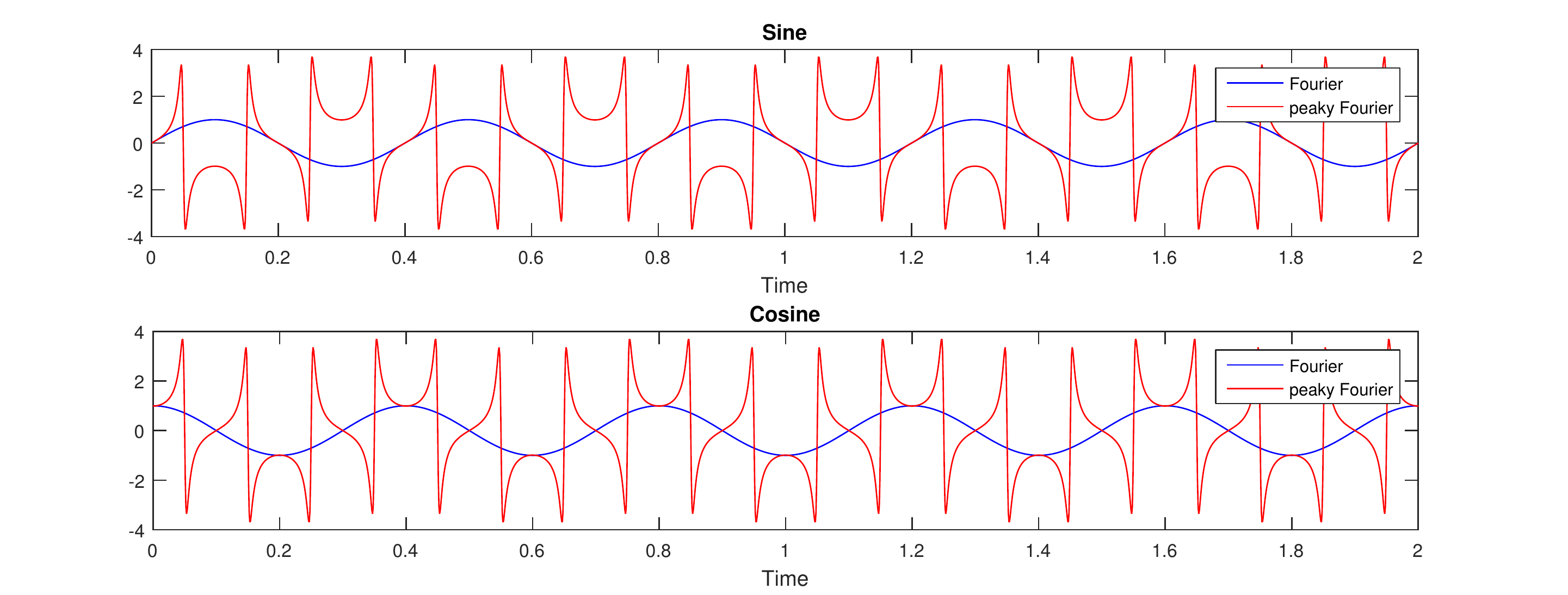}
	\caption{Upper plot: Sine Fourier mode (blue) and its peaky version (red). Lower plot: Same as upper plot but for cosine Fourier mode. Notice that peaky Fourier modes are actually smooth but they do have abrupt changes. The advantage of peaky Fourier as test functions is that they ``extract'' information in an adequate manner as they capture variations more accurately.}
	\label{disc:fourier:fig}
\end{figure}

In the main text, we highlight that the crucial decision is related with the definition of the test functions. Figure~\ref{OUprocess:fig:more:tf} presents the results of USDL algorithm when $M=81$ Fourier modes are deployed. Both precision and recall are about 90\%. As in all other experiments, we set USDL algorithm's threshold according to the optimal F1 score shown in Figure~\ref{F1score:OUprocess:Fourier}. Results from SINDy algorithm are shown for comparison purposes.
Looking in depth why perfect reconstruction was not successful, we found out that the problematic cases had the following pattern. When an interaction of the form $x_n\rightarrow x_{n'}$ exist in the connectivity matrix then OMP inferred two interactions; the correct one as well as that $x_{n'} \rightbararrow x_n$. These two types of interaction are closely related since they both assert that $x_n$ precedes $x_{n'}$. The problematic inference arises because the degradation of $x_n$, which is represented as $x_{n} \rightbararrow x_n$ is not enough to explain the dynamics of $x_n$ hence additional interactions are inferred. These additional edges are not random but they represent variables with strong time correlations. A way to separate these additional interactions is to observe that the time cross-correlations between one variable and the other variables are ordered based on which are the driving forces. Thus, we need to define another type of test functions with the property of sharp changes which will be able to separate between small time-differences. For instance, multiplying the Fourier modes with sawtooth functions would work. However, sawtooth function is not smooth thus we propose to use the following test functions which we call peaky Fourier modes
\begin{equation}
\phi_{2m-1}(t) = \frac{cos(4\pi m t/T)}{0.01+cos(4\pi m t/T)^2}  \cos(2\pi m t/T)
\end{equation}
and
\begin{equation}
\phi_{2m}(t) = \frac{cos(4\pi m t/T)}{0.01+cos(4\pi m t/T)^2}  \sin(2\pi m t/T)
\end{equation}
with $m=1,...,(M-1)/2$. We remark that the above functions are infinitely smooth but they have abrupt edges. Figure~\ref{disc:fourier:fig} show the sine and cosine Fourier modes for $m=5$ (blue lines) as well as the respective sine and cosine peaky Fourier modes (red lines).

\section{Sparse Learning of the Lorenz96 Model}

\begin{figure}[ht]
	\centering
	\subfigure[Lorenz96 system \ \ \ \ \ \ \ \ \ \ ]{
	        	\includegraphics[width=.4\textwidth]{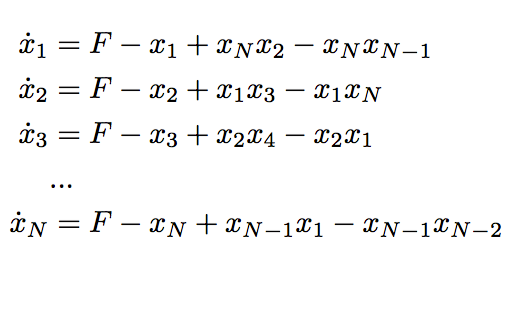}
		\label{lorenz96:ode:fig}}
	\quad
	\subfigure[Lorenz96 trajectories]{
	        	\includegraphics[width=.4\textwidth]{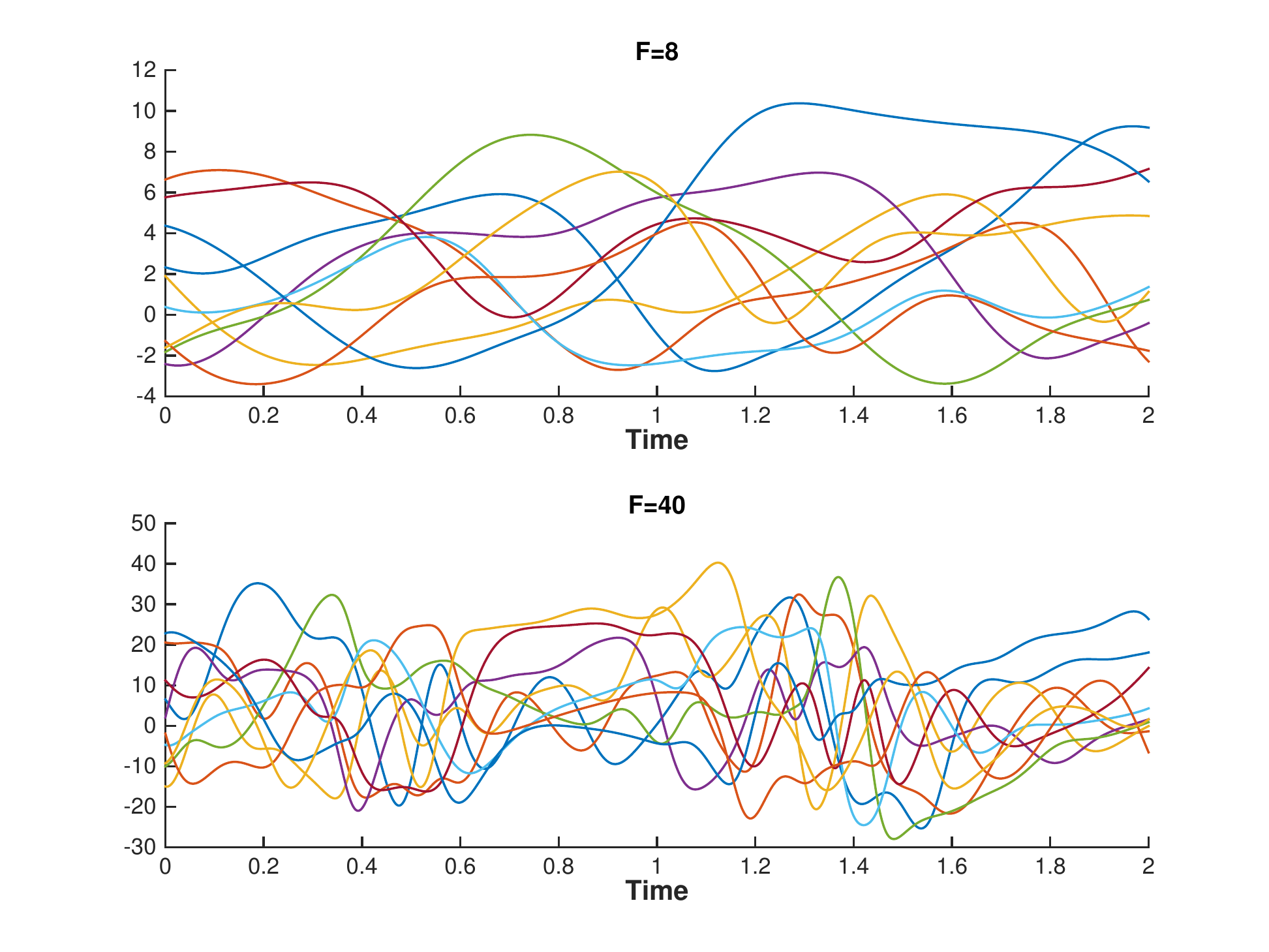}
		\label{lorenz96:ode:traj:fig}}
	\caption{(a) The nonlinear system of equations with periodic indexing. $F$ is the driving force, the linear term corresponds to the dissipation of energy while the quadratic terms correspond to the convection (mixing). Force values above 8 results in chaotic behavior \cite{Lorenz1996}. (b) Time-series (or trajectories) of Lorenz96 for two values of $F$. Upper plot shows the time-series under weak force ($F=8$) while the lower plot shows the time-series under strong force ($F=40$).
	}
	\label{lorenz96:fig}
\end{figure}

Lorenz96 \cite{Lorenz1996} is an idealized deterministic climate model  whose nonlinear dynamical set of equations is given in Figure~\ref{lorenz96:fig}(a). We set $N=10$ and for large enough force ($F \ge 8$) the system is chaotic (Figure~\ref{lorenz96:fig}(b)). The time-series of Lorenz96 system are generated through numerical integration. We utilized a third-order Runge-Kutta scheme with time step set to $0.001$. The initial values for each time-series are randomly and uniformly sampled in the interval $[-F/2, F/2]$. The sampling rate is set to $1000Hz$ or, equivalently, the sampling time is $0.001$ which is low enough so as to assume that the complete time-series is measured.

In order to perform sparse learning inference, the dictionary elements have to be determined. The chosen dictionary, $\psi(x)$, contains the constant term, the linear terms and all quadratic combinations resulting in $Q=66$ dictionary atoms (or constructed features) in total. Then, the elements of vectors $z_n$ with $n=1,...,N$ and matrix $\Psi$ are numerically estimated using the trapezoidal rule. Even though this is a deterministic system and should fall to the noiseless SSR case, this is not true since there are errors, first, in the construction/discretization of the time-series due to the numerical scheme and, second, in the numerical evaluation of the integrals. We tested two cases one with the force taking a low value ($F=8$) and another case taking a high value ($F=40$). In total, $M=41$ Fourier modes which constitutes of the constant function, 20 sines and 20 cosines in the interval $[0,2]$ were defined as test functions.

\begin{figure}[!htb]
	\centering
	\subfigure[Precision and recall]{
	        	\includegraphics[width=.4\textwidth]{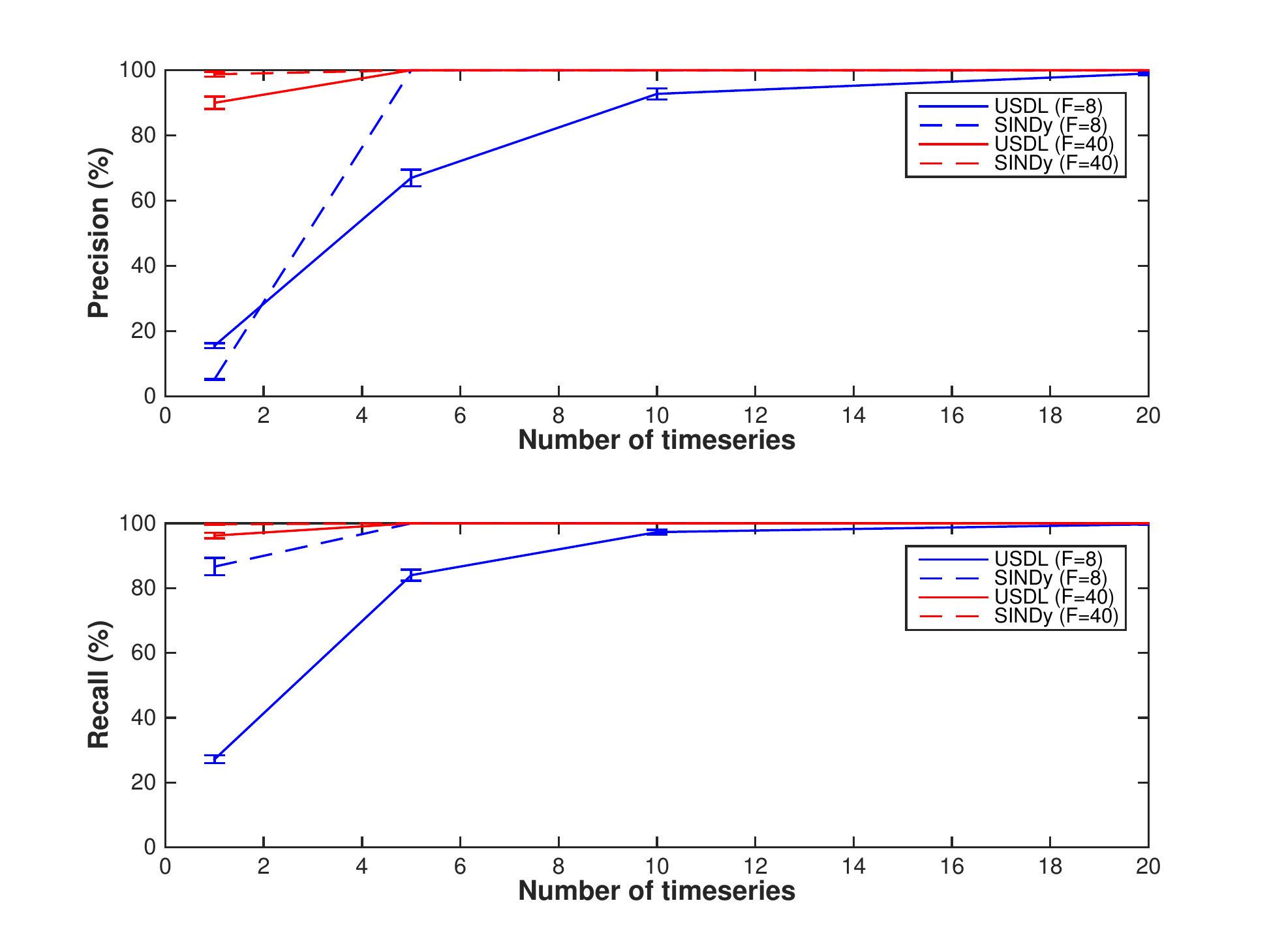}
		\label{pre:rec:lorenz96:basic:fig}}
	\quad
	\subfigure[RMSE]{
		\includegraphics[width=.4\textwidth]{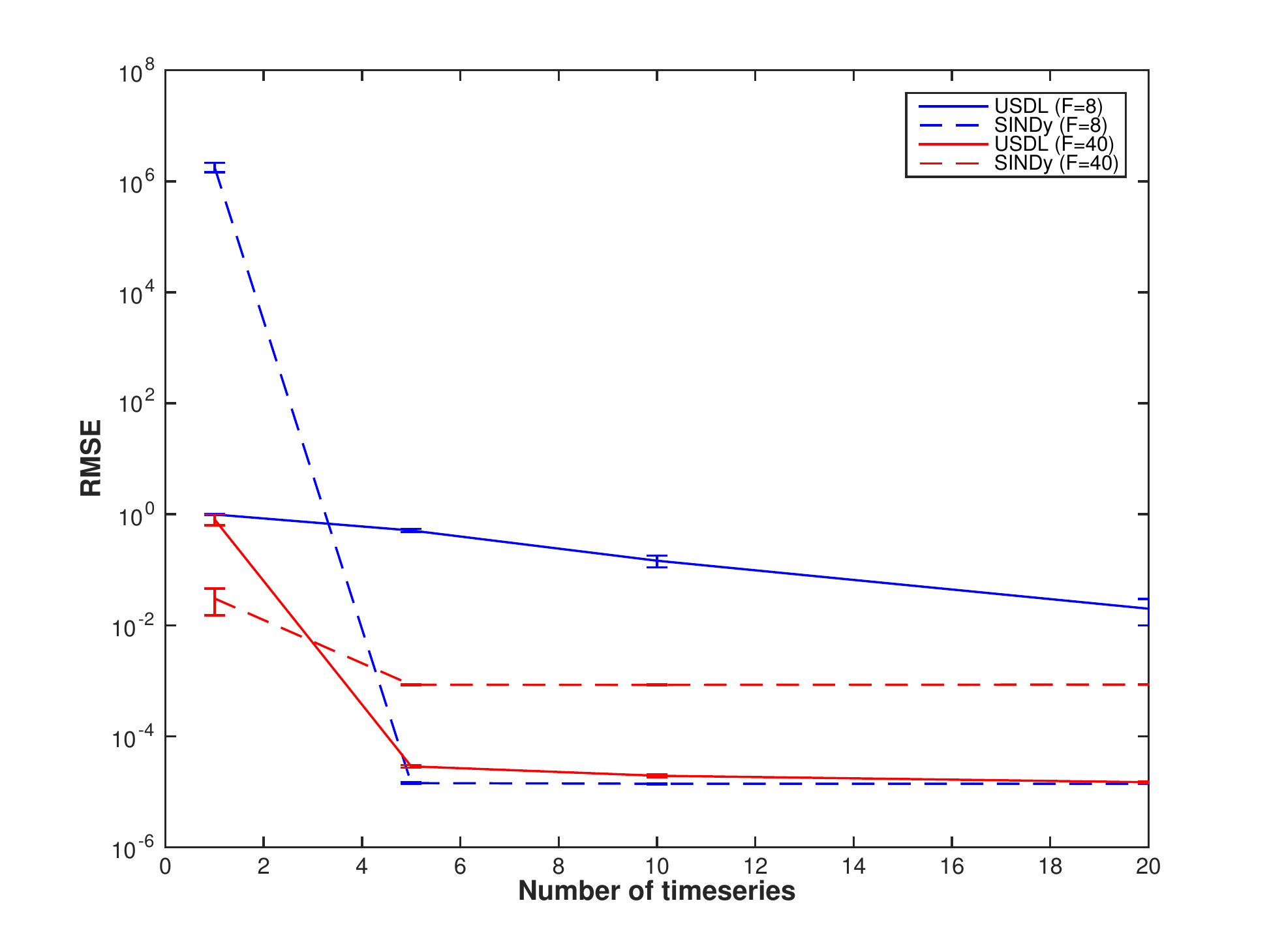}
		\label{rmse:lorenz96:basic:fig}}
		\quad	
	\subfigure[MIP and ERC]{
	        	\includegraphics[width=.4\textwidth]{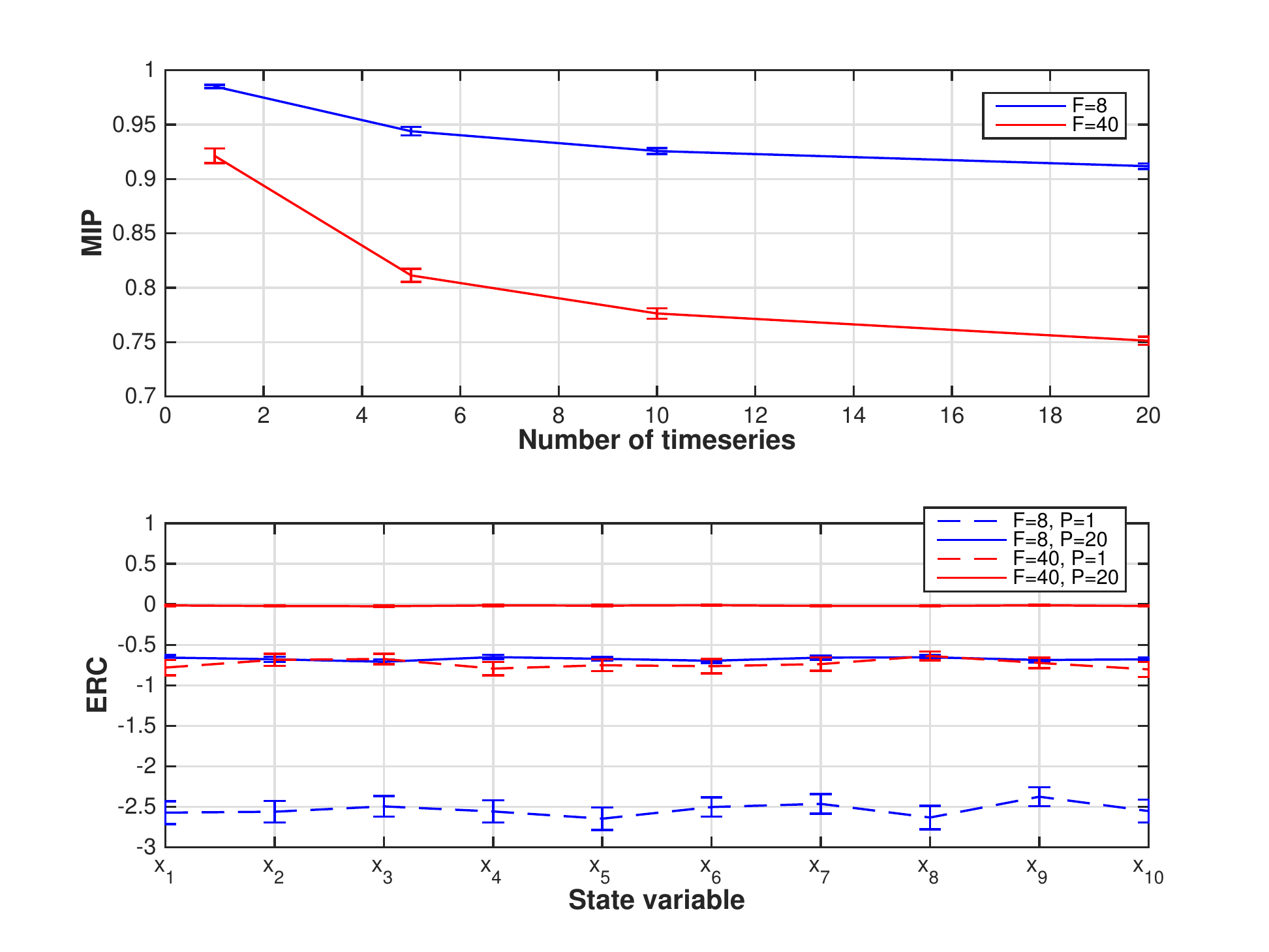}
		\label{mip:erc:lorenz96:basic:fig}}
	\caption{Performance analysis of USDL algorithm for the nonlinear Lorenz96 climate model. (a) Precision and recall curves under two different parameter regimes; weak force (blue) and strong force (red). Results from both USDL (solid lines) and SINDy \cite{Brunton2016} (dashed lines) are presented. Perfect reconstruction of the dynamical system is achieved when enough time-series are measured for both parameter regimes and both inference algorithms. However, SINDy algorithm requires less time-series in order to achieve perfect reconstruction in both regimes revealing that there are cases where iterative thresholding algorithm which is part of SINDy outperforms the greedy OMP algorithm which is part of the proposed approach. Moreover, the driving force $F$ affects the performance of both algorithms for the same number of time-series. Higher value of $F$ which implies more chaotic behavior results in higher inference accuracy. (b) RMSE as a function of the number of time-series for USDL (solid lines) and SINDy (dashed lines). Except when $F=8$ and USDL algorithm (blue solid line), RMSE quickly reaches a plateau whose value quantifies the total discretization error. (c) ERC value for each state variable supplemented with confidence intervals. Weak (blue) and strong (red) forces are considered while one (dashed) or twenty time-series are feed to the algorithm. Greater values of ERC are observed under strong forcing revealing that chaos assists the correct inference of the dynamics. True recovery is theoretically guaranteed only for the strong forcing case and when at least 20 time-series are measured since then ERC is positive.
	}
	\label{lorenz96:basic:fig}
\end{figure}

Figure~\ref{lorenz96:basic:fig} summarizes the performance of USDL approach as a function of the number of time-series denoted by $P$. Threshold value for OMP algorithm was set to $\alpha=0.1$ while the maximum number of non-zero components was set to $K=7$ which is larger than the true value which is 4. As quantified from the precision-recall subplots (see Figure~\ref{pre:rec:lorenz96:basic:fig}), the performance in terms of both precision and recall is improved as the number of measured time-series is increased. It actually reaches perfect reconstruction when $P=5$ for the strong forcing case and when $P=20$ for the weak forcing case. In general, stronger forces which result in more chaotic behavior and stronger mixing are helpful in identifying the true model as it is evident from the fact that red curves outperformed the blue ones. Figure~\ref{pre:rec:lorenz96:basic:fig} presents also the reconstruction accuracy of SINDy algorithm \cite{Brunton2016} (dashed lines) with its hyperparameter value being set to $\lambda=0.1$. We employ the central difference scheme which is a second order method for the numerical estimation of the derivatives. Interestingly, SINDy requires less time-series in order to achieve perfect reconstruction in both parameter regimes showing that Lasso is a competitive alternative for solving the SSR problem.

Figure~\ref{rmse:lorenz96:basic:fig} shows the RMSE between the true connectivity matrix and the estimated one for both algorithms and both parameter regimes. As expected, the RMSE decreases with the increase of the data size until it hits a plateau which stems from the discretization error of the various numerical integrations. Evidently, RMSE performance of USDL is better compared to SINDy at the strong force regime while the opposite is true at the weak force regime.
Moreover, SSR metrics such as MIP and ERC correlate well with the performance of USDL algorithm as depicted in the subplots of Figure~\ref{mip:erc:lorenz96:basic:fig}. ERC values per variable are (statistically) equal as expected because of the symmetry in the dynamical system's state variables. We remark also that theoretical guarantees on perfect reconstruction are satisfied only under strong forcing and at least $P=20$ time-series of length 2 are taken into account since then ERC is positive for all state variables. In contract, MIP is not small enough --it should be below $0.143$-- so as to theoretically guarantee perfect reconstruction.

\begin{figure}[!htb]
	\centering
	\includegraphics[width=.8\textwidth]{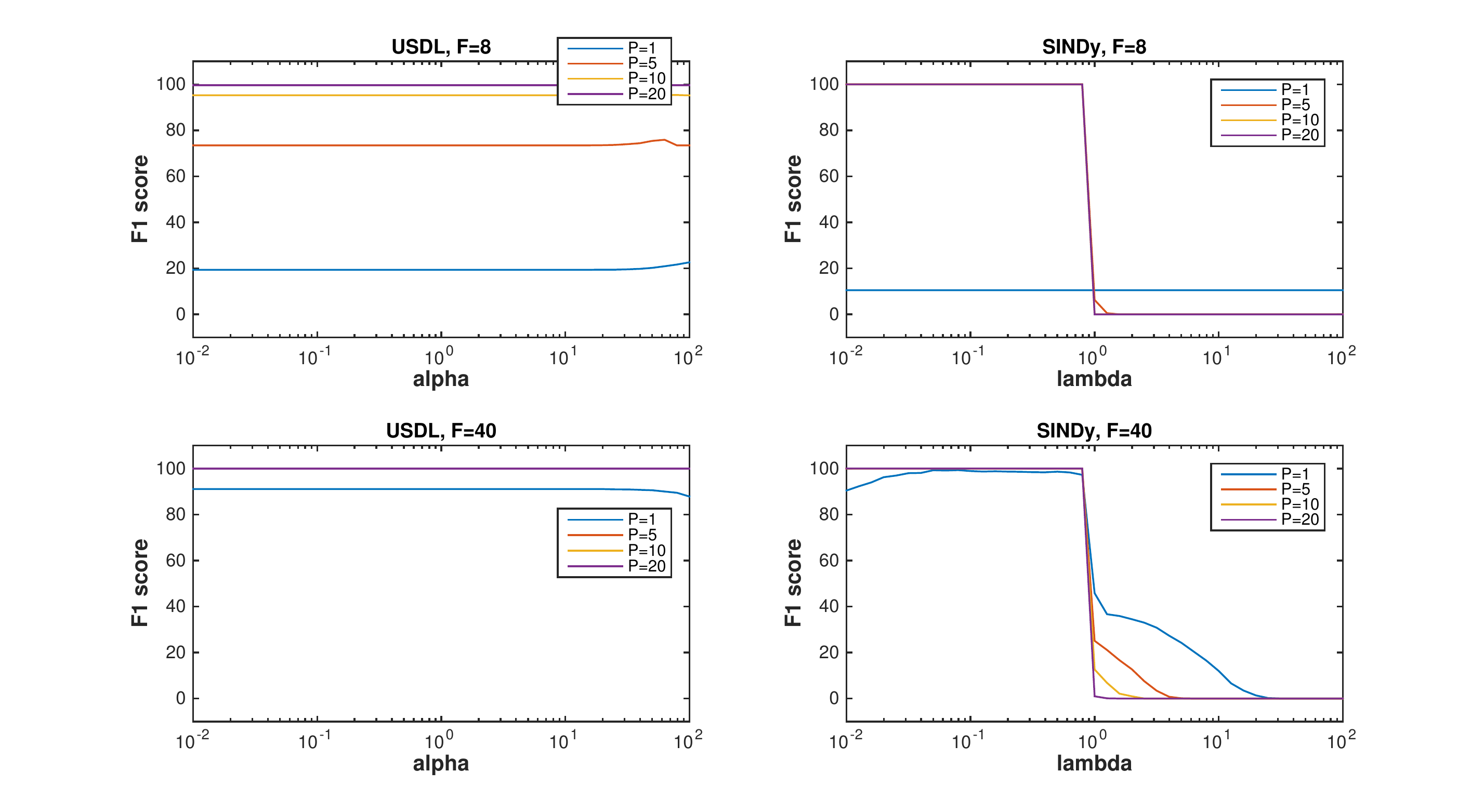}
	\caption{The harmonic mean between averaged precision and averaged recall (i.e., the F1 score) as a function of the hyperparameter for both USDL (left column) and SINDy (right column) and various number of time-series and forces. There is a large parametric region where optimal values are achieved for both algorithms.}
	\label{F1:score:Lorenz96:fig}
\end{figure}

Finally, we comment on the fine tuning of the hyperparameter for each approach. Each subplot of Figure~\ref{F1:score:Lorenz96:fig} presents the F1 score (i.e., the harmonic mean of precision and recall) as a function of the hyperparameter value for both algorithms and forcing values. To avoid data leakage, the F1 score is estimated on a different set of data and not on the time-series used for the creation of Figure~\ref{lorenz96:basic:fig}. We ubiquitously observe that  the maximum F1 score increases as the number of time-series increases. Interestingly, the region of optimal hyperparameter values is wide showing the robustness of the sparse inference methods on the hyperparameter values.% It is also evident from the F1 score that SINDy algorithm has better accuracy than USDL algorithm in the small data regime.

%\newpage

\subsection{Further Experimentation}

%% FIRST DEMONSTRATION
\begin{figure}[!htb]
	\centering
	\subfigure[Precision and recall]{%
		\includegraphics[width=.4\textwidth]{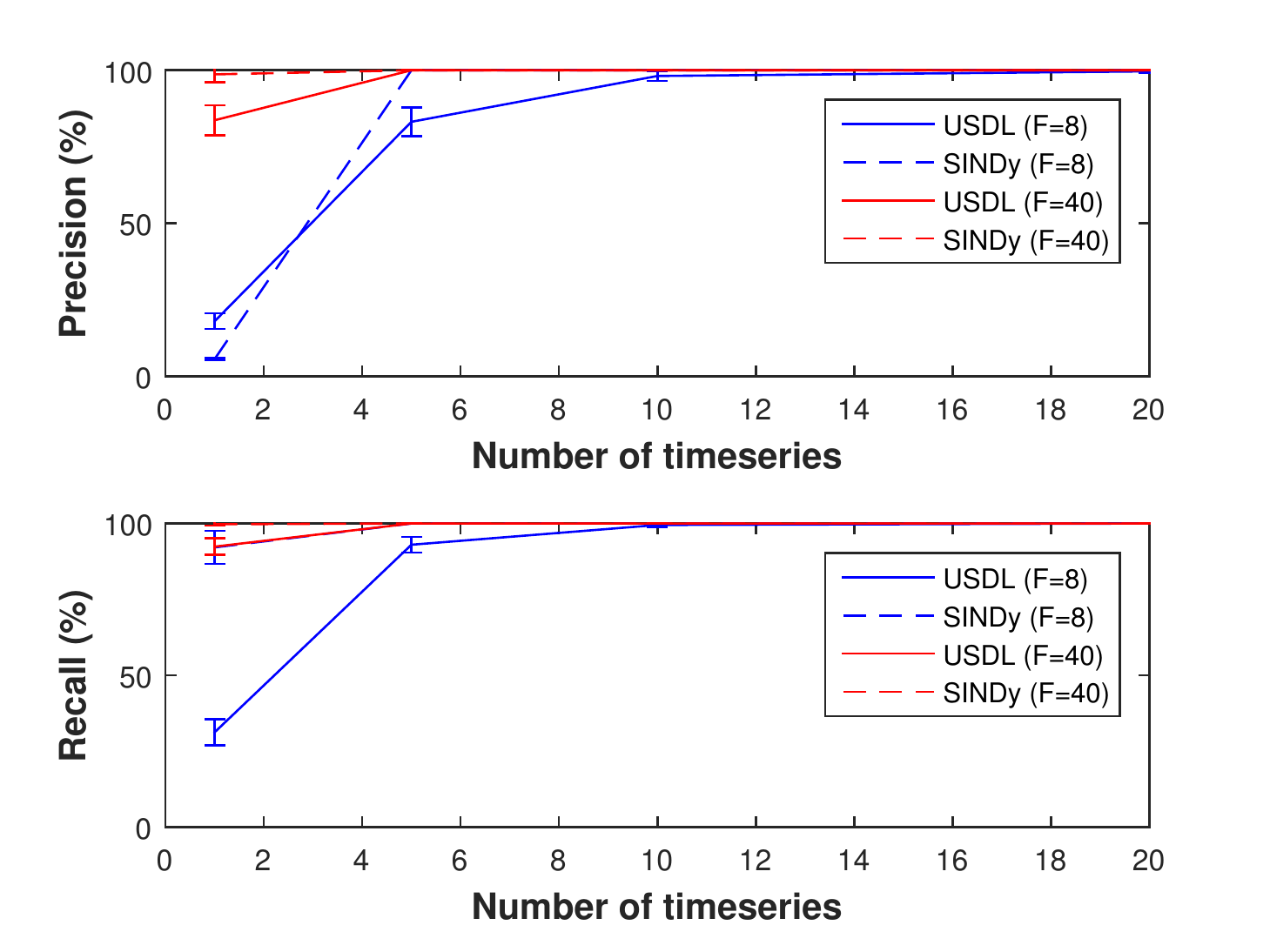}
		\label{pre:rec:lorenz96:spline}}
	\quad
	\subfigure[RMSE]{%
		\includegraphics[width=.4\textwidth]{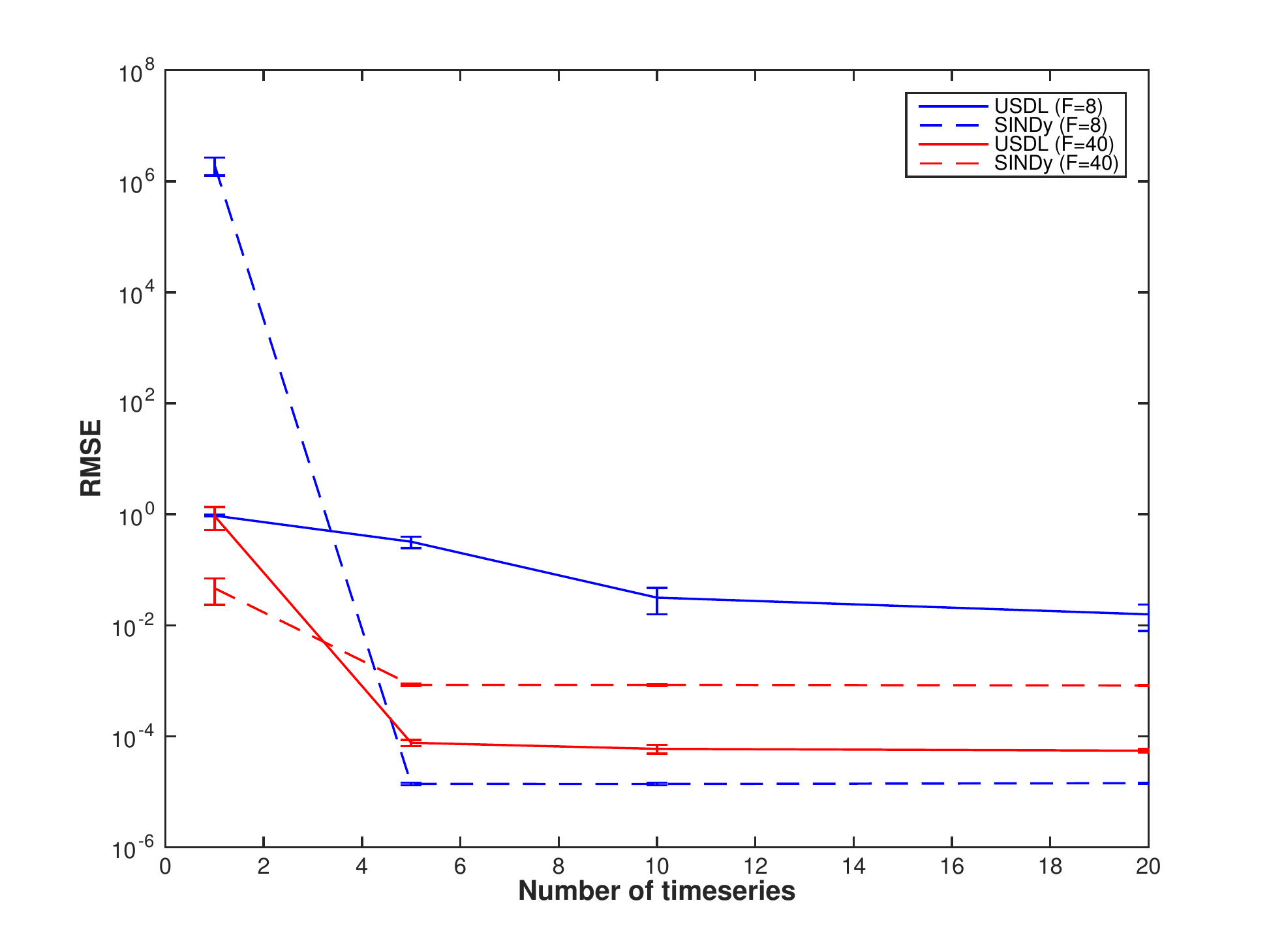}
		\label{rmse:lorenz96:spline}}
	\quad
	\subfigure[MIP and ERC]{%
		\includegraphics[width=.4\textwidth]{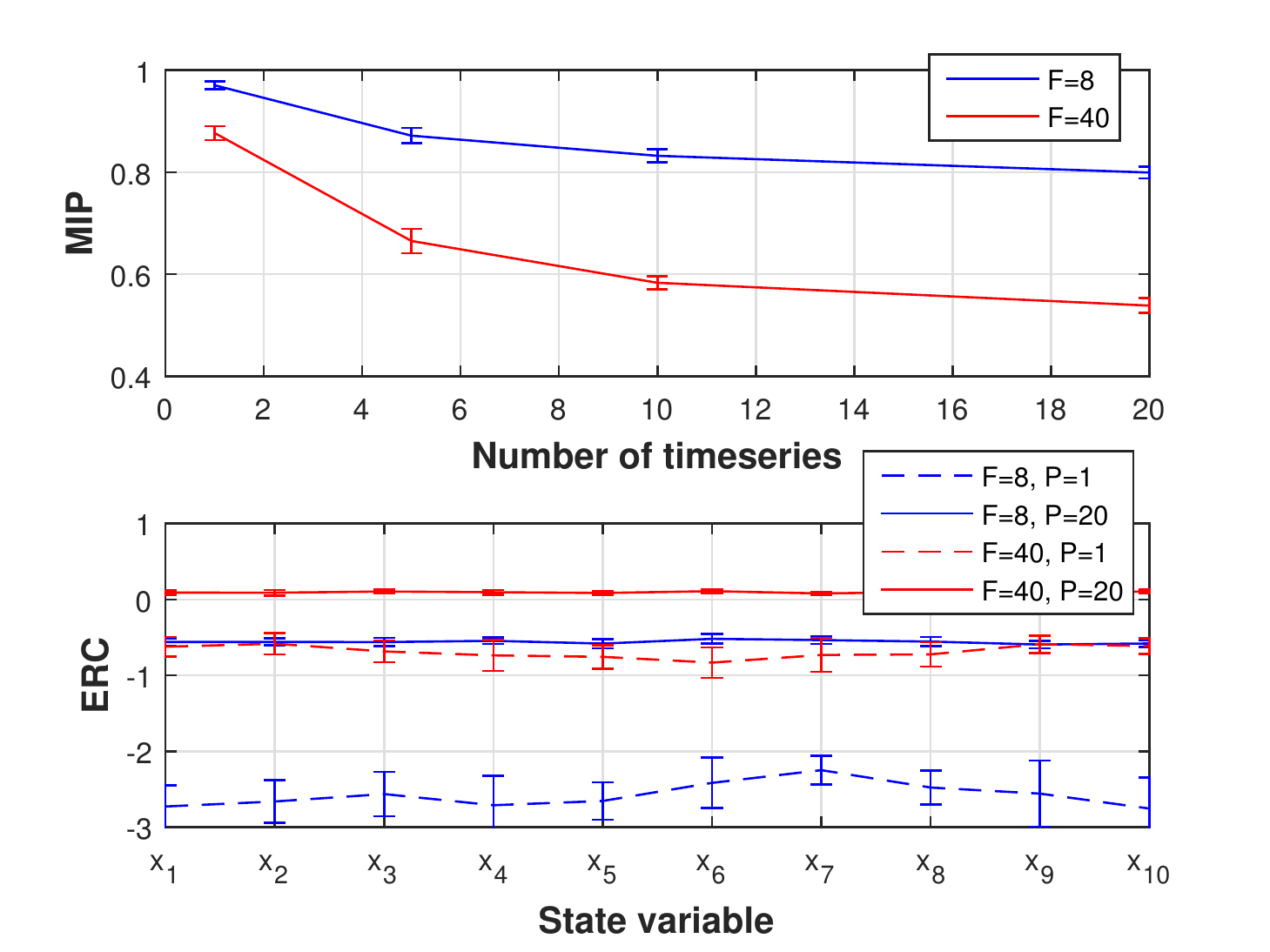}
		\label{mip:erc:lorenz96:spline}}
	\caption{Results for the Lorenz96 model when $M=40$ and second-order basis spline functions are used as test functions. (a) Precision (upper panel) and recall (lower panel) curves as a function of $P$, i.e., the number of trajectories for both USDL (solid lines) and SINDy (dashed lines) algorithms. Performance of USDL algorithm is overall similar to the performance when Fourier modes are used. (b) RMSE as a function of the number of time-series for USDL (solid lines) and SINDy (dashed lines). It frequently occurs in our experiments that the RMSE for USDL algorithm when $F=8$ (blue solid) is worse than when $F=40$ (red solid) while the opposite frequently happens for the SINDy algorithm. (c) MIP (upper panel) and ERC per state variable (lower panel).
	}
	\label{lorenz96:fig:spline}
\end{figure}

\begin{figure}[!htb]
	\centering
	\subfigure[Precision and recall]{%
		\includegraphics[width=.4\textwidth]{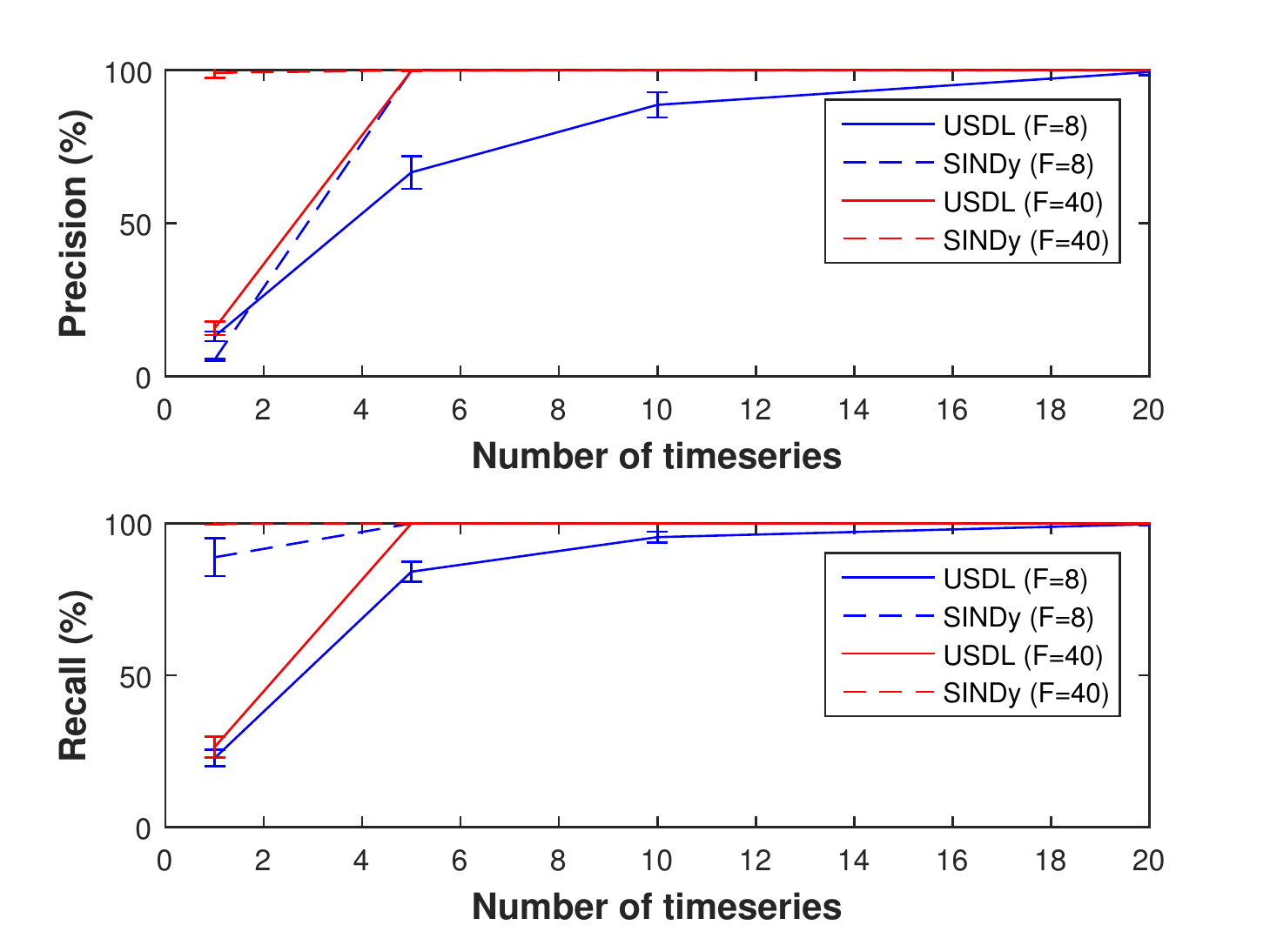}
		\label{pre:rec:lorenz96:L_11}}
	\quad
	\subfigure[RMSE]{%
		\includegraphics[width=.4\textwidth]{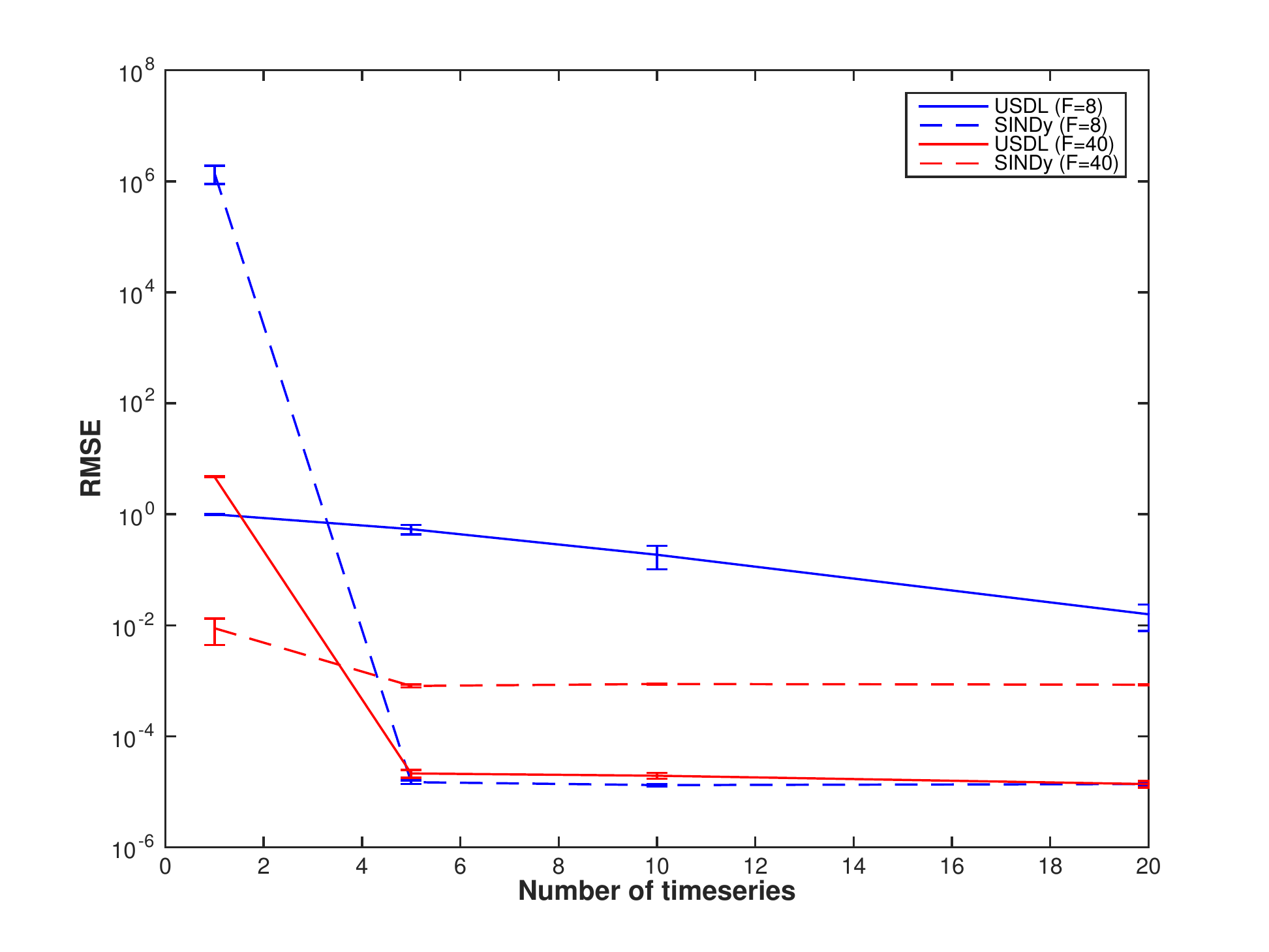}
		\label{rmse:lorenz96:L_11}}
	\quad
	\subfigure[MIP and ERC]{%
		\includegraphics[width=.4\textwidth]{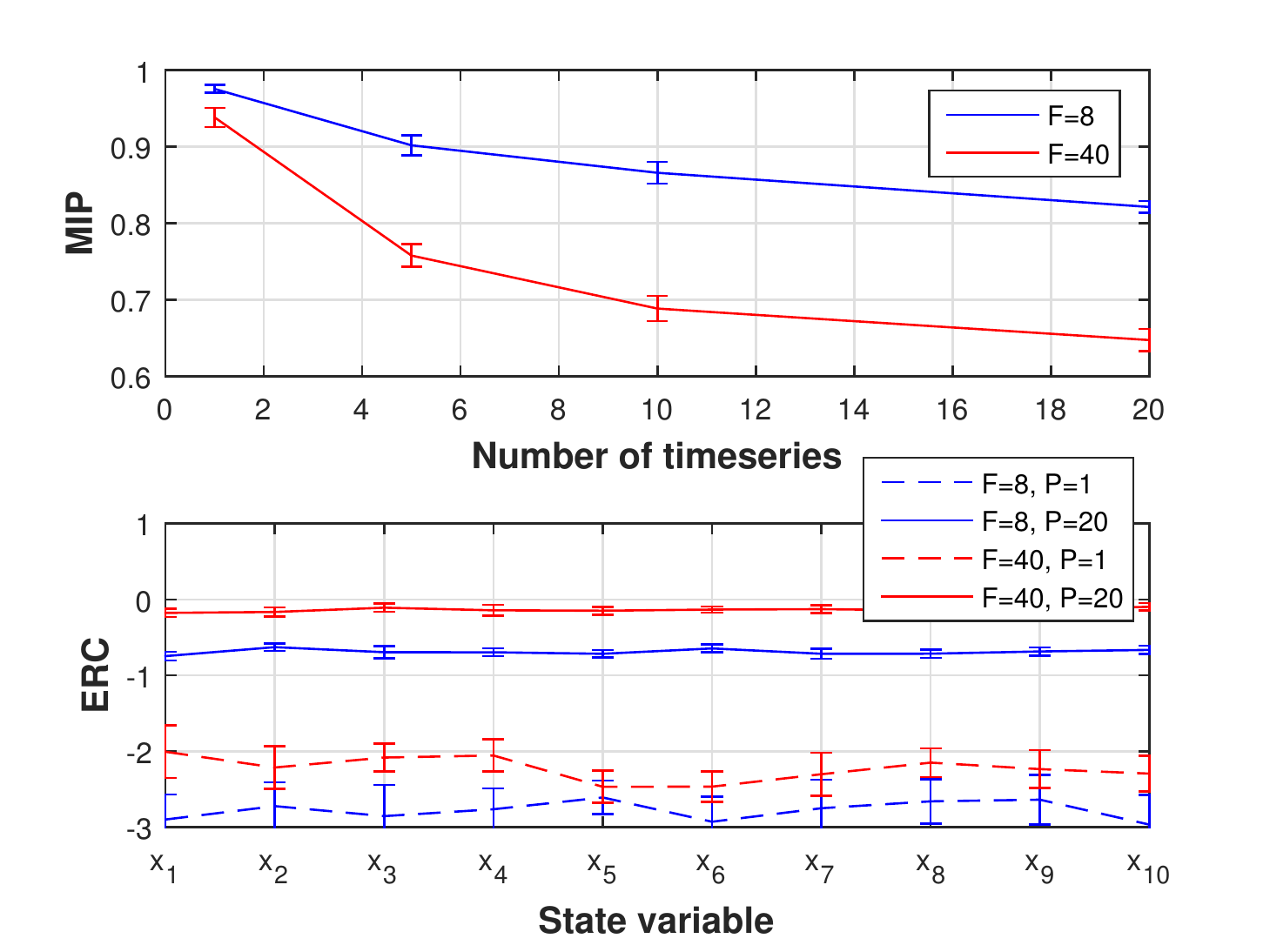}
		\label{mip:erc:lorenz96:L_11}}
	\caption{Results for the Lorenz96 model when $M=11$ Fourier modes are used as test functions. (a) Precision (upper panel) and recall (lower panel) curves as a function of $P$, i.e., the number of trajectories for both USDL (solid lines) and SINDy (dashed lines) algorithms. Even though four times less test functions are used, there is no or little deterioration of the performance in most of the cases ($P\ge 5$). (b) RMSE as a function of the number of time-series for USDL (solid lines) and SINDy (dashed lines). RMSE performance is similar to the previous experiment. (c) MIP (upper panel) and ERC per state variable (lower panel).
	}
	\label{lorenz96:fig:L_11}
\end{figure}

We present a series of comparisons under various setups and assess the performance of the USDL algorithm. The first variation uses a different family of test functions keeping all the other hyperparameters fixed. A set of $M=40$, equally-spaced basis spline (B-splines) functions of second-order are employed as test functions. Figure~\ref{lorenz96:fig:spline} presents the performance of the proposed approach in terms of precision and recall curves (upper left plot), in terms of RMSE (upper right plot) as well as metrics from SSR theory (lower plot). Results with B-splines are overall similar as in the case of Fourier modes. There is a minor difference with spline functions with USDL performing slightly better in terms of precision for weak forcing (blue in upper curve of Figure~\ref{pre:rec:lorenz96:spline}) when the number of trajectories is in the range of 3-10. For comparison purposes, we also present the accuracy performance for SINDy algorithm which are the same as in Figure~\ref{pre:rec:lorenz96:basic:fig}.

The second variation utilizes less test functions. Using less test functions is expected to weaken the strength of the inference method since less information is fed to the algorithm. Figure~\ref{lorenz96:fig:L_11} shows the same quantities as in the first variation when only $M=11$ Fourier modes are computed. Interestingly, the performance is similar to the case with 41 Fourier modes when at least $P=5$ trajectories are provided for both weak and strong forces. The performance deteriorates significantly both in terms of precision and recall when the force is strong and one or two trajectories are measured. This behavior is in accordance with ERC which is significantly dropped from $-0.8$ to below $-2$ as lower panel of Figure~\ref{mip:erc:lorenz96:L_11} shows (dashed red line).

%% SECOND DEMONSTRATION
\begin{figure}[!htb]
	\centering
	\subfigure[Precision and recall]{
		\includegraphics[width=.4\textwidth]{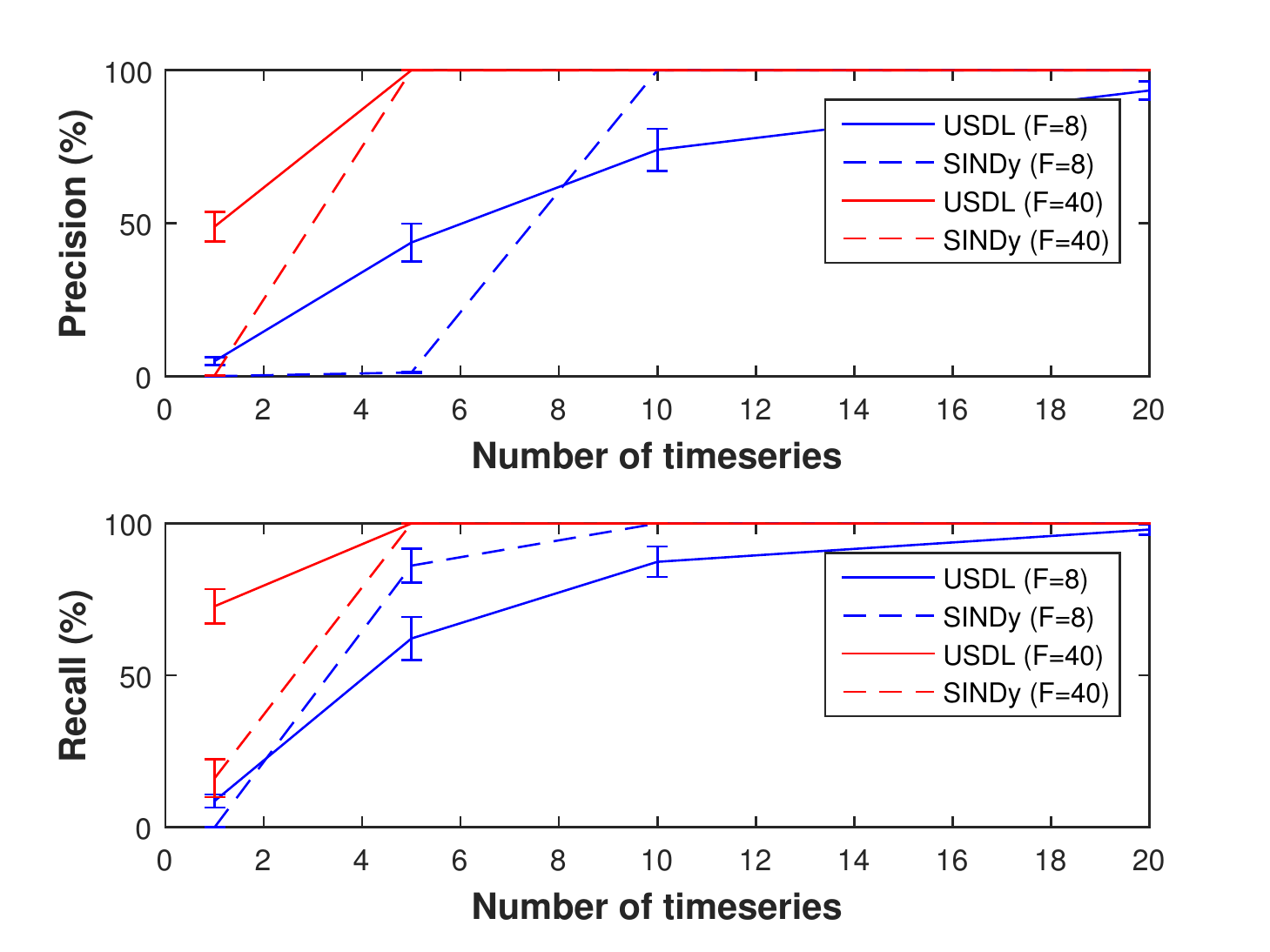}
		\label{pre:rec:lorenz96:cubic}}
	\quad
	\subfigure[RMSE]{
		\includegraphics[width=.4\textwidth]{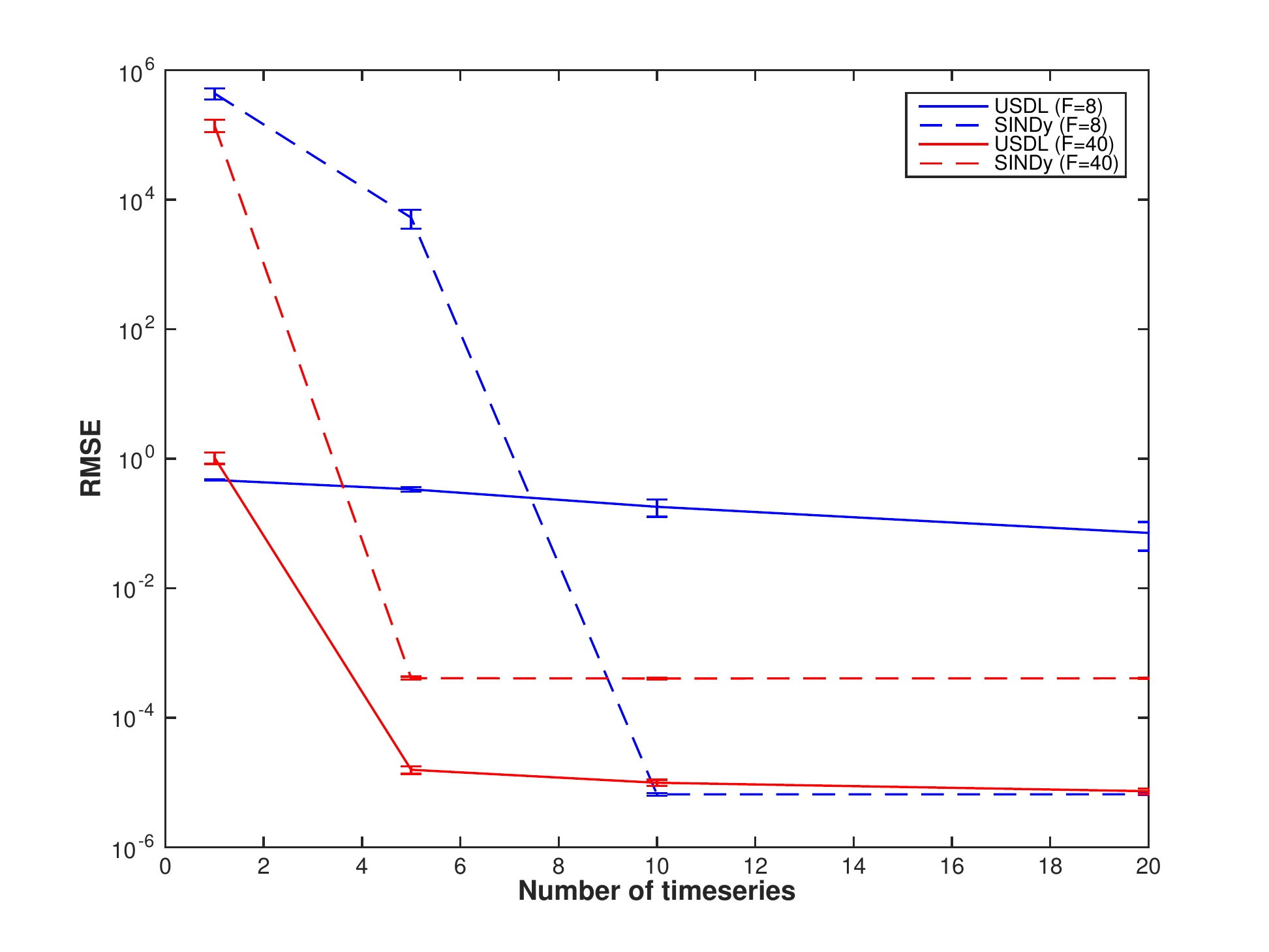}
		\label{rmse:lorenz96:cubic}}
	\quad
	\subfigure[MIP and ERC]{%
		\includegraphics[width=.4\textwidth]{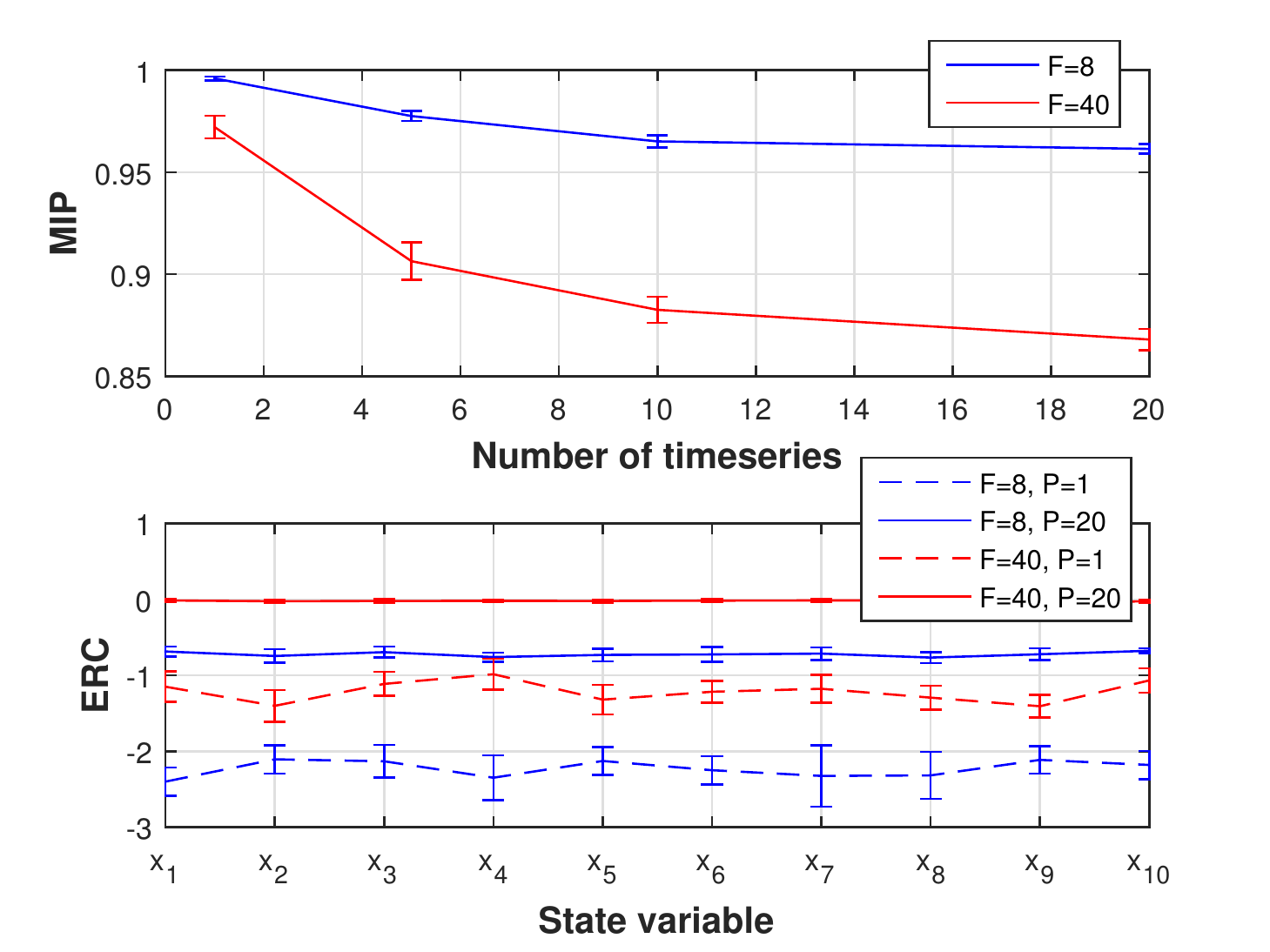}
		\label{mip:erc:lorenz96:cubic}}
	\caption{Results for the Lorenz96 model when cubic terms are added to the dictionary. The size of the dictionary is approximately five times larger. (a) Precision (upper panel) and recall (lower panel) curves as a function of $P$, i.e., the number of trajectories for both USDL (solid lines) and SINDy (dashed lines) algorithms. Performance is slightly worse for the weak forcing case (blue) while perfect reconstruction is achieved when the force is strong (red) and more than $P=5$ time-series are measured for both algorithms. Despite being less accurate for small $P$, SINDy algorithm is able to perfectly reconstruct the dynamical system under the weak force when $P=10$ trajectories are provided. (b) RMSE as a function of the number of time-series for USDL (solid lines) and SINDy (dashed lines). RMSE results are in accordance with the precision and recall curves. (c) MIP (upper panel) and ERC per state variable (lower panel).
	}
	\label{lorenz96:fig:cubic}
\end{figure}

%% THIRD DEMONSTRATION

\begin{figure}[!htb]
	\centering
	\subfigure[Precision and recall]{
		\includegraphics[width=.4\textwidth]{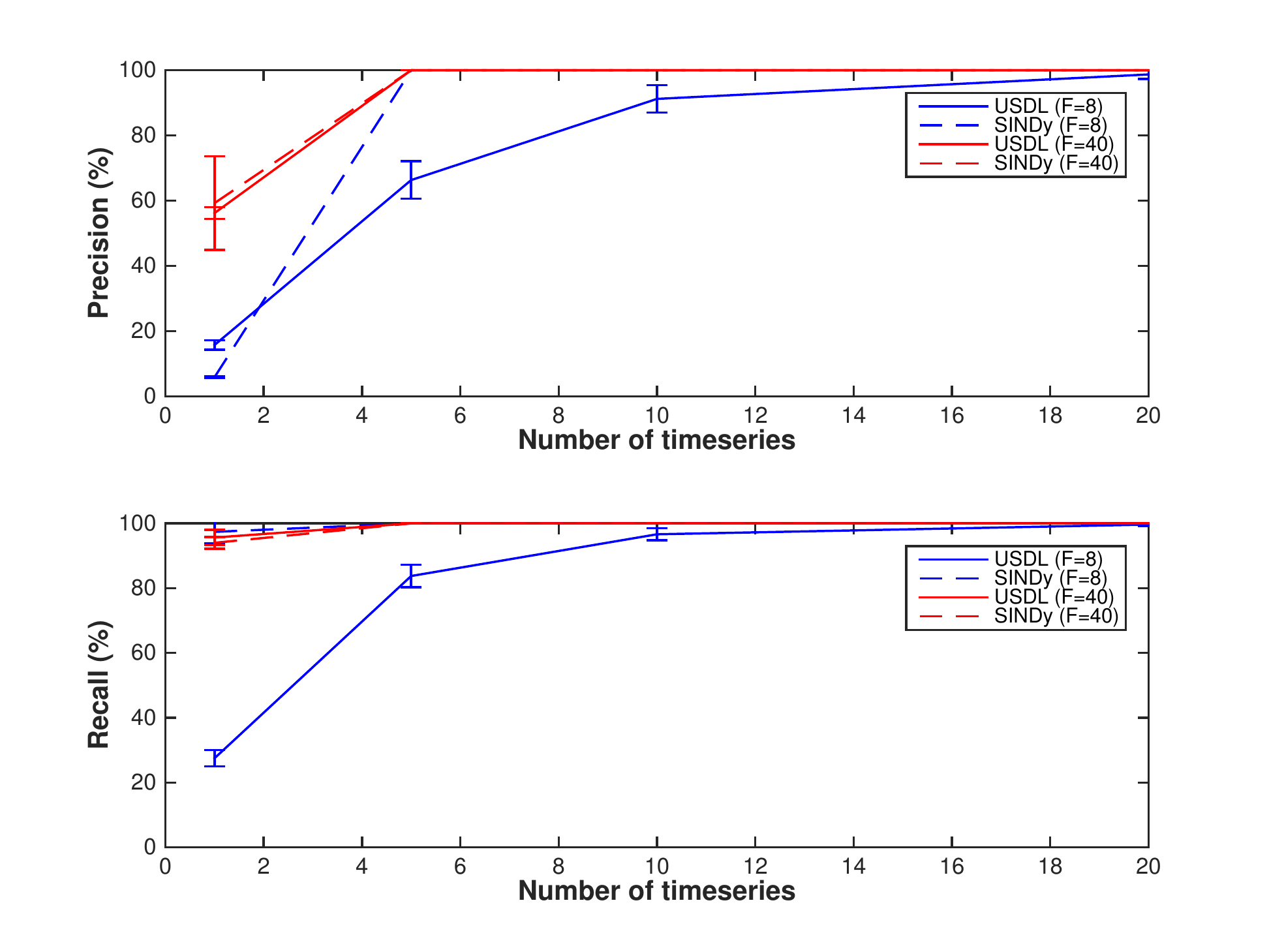}
		\label{pre:rec:lorenz96:Dt:0:01}}
	\quad
	\subfigure[RMSE]{
		\includegraphics[width=.4\textwidth]{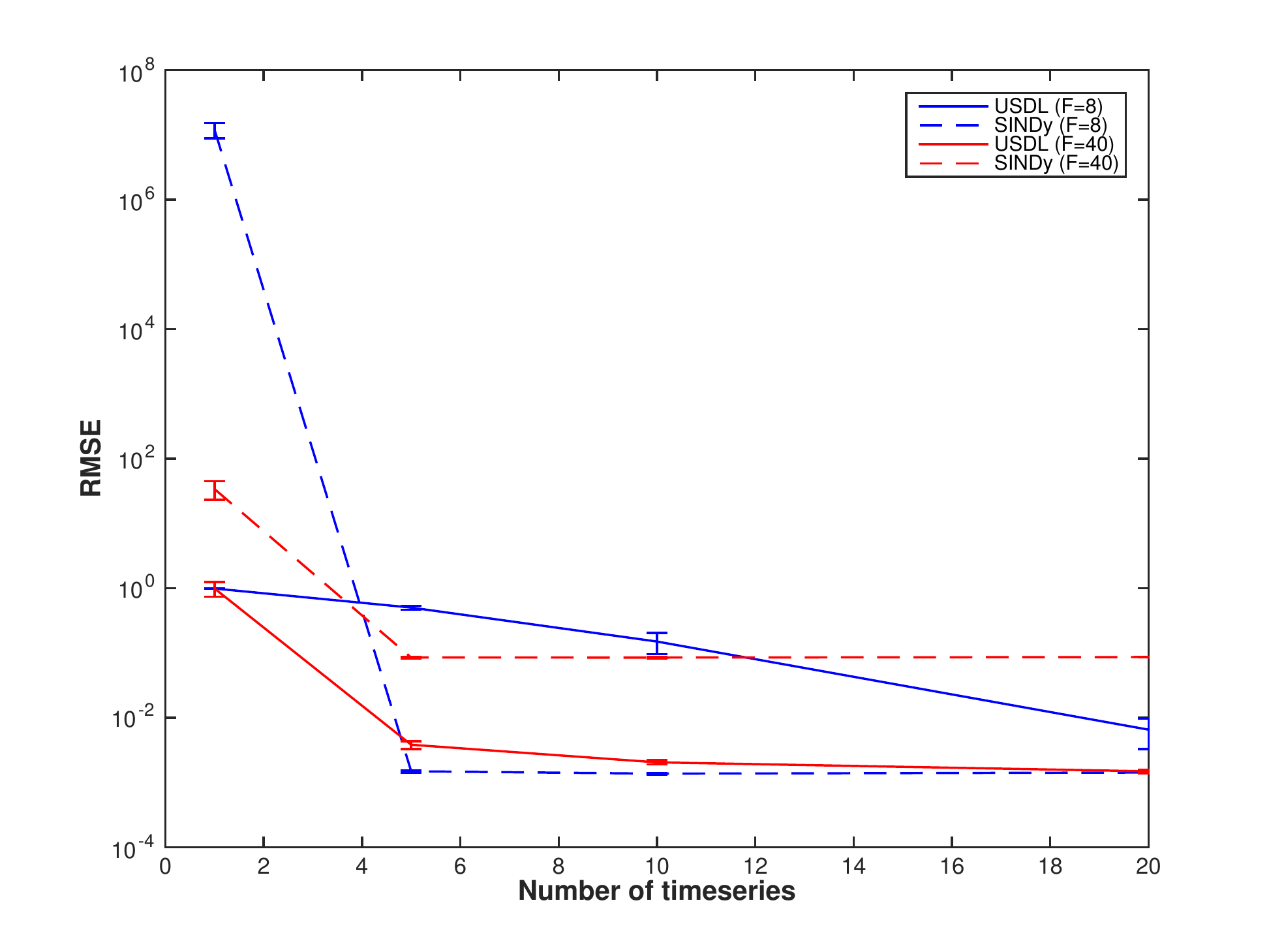}
		\label{rmse:lorenz96:Dt:0:01}}
	\quad
	\subfigure[MIP and ERC]{%
		\includegraphics[width=.4\textwidth]{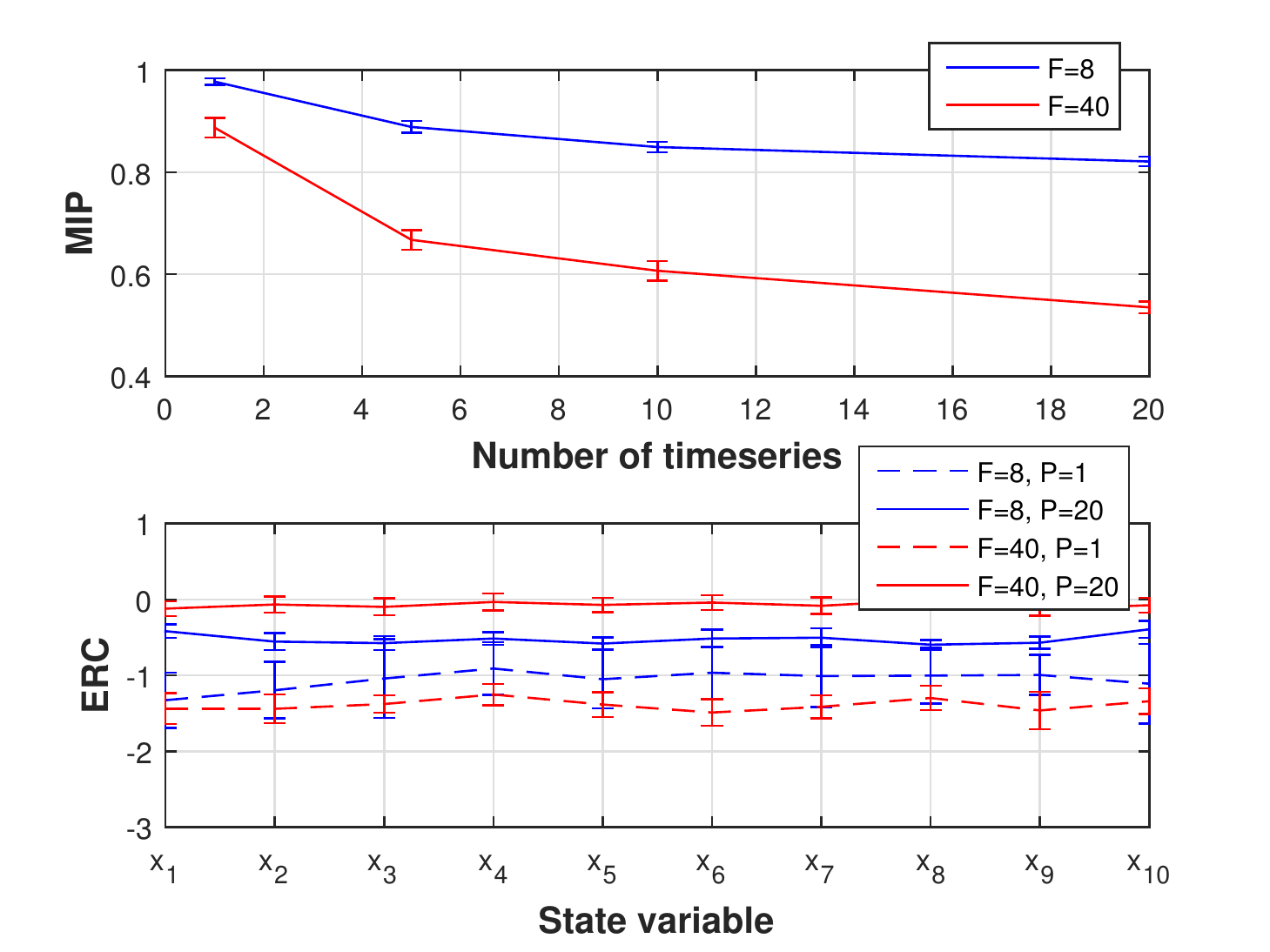}
		\label{mip:erc:lorenz96:Dt:0:01}}
	\caption{Results for the Lorenz96 model when time-series are sampled at the rate of $100 Hz$ instead of $1000 Hz$. (a) Precision (upper panel) and recall (lower panel) curves as a function of $P$, i.e., the number of trajectories for both USDL (solid lines) and SINDy (dashed lines) algorithms. Minor reduction in the performance of both USDL and SINDy algorithms is observed. (b) RMSE as a function of the number of time-series for USDL (solid lines) and SINDy (dashed lines). Results are qualitatively similar to the case where the rate is $1000 Hz$ (shown in Figure~\ref{lorenz96:basic:fig}) while from a quantitative perspective RMSE is now two orders of magnitude worse compared to Figure~\ref{lorenz96:basic:fig}. (c) MIP (upper panel) and ERC per state variable (lower panel) .
	}
	\label{lorenz96:fig:Dt:0:01}
\end{figure}

\begin{figure}[!htb]
	\centering
	\subfigure[Precision and recall]{
		\includegraphics[width=.4\textwidth]{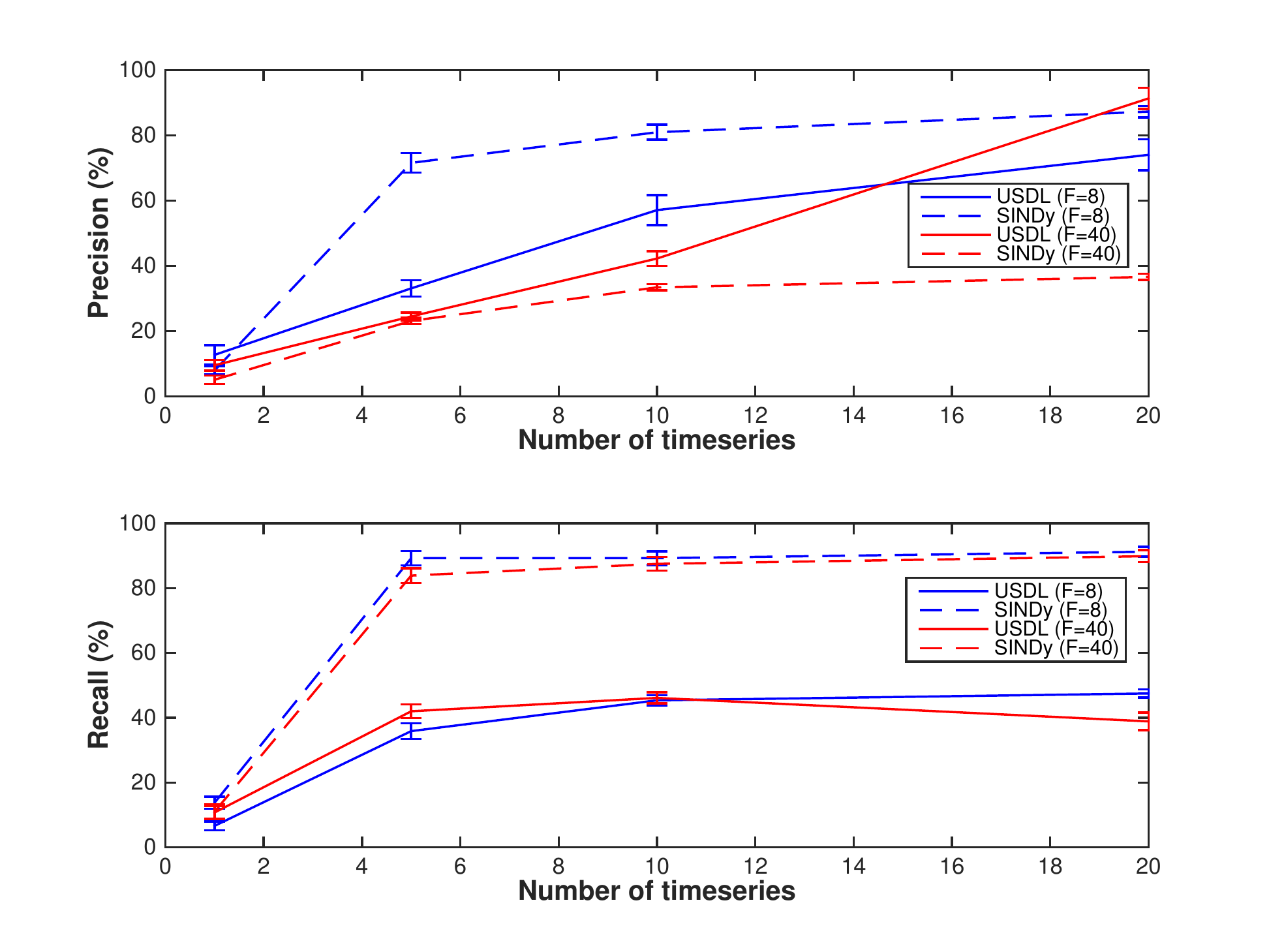}
		\label{pre:rec:lorenz96:Dt:0:1}}
	\quad
	\subfigure[RMSE]{
		\includegraphics[width=.4\textwidth]{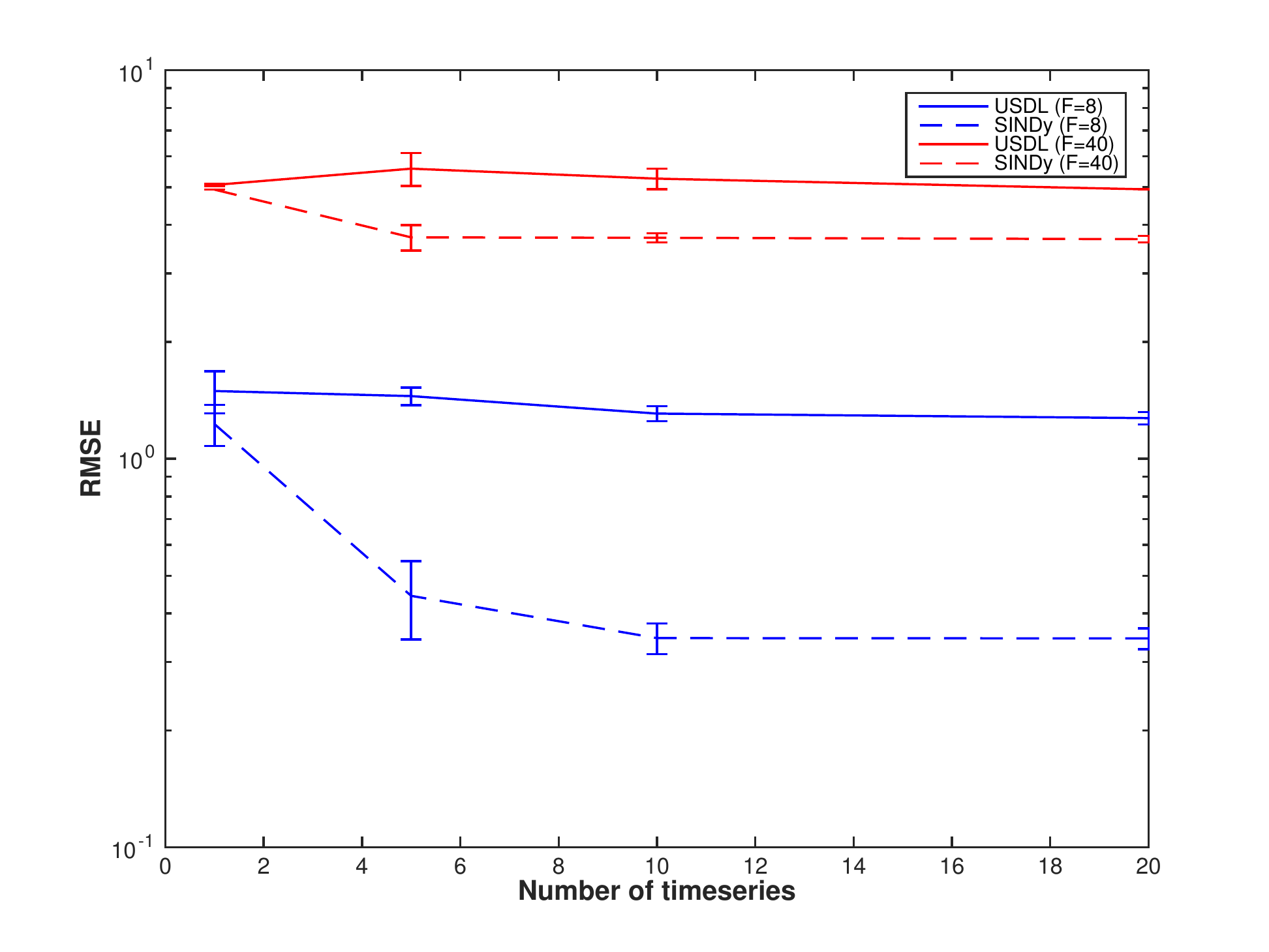}
		\label{rmse:lorenz96:Dt:0:1}}
	\quad
	\subfigure[MIP and ERC]{%
		\includegraphics[width=.4\textwidth]{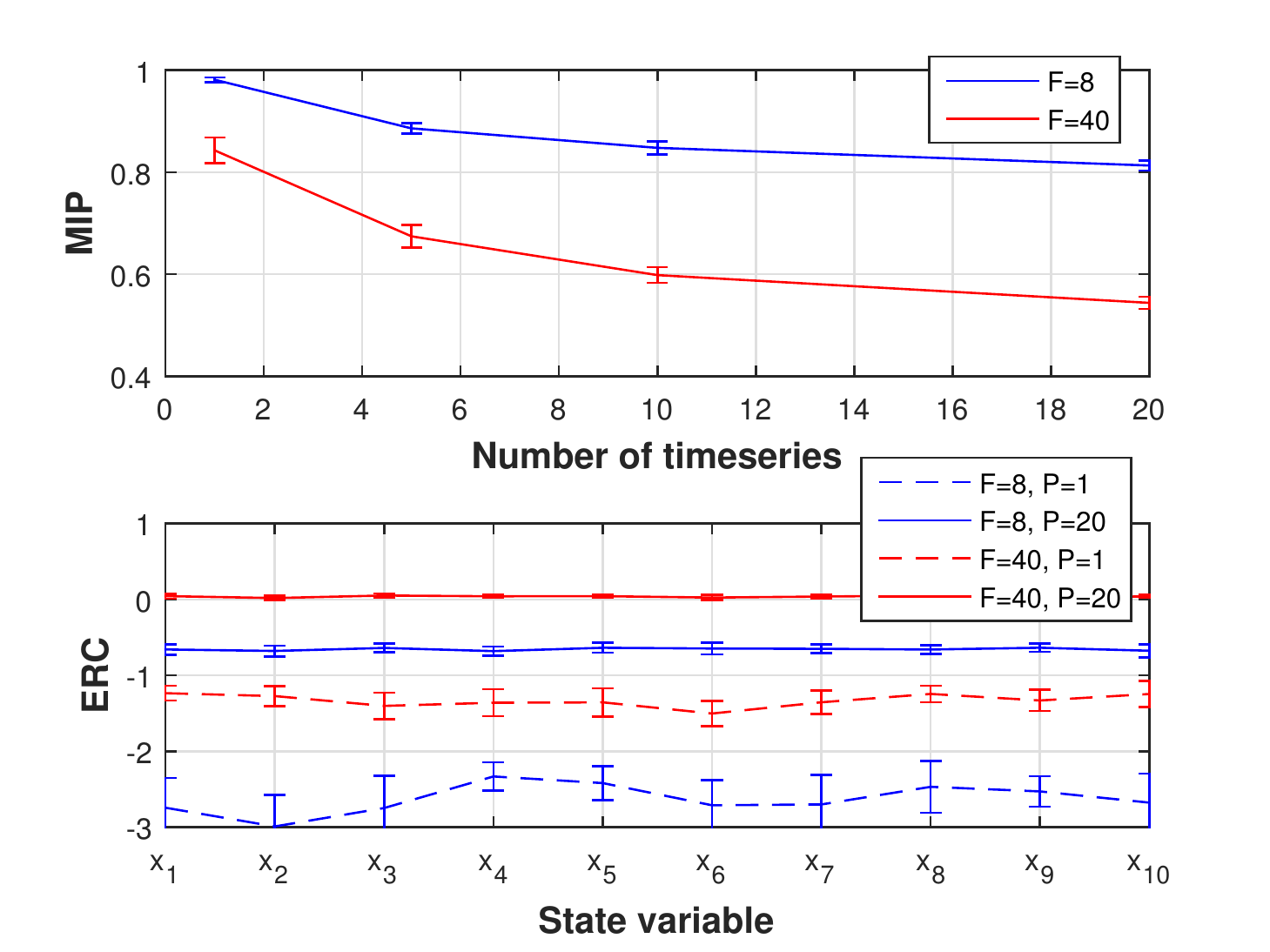}
		\label{mip:erc:lorenz96:Dt:0:1}}
	\caption{Results for the Lorenz96 model when time-series are sampled at the rate of $10 Hz$ instead of $1000 Hz$. (a) Precision (upper panel) and recall (lower panel) curves as a function of $P$, i.e., the number of trajectories for both USDL (solid lines) and SINDy (dashed lines) algorithms. The impact of sub-sampling by a factor of 100 is huge, especially for the USDL algorithm, resulting in bad performance of the inference methods primarily in terms of precision. As expected, inference is more accurate under weak forcing due to insufficient sampling in the strong forcing case. (b) RMSE as a function of the number of time-series for USDL (solid lines) and SINDy (dashed lines). Stronger forces results in deterioration of RMSE performance for both algorithms. (c) MIP (upper panel) and ERC per state variable (lower panel).
	}
	\label{lorenz96:fig:Dt:0:1}
\end{figure}

\begin{figure}[!htb]
	\centering
	\subfigure[F1 score with sampling frequency at $100Hz$.]{
	        	\includegraphics[width=.47\textwidth]{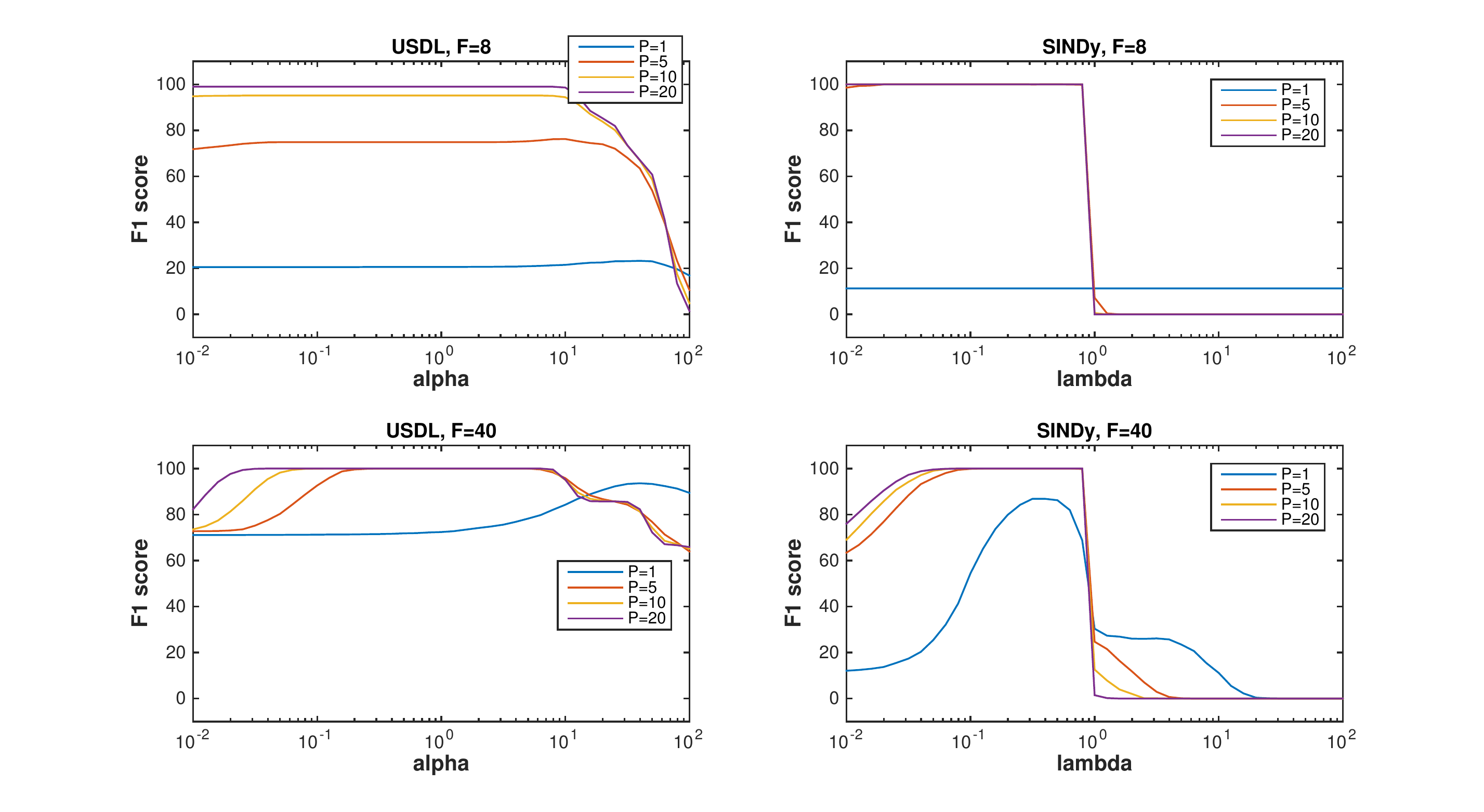}
		\label{lorenz96:tuning:Dt:0:01}}
	\quad
	\subfigure[F1 score with sampling frequency at $10Hz$.]{
		\includegraphics[width=.47\textwidth]{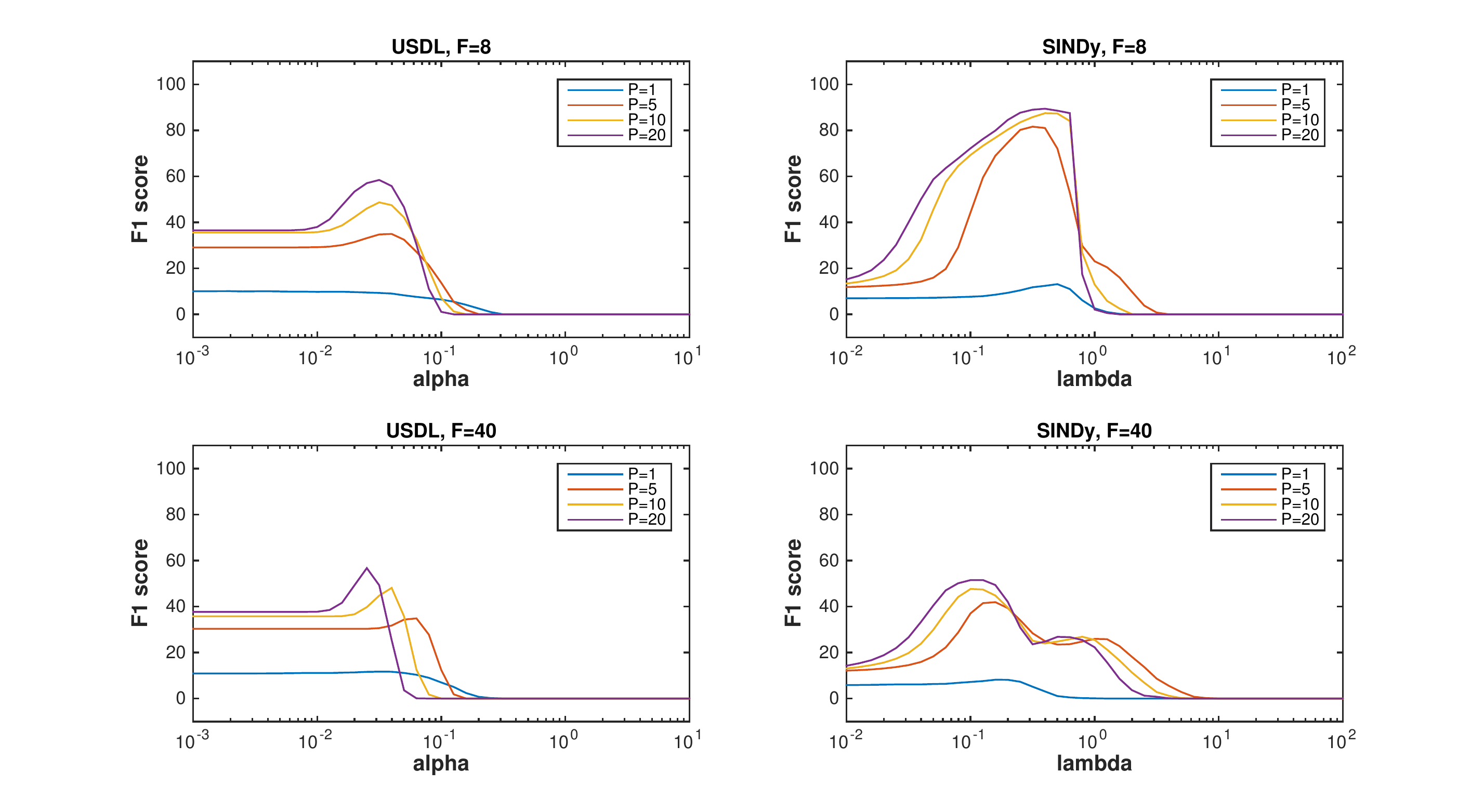}
		\label{lorenz96:tuning:Dt:0:1}}
	\caption{(a) The F1 score as a function of the hyperparameter for both USDL (left column) and SINDy (right column) when the sampling rate of the time-series is at $100Hz$. The region of optimal values for the hyperparameter has been decreased. (b) Same as (a) with sampling frequency at $10Hz$. F1 score deteriorates significantly in all cases.}
	\label{lorenz96:tuning:fig}
\end{figure}

The third variation enriches the dictionary (i.e., $\psi(x)$) by considering additional candidate functions as driving forces of the dynamical system. Thus, apart from linear and quadratic terms, we add the complete set of cubic combinations resulting in $Q=286$ dictionary atoms for the case of $N=10$ state variables. Figure~\ref{lorenz96:fig:cubic} presents the performance of both USDL (solid lines) and SINDy (dashed lines) algorithms upper plots) under this setup as well as the incoherence metrics (lower plots). Starting with the metrics of SSR theory, it is evident that MIP is increased for both weak and strong forcing reflecting the fact that the additional atoms increase the collinearity of the measurement matrix, $\Psi$. In contrast, ERC remains almost the same. In terms of performance, both precision and recall deteriorate for the case of weak forcing (blue curves in Figure~\ref{pre:rec:lorenz96:cubic}). However, as the number of time-series is increased, the performance of both algorithms is greatly improved from almost 0 to almost 100 percent. For the strong forcing case (red curves in Figure~\ref{pre:rec:lorenz96:cubic}), apart from the case where very few trajectories are provided, perfect reconstruction of the dynamical system is observed showing once again that strong chaos assists the dynamical model inference. In contrast to the previous experiments, USDL algorithm performs better when the number of trajectories is low while SINDy algorithm performs better when the number of trajectories is above $P=10$ where we observe perfect reconstruction of the dynamical system even for the weak force case (dashed blue lines in Figure~\ref{pre:rec:lorenz96:cubic}).

In all above variations, the RMSE performance were minimally affected by the different setups. The observed quantitative differences were in accordance with the precision-recall results. Moreover, we repeated the fine-tuning procedure for the hyperparameter values and the results were similar to Figure~\ref{F1:score:Lorenz96:fig} therefore we keep the same values for both algorithms' hyperparameter.

The final comparisons demonstrate the effect of the sampling frequency on the sparse inference algorithms. Up to now, we considered a sampling rate of $1000Hz$ meaning that we sampled the time-series every $0.001$ time units. Such a high sampling rate enabled us to assume that the time-series are accurate and merely continuous. In real applications though, high sampling frequency results in increased costs and typically there is a trade-off between expenses and sampling rate. Figure~\ref{lorenz96:fig:Dt:0:01} presents various performance metrics as well as both SSR incoherence metrics when sampling rate is dropped from $1000Hz$ to $100Hz$. Results are similar except for the strong forcing case with one trajectory where precision's performance decreases (red curves in upper panel of Figure~\ref{pre:rec:lorenz96:Dt:0:01}). The relative change of ERC (dashed red line in lower panel of Figure~\ref{mip:erc:lorenz96:Dt:0:01}) captures this deterioration. Notice that such deterioration is expected since strong forces result in larger and sharper modulations of the time-series. Again, SINDy algorithm achieves perfect reconstruction with less time-series compared to USDL algorithm. Overall, the performance is satisfactory when sampling frequency drops by a factor of 10.
However, if the sampling frequency drops by a factor of 100, performance is severely reduced for USDL algorithm (solid lines) and moderately reduced for SINDy algorithm (dashed lines) as Figure~\ref{pre:rec:lorenz96:Dt:0:1} asserts. The increase of RMSE shown in Figure~\ref{rmse:lorenz96:Dt:0:1} further confirms the deterioration of the inference methods. There are two major factors contributing to this behavior. First, numerical integration produce larger error and, second, the time-series are sampled below the Nyquist frequency which resulted in aliasing and heavy distortion of the signals. Consequently, the dynamical system inference failed in this case revealing the sensitivity of the proposed method to adequate sampling. Finally, we would like to remark that we optimized over the hyper-parameter values using F1 score as a measure of goodness. The results on F1 score for both cases are shown in Figure~\ref{lorenz96:tuning:fig} where it is also evidence the severe effects of over-downsampling (right plots). Moreover, the optimal parameter region has been shrunk implying that the inference methods are more sensitive to the hyperparameter value.


\begin{thebibliography}{10}

\bibitem{Newman2014}
M~E~J Newman.
\newblock {Networks: An introduction}.
\newblock {\em Oxford University}, pages 163--186, 2014.

\bibitem{Bonneau2006}
R.~Bonneau, D.~J. Reiss, P.~Shannon, M.~Facciotti, L.~Hood, N.~S. Baliga, and
  V.~Thorsson.
\newblock The inferelator: an algorithm for learning parsimonious regulatory
  networks from systems-biology data sets de novo.
\newblock {\em Genome biology}, 7(5):R36, 2006.

\bibitem{Gennemark2014}
P~Gennemark and D~Wedelin.
\newblock {ODEion — A SOFTWARE MODULE FOR STRUCTURAL IDENTIFICATION OF
  ORDINARY DIFFERENTIAL EQUATIONS}.
\newblock {\em Journal of Bioinformatics and Computational Biology},
  12(01):1350015, feb 2014.

\bibitem{Brunton2016}
Steven~L Brunton, Joshua~L Proctor, and J~Nathan Kutz.
\newblock {Discovering governing equations from data by sparse identification
  of nonlinear dynamical systems.}
\newblock {\em Proceedings of the National Academy of Sciences of the United
  States of America}, 113(15):3932--7, apr 2016.

\bibitem{Mangan2016}
Niall~M. Mangan, Steven~L. Brunton, Joshua~L. Proctor, and J.~Nathan Kutz.
\newblock {Inferring Biological Networks by Sparse Identification of Nonlinear
  Dynamics}.
\newblock {\em IEEE Transactions on Molecular, Biological and Multi-Scale
  Communications}, 2(1):52--63, jun 2016.

\bibitem{Klimovskaia2016}
Anna Klimovskaia, Stefan Ganscha, and Manfred Claassen.
\newblock {Sparse Regression Based Structure Learning of Stochastic Reaction
  Networks from Single Cell Snapshot Time Series}.
\newblock {\em PLOS Computational Biology}, 12(12):e1005234, dec 2016.

\bibitem{Tropp2006}
Joel~A. Tropp.
\newblock {Just relax: convex programming methods for identifying sparse
  signals in noise}.
\newblock {\em Information Theory, IEEE Transactions on}, 52(3):1030--1051,
  2006.

\bibitem{Tibshirani1994}
Robert Tibshirani.
\newblock Regression shrinkage and selection via the lasso.
\newblock {\em Journal of the Royal Statistical Society, Series B},
  58:267--288, 1994.

\bibitem{Gustafsson2009}
Mika Gustafsson, Michael H{\"{o}}rnquist, Jesper Lundstr{\"{o}}m, Johan
  Bj{\"{o}}rkegren, and Jesper Tegn{\'{e}}r.
\newblock {Reverse engineering of gene networks with LASSO and nonlinear basis
  functions.}
\newblock {\em Annals of the New York Academy of Sciences}, 1158:265--75, mar
  2009.

\bibitem{August2009}
Elias August and Antonis Papachristodoulou.
\newblock {Efficient, sparse biological network determination}.
\newblock {\em BMC Systems Biology}, 3(1):25, 2009.

\bibitem{Friedman2008}
Jerome Friedman, Trevor Hastie, and Robert Tibshirani.
\newblock {Sparse inverse covariance estimation with the graphical lasso}.
\newblock {\em Biostatistics}, 9(3):432--441, 2008.

\bibitem{Charbonnier2010}
Camille Charbonnier, Julien Chiquet, and Christophe Ambroise.
\newblock {Weighted-LASSO for structured network inference from time course
  data.}
\newblock {\em Statistical applications in genetics and molecular biology},
  9(1):Article 15, 2010.

\bibitem{Bolstad2011}
Andrew Bolstad, Barry~D. {Van Veen}, and Robert Nowak.
\newblock {Causal network inference via group sparse regularization}.
\newblock {\em IEEE Transactions on Signal Processing}, 59(6):2628--2641, 2011.

\bibitem{Friston2003}
K~J Friston, L~Harrison, and W~Penny.
\newblock {Dynamic causal modelling.}
\newblock {\em NeuroImage}, 19(4):1273--302, aug 2003.

\bibitem{Daniels2015}
B~C Daniels and I~Nemenman.
\newblock {Automated adaptive inference of phenomenological dynamical models}.
\newblock {\em Nature Communications}, 6:8, 2015.

\bibitem{Bento2010}
Jos\'{e} Bento, Morteza Ibrahimi, and Andrea Montanari.
\newblock Learning networks of stochastic differential equations.
\newblock In J.~D. Lafferty, C.~K.~I. Williams, J.~Shawe-Taylor, R.~S. Zemel,
  and A.~Culotta, editors, {\em Advances in Neural Information Processing
  Systems 23}, pages 172--180. Curran Associates, Inc., 2010.

\bibitem{DiStefano2015}
Joseph~J. DiStefano.
\newblock {\em {Dynamic systems biology modeling and simulation}}.
\newblock Academic Press, 2015.

\bibitem{Strang2008}
Gilbert Strang and George~J. Fix.
\newblock {\em {An analysis of the finite element method}}.
\newblock Wellesley-Cambridge Press, 2nd edition, 2008.

\bibitem{Davis1984}
M~E Davis.
\newblock {\em Numerical Methods and Modeling for Chemical Engineers}.
\newblock John Wiley \& Sons, 1984.

\bibitem{Candes2006}
Emmanuel~J. Cand{\`{e}}s, Justin~K. Romberg, and Terence Tao.
\newblock {Stable signal recovery from incomplete and inaccurate measurements}.
\newblock {\em Communications on Pure and Applied Mathematics},
  59(8):1207--1223, aug 2006.

\bibitem{Bruckstein2009}
Alfred~M. Bruckstein, David~L. Donoho, and Michael Elad.
\newblock {From Sparse Solutions of Systems of Equations to Sparse Modeling of
  Signals and Images}.
\newblock {\em SIAM Review}, 51(1):34--81, feb 2009.

\bibitem{Donoho2006}
D.L. Donoho.
\newblock {Compressed sensing}.
\newblock {\em IEEE Transactions on Information Theory}, 52(4):1289--1306,
  2006.

\bibitem{Foucart2013}
Simon Foucart and Holger Rauhut.
\newblock {\em {A Mathematical Introduction to Compressive Sensing}}.
\newblock Applied and Numerical Harmonic Analysis. Springer New York, New York,
  NY, 2013.

\bibitem{Pati1993}
Y.C. Pati, R.~Rezaiifar, and P.S. Krishnaprasad.
\newblock {Orthogonal matching pursuit: recursive function approximation with
  applications to wavelet decomposition}.
\newblock In {\em Proceedings of 27th Asilomar Conference on Signals, Systems
  and Computers}, pages 40--44. IEEE Comput. Soc. Press, 1993.

\bibitem{Davis1997}
G.~Davis, S.~Mallat, and M.~Avellaneda.
\newblock {Adaptive greedy approximations}.
\newblock {\em Constructive Approximation}, 13(1):57--98, mar 1997.

\bibitem{Tropp2007}
Joel~A. Tropp and Anna~C. Gilbert.
\newblock {Signal Recovery From Random Measurements Via Orthogonal Matching
  Pursuit}.
\newblock {\em IEEE Transactions on Information Theory}, 53(12):4655--4666, dec
  2007.

\bibitem{Donoho2001}
D.L. Donoho and X.~Huo.
\newblock {Uncertainty principles and ideal atomic decomposition}.
\newblock {\em IEEE Transactions on Information Theory}, 47(7):2845--2862,
  2001.

\bibitem{Sachs2005}
K~Sachs, O~Perez, D~Pe'er, DA~Lauffenburger, and GP~Nolan.
\newblock Causal protein-signaling networks derived from multiparameter
  single-cell data.
\newblock {\em Science}, 308:523--529, 2005.

\bibitem{Lorenz1996}
Edward~N. Lorenz.
\newblock Predictability: a problem partly solved.
\newblock In {\em Seminar on Predictability}, volume~1, pages 1--18, Shinfield
  Park, Reading, 1996. ECMWF.

\bibitem{Evans1998}
Lawrence~C Evans.
\newblock {\em {Partial Differential Equations}}, volume~19.
\newblock American Mathematical Society, 1998.

\bibitem{Lente2015}
G{\'{a}}bor Lente.
\newblock {\em {Deterministic Kinetics in Chemistry and Systems Biology}}.
\newblock SpringerBriefs in Molecular Science. Springer International
  Publishing, 2015.

\bibitem{Oksendal1985}
B.~Oksendal.
\newblock {\em Stochastic Differential Equations: {A}n introduction with
  applications}.
\newblock Springer-Verlag, 1985.

\bibitem{Ramsay2007}
J.~O. Ramsay, G.~Hooker, D.~Campbell, and J.~Cao.
\newblock {Parameter estimation for differential equations: A generalized
  smoothing approach}.
\newblock {\em Journal of the Royal Statistical Society. Series B: Statistical
  Methodology}, 69(5):741--796, 2007.

\bibitem{Zhan2011}
Choujun Zhan and Lam~F Yeung.
\newblock {Parameter estimation in systems biology models using spline
  approximation.}
\newblock {\em BMC systems biology}, 5(1):14, 2011.

\bibitem{Craven1978}
Peter Craven and Grace Wahba.
\newblock {Smoothing noisy data with spline functions - Estimating the correct
  degree of smoothing by the method of generalized cross-validation}.
\newblock {\em Numerische Mathematik}, 31(4):377--403, 1978.

\bibitem{Peifer2007}
M~Peifer and J~Timmer.
\newblock {Parameter estimation in ordinary differential equations for
  biochemical processes using the method of multiple shooting.}
\newblock {\em IET systems biology}, 1(2):78--88, mar 2007.

\bibitem{Mallat1993}
S.G. Mallat and {Zhifeng Zhang}.
\newblock {Matching pursuits with time-frequency dictionaries}.
\newblock {\em IEEE Transactions on Signal Processing}, 41(12):3397--3415,
  1993.

\bibitem{Tropp2004}
J.A. Tropp.
\newblock {Greed is Good: Algorithmic Results for Sparse Approximation}.
\newblock {\em IEEE Transactions on Information Theory}, 50(10):2231--2242, oct
  2004.

\bibitem{Cai2011}
T.~Tony Cai and Lie Wang.
\newblock {Orthogonal matching pursuit for sparse signal recovery with noise}.
\newblock {\em IEEE Transactions on Information Theory}, 57(7):4680--4688,
  2011.

\bibitem{Candes2005}
E.J. Candes and T.~Tao.
\newblock {Decoding by Linear Programming}.
\newblock {\em IEEE Transactions on Information Theory}, 51(12):4203--4215, dec
  2005.

\bibitem{Krishnaswamy2014}
Smita Krishnaswamy, Matthew~H. Spitzer, Michael Mingueneau, Sean~C. Bendall,
  Oren Litvin, Erica Stone, Dana Pe'er, and Garry~P. Nolan.
\newblock {Conditional density-based analysis of T cell signaling in
  single-cell data}.
\newblock {\em Science}, 346(6213), 2014.

\bibitem{Bodenmiller2012}
Bernd Bodenmiller, Eli~R Zunder, Rachel Finck, Tiffany~J Chen, Erica~S Savig,
  Robert~V Bruggner, Erin~F Simonds, Sean~C Bendall, Karen Sachs, Peter~O
  Krutzik, and Garry~P Nolan.
\newblock {Multiplexed mass cytometry profiling of cellular states perturbed by
  small-molecule regulators.}
\newblock {\em Nature biotechnology}, 30(9):858--67, 2012.

\bibitem{Kanehisa2000}
M~Kanehisa and S~Goto.
\newblock {KEGG: kyoto encyclopedia of genes and genomes.}
\newblock {\em Nucleic acids research}, 28(1):27--30, jan 2000.

\bibitem{Papoutsoglou2017}
G~Papoutsoglou, G~Athineou, V~Lagani, I~Xanthopoulos, A~Schmidt, S~{\'{E}}lias,
  J~Tegn{\'{e}}r, and I~Tsamardinos.
\newblock Scenery: a web application for (causal) network reconstruction from
  cytometry data.
\newblock {\em Nucleic Acids Research}, 45:W270--W275, 2017.

\bibitem{Bartoszek2012}
Krzysztof Bartoszek, Jason Pienaar, Petter Mostad, Staffan Andersson, and
  Thomas~F. Hansen.
\newblock {A phylogenetic comparative method for studying multivariate
  adaptation}.
\newblock {\em Journal of Theoretical Biology}, 314:204--215, 2012.

\bibitem{Gardiner2004}
Crispin Gardiner.
\newblock {\em {Handbook of stochastic methods: for physics, chemistry {\&} the
  natural sciences}}.
\newblock Springer-Verlag, 2004.

\bibitem{Gardiner2009}
Crispin Gardiner.
\newblock {Stochastic Methods: A Handbook for the Natural and Social Sciences},
  2009.

\bibitem{Guyon2003}
Isabelle Guyon, Andr{\'{e}} Elisseeff, and Andre@tuebingen~Mpg De.
\newblock {An Introduction to Variable and Feature Selection}.
\newblock {\em Journal of Machine Learning Research}, 3:1157--1182, 2003.

\bibitem{Bar-Joseph2012}
Ziv Bar-Joseph, Anthony Gitter, and Itamar Simon.
\newblock {Studying and modelling dynamic biological processes using
  time-series gene expression data}.
\newblock {\em Nature Reviews Genetics}, 13(8):552--564, jul 2012.

\bibitem{Mikosch2003}
Thomas Mikosch.
\newblock {\em {Elementary stochastic calculus with finance in view}}.
\newblock World Scientific, 2003.

\bibitem{Natarajan1995}
B.~K. Natarajan.
\newblock {Sparse Approximate Solutions to Linear Systems}.
\newblock {\em SIAM Journal on Computing}, 24(2):227--234, apr 1995.

\bibitem{Zhang2015}
Zheng Zhang, Yong Xu, Jian Yang, Xuelong Li, and David Zhang.
\newblock {A Survey of Sparse Representation: Algorithms and Applications}.
\newblock {\em IEEE Access}, 3:490--530, 2015.

\bibitem{Tibshirani1996}
Robert Tibshirani.
\newblock {Regression selection and shrinkage via the lasso}.
\newblock {\em Journal of the Royal Statistical Society B}, 58(1):267--288,
  1996.

\bibitem{Efron2004}
B.~Efron, T.~Hastie, I.~Johnstone, and R.~Tibshirani.
\newblock {Least angle regression}.
\newblock {\em Annals of Statistics}, 32(2):407--499, 2004.

\bibitem{Gutenkunst2007}
R.~N. Gutenkunst, J.~J. Waterfall, F.~P. Casey, K.~S. Brown, C.~R. Myers, and
  J.~P. Sethna.
\newblock Universally sloppy parameter sensitivities in systems biology models.
\newblock {\em PLOS Computational Biology}, 3, 2007.

\end{thebibliography}
\end{document}